\documentclass[prd,tightenlines,nofootinbib,showpacs,preprintnumbers,superscriptaddress, longbibliography, notitlepage]{revtex4-1}

\usepackage{amsfonts,amsmath,amssymb,amsthm,bbm,hyperref}
\usepackage{graphicx}
\usepackage{color}

\newcommand{\R}{{\mathbb R}}

\newcommand{\be}{\begin{equation}}
\newcommand{\ee}{\end{equation}}
\newcommand{\beq}{\begin{eqnarray}}
\newcommand{\eeq}{\end{eqnarray}}
\newcommand{\nn}{\nonumber}

\newcommand{\la}{\label}
\newcommand{\bx}{\textbf{x}}
\newcommand{\tb}{\textbf}

\usepackage{amssymb}

\extrafloats{100}

\DeclareMathOperator\arctanh{arctanh}

\DeclareUnicodeCharacter{2212}{-}

\begin{document}

\title{Quantum cosmology with third quantisation}
\author{Salvador J. Robles-Pérez}
\email{sarobles@math.uc3m.es}
\affiliation{Departamento de Matemáticas, Universidad Carlos III de Madrid. Avda. de la Universidad 30,  28911 Leganés, Spain.}

\date{\today}% It is always \today, today,

\begin{abstract}
We review  the canonical quantisation of the geometry of the spacetime in the cases of a simply and a non-simply connected manifold. In the former, we analyse the information contained in the solutions of the Wheeler-DeWitt equation and interpret them in terms of the customary boundary conditions that are typically imposed on the semiclassical wave functions. In particular, we review three different paradigms for the quantum creation of a homogeneous and isotropic universe. For the quantisation of a non-simply connected manifold the best framework is the so-called third quantisation formalism, in which the wave function of the universe is seen as a field that propagates in the space of Riemannian $3$-geometries, which turns out to be isomorphic to a (part of a) $1+5$ Minkowski spacetime. Thus, the quantisation of the wave function follows the customary formalism of a quantum field theory. A general review of the formalism is given and it is analysed the creation of the universes, their initial expansion and the appearance of matter after inflation. These features are presented in more detail in the case of a homogeneous and isotropic universe. The main conclusion in both cases is that the most natural way in which the universes should be created is in entangled universe-antiuniverse pairs.
\end{abstract}


\maketitle

% Keywords
%\keyword{quantum cosmology; multiverse; superspace; third quantisation; universe-antiuniverse pair;} 

\tableofcontents

\section{Introduction}\label{sec01}

Quantum cosmology is the application of the quantum theory to the universe as a whole. However, it was clear from the beginning that the customary formalism of the Copenhagen interpretation cannot be applied to the quantisation of the universe because the Schrödinger equation and the measurement process in which the Copenhagen formalism is based on cannot be fundamental elements of a quantum theory of the spacetime. Let us notice that if the quantum theory must represent the quantum state of the spacetime, then, as Wheeler showed \cite{Wheeler1957},  its quantum state must be at the Planck length a superposition of spacetime geometries that is impossible to visualise or represent and where one cannot even \emph{use the word ''observation'' at all} (cf. \cite{Wheeler1957}). However, as we approach the macroscopic scale the quantum state of the universe must represent the approximately stable spacetime where we live and  perform measurements of particles and other matter fields. It means that the description of the universe that we observe must be an emergent feature of the quantum representation of the universe.

In this paper, we review the canonical formulation of quantum cosmology. We start from the foliation of the spacetime into space and time that allows us to express the Einstein-Hilbert action as a functional action of the components of the $3$-metric of the spatial sections of the spacetime. The evolution of the universe turns out to be then a trajectory in the space of $3$-Riemannian metrics, $M$, and its quantum state is represented by a wave function that is the solution of the Wheeler-DeWitt equation that, in principle, contains all the information about the spacetime and the matter fields that propagate therein. However, as we have already said, the full quantum state is a superposition of solutions that correspond to different paths, i.e. different evolutions, in the space $M$. It is only in the semiclassical regime where a particular kind of solutions emerge by a decoherence process\footnote{We shall not deal here with such processes of decoherence, which can be seen in the bibliography \cite{Kiefer1987, Halliwell1989, Hartle1990, GellMann1990, Kiefer1992}.}. This kind of solutions are the semiclassical solutions that represent a fixed classical spacetime background with matter fields propagating therein, and in which the Schrödinger equation appears as an approximated equation  at order $\hbar^1$. In particular, we analyse the semiclassical wave function of a homogeneous and isotropic universe with small inhomogeneities that can be treated as perturbations. In that case, explicit solutions can be given for which it is easier to analyse the different boundary conditions that can be imposed on the state of the universe. They give rise to different scenarios for the creation of the universe that are analysed in detail.

On the other hand, the space of $3$-dimensional space-like metrics, $M$, defined at any point in the space, turns out to be isomorphic to a $1+5$ dimensional Minkowski spacetime\footnote{More exactly, it is isomorphic to a Milne spacetime, which is a particular coordination of the light cones of the Minkowski spacetime.}. This analogy between the space $M$ and the spacetime allows us to consider the wave function of the universe as a field that propagates in $M$, and the Wheeler-DeWitt equation as the field equation. In that case, a procedure of quantisation called \emph{third quantisation} can formally be performed in a similar way as it is done in a quantum field theory. For instance, we can define quantum operators representing the creation and annihilation of particular modes of the spacetime, i.e. different universes, and the corresponding Fock space will then allow us to represent the quantum state of  a multiply connected spacetime manifold. It turns out to be then the appropriate framework to describe the quantum state of the multiverse. Moreover, as it happens in a quantum field theory, the isotropy of the background space implies that the creation of universes must be in pairs with opposite values of the components of the momentum conjugated to the configuration variables. We shall analyse the charge and parity relation between the matter fields that propagate in one of these pairs to see that the matter content of one of the universes must be CP inversely related with the content of the other universe. They thus form a universe-antiuniverse pair. We analyse this pair creation in detail in the case of a homogeneous and isotropic universe where the period of reheating after inflation is investigated. The decay of the inflaton field into the particles of the Standard Model is produced in a CP conjugated way in the two universes so any excess of matter over antimatter in one of the universes of the entangled pair is balanced with the excess of the antimatter over matter in the partner universe, having always these two concepts (matter and antimatter) a relative meaning, i.e. an internal observer in any of the universe always interprets the content of his/her universe as matter.

This proposal is however still far from being directly testable. The effects of quantum cosmology and, in particular, the effects of the existence of an entangled universe are mainly restricted to the very early stage of the universe. Perhaps with the future advances in the detection of gravitational waves or of a cosmic neutrino background we will be able to test the pre-inflationary stage of the universe where the effects of quantum gravity may be significant. Moreover, the third quantisation formalism can also be a proposal for the quantisation of the spacetime and thus a better understanding of the formalism and its application to other gravitational scenarios can provide us with a new line of research for the search of a quantum theory of gravity.

%%%%%%%%%%%%%%%%%%%%%%%%%%%%%%%%%%%%%%%%%%

\section{Quantisation of a simply connected spacetime manifold}\label{sec02}

\subsection{Quantisation of the spacetime geometry}\label{sec0201}

\begin{figure}
\centering
\includegraphics[width=6 cm]{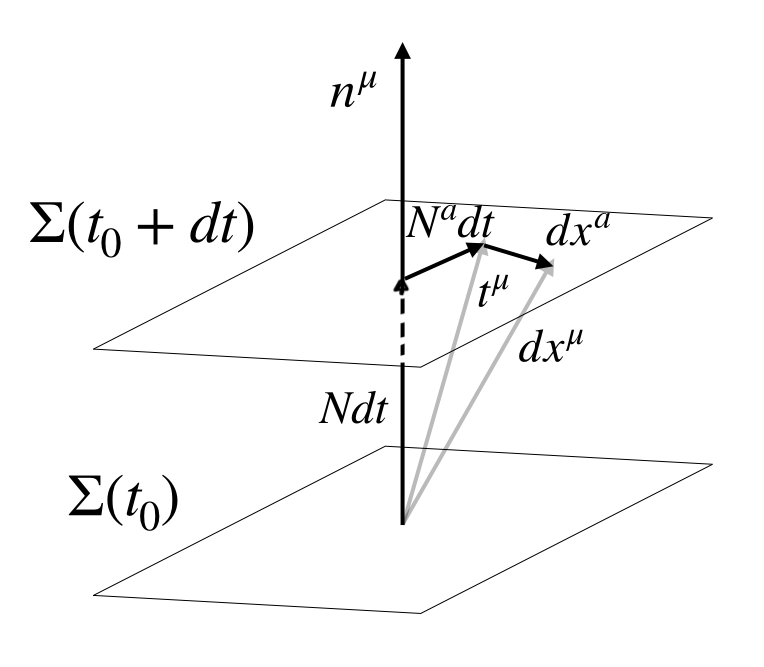}
\caption{Splitting the spacetime into space and time.}
\label{figure101}
\end{figure}

Following the customary approach\footnote{I shall closely follow Refs. \cite{Kiefer2007, Wiltshire2003}.}, the spacetime can be foliated in space and time by assuming a global time function $t$ such that each surface $t=constant$ is a spacelike Cauchy hypersurface, $\Sigma_t$. The proper distance between the point $x_0$ in $\Sigma_t$ and the point $x_0+dx$ of $\Sigma_{t+dt}$ is given by \cite{Wiltshire2003, Kiefer2007} (see, Fig. \ref{figure101})
\be\label{STfol}
ds^2 = g_{\mu \nu} dx^\mu dx^\nu = \left( N_a N^a - N^2 \right) dt^2  + 2  N_a  dx^a  dt + h_{ab} dx^a dx^b ,
\ee
where $N_a N^a = h_{ab} N^a N^b$, and $h_{ab}$ is the three-dimensional metric induced on each hypersurface $\Sigma_t$, with unit normal $n_\mu$, satisfying, $n^\mu n_\mu = -1$. The functions $N$ and $N^a$ are called the lapse and the shift functions, respectively. They are the normal and tangential components of the vector field $t^\mu$, which is the vector field that transport the point $x_0$ from $\Sigma_t$ to $\Sigma_{t+dt}$.

\begin{figure}
\centering
\includegraphics[width=12 cm]{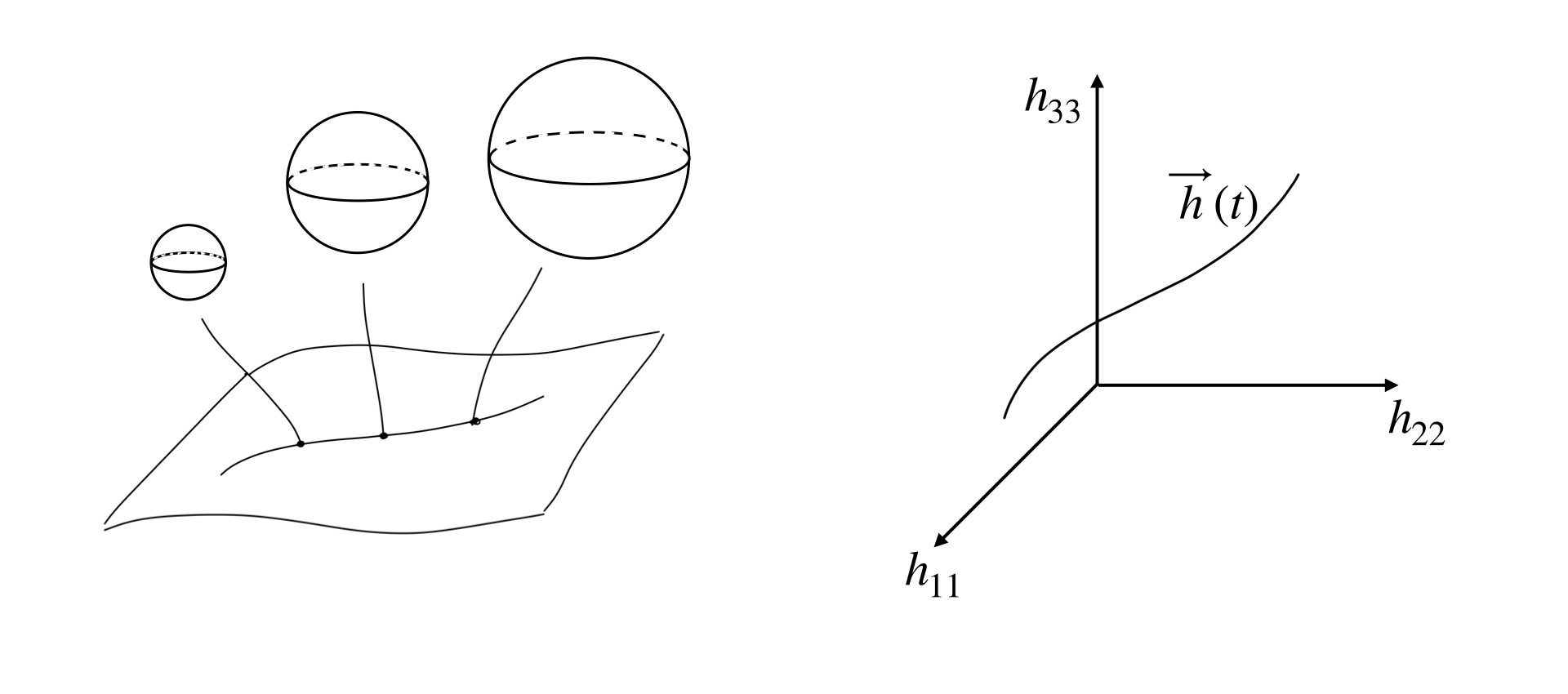}
\caption{Left: the evolution of the universe can be seen as a path in the abstract space spanned by the component of the spatial metric tensor, $h_{ab}$. Right: an example of evolution is depicted for the case that the spatial metric is diagonal, i.e. given by, $h_{ab}(t) = {\rm diag}(h_{11}(t), h_{22}(t), h_{33}(t))$.}
\label{figure102}
\end{figure}

With the split of the spacetime in space and time the spacetime can be seen as a spacelike hypersurface evolving in time. The geometry of the hypersurface $\Sigma_t$ at a given time $t_0$ is determined by the metric tensor $h_{ab}(t_0)$, so eventually the evolution of the universe is encoded in the evolution of the metric $h_{ab}(t)$. It is then remarkable that in the end, as Wheeler says \cite{Wheeler1968}, \emph{Eintein's geometrodynamics deals with the dynamics of $3$-geometry, not $4$-geometry!} (emphasis his). From this $3+1$ viewpoint of the spacetime, we can cast the Einstein-Hilbert action\footnote{Following Ref. \cite{DeWitt1967}, we are going to use throughout units in which, $\hbar = c = 16 \pi G = 1$, although we will leave the constant $\hbar$ in some expressions to remark their quantum character.} \cite{Kiefer2007}
\be\label{EHA01}
S_{EH}  \int_{\mathcal{M}} d^4x \sqrt{-g} \left( ^{4}R - 2 \Lambda \right) - 2 \int_{\partial\mathcal{M}} d^3x \sqrt{h} K  ,
\ee
and the action of the matter fields\footnote{For the matter fields we shall generally consider a scalar field.}
\be\label{SMA01}
S_{\rm matter} = \int_\mathcal{M} d^4x \sqrt{- g} \left( \frac{1}{2} g^{\mu\nu} \partial_\mu \varphi \partial_\nu \varphi - V(\varphi) \right) ,
\ee
into the standard Lagrangian form, 
\be\label{ACT01}
S = \int dt \, L(q^i, \dot q^i, t) ,
\ee
where $q_i$ and $\dot q_i$ will be here the spatial metric, $h_{ab}(x)$, the matter field(s), encoded in the variable $\varphi(x)$, and their corresponding velocities. In \eqref{EHA01}, $^{4}R$ is the Ricci scalar associated to the $4$-dimensional metric $g_{\mu\nu}$, $\Lambda$ is the cosmological constant, $K = h_{ab} K^{ab}$, is the trace of the extrinsic curvature, which can be written as,
\be\label{EC01}
K_{ab} = \frac{1}{2 N} \left( \dot h_{ab} - D_a N_b - D_b N_a \right) ,
\ee
where, $D_a$ is the covariant derivative on the spatial section $\Sigma_t$, and $\partial \mathcal M$ is the boundary of the manifold $\mathcal M$.  After some manipulation (see Ref. \cite{Kiefer2007}), one finds that the Einstein-Hilbert action (\ref{EHA01}) can be written as
\be\label{EHA02}
S_{EH} =  \int_\mathcal{M} dt d^3x N \left( G^{abcd} K_{ab} K_{cd} + \sqrt{h} \left( {}^3R -  2 \Lambda \right)  \right) ,
\ee
where $h$ is the determinant of the spatial metric $h_{ab}$, and
\be\label{DWM02}
G^{abcd} = \frac{\sqrt{h}}{2} \left( h^{ac} h^{bd} + h^{ad} h^{bc} - 2 h^{ab} h^{cd} \right) ,
\ee
is the so-called DeWitt's metric \cite{DeWitt1967}. The structure of the action (\ref{EHA02}) is very interesting. First, it is of the standard form \eqref{ACT01},
\be 
S_{EH} = \int dt \ L_{EH} = \int dt d^3x \ \mathcal  L_{EH} ,
\ee
with $L_{EH}$ the Lagrangian associated to the Einstein-Hilbert action (\ref{EHA02}) and $\mathcal L_{EH}$ the \emph{Lagrangian density}. Second, from (\ref{EC01}) one can see that the extrinsic curvature $K_{ab}$ contains the time derivative of the metric tensor $h_{ab}$, and  ${}^3R$ only depends on $h_{ab}$. Thus, the action \eqref{EHA02} presents the customary structure of a kinetic term that is quadratic in the \emph{velocities} plus a potential term. Furthermore, the supermetric \eqref{DWM02} defines a metric structure on the space of spacelike metrics, called the superspace\footnote{We shall be more precise later on.}. Thus, the action \eqref{EHA02} looks like the action of a particle that moves in a curved space, the coordinates of the ''particle'' are the time dependent values of the components of the metric tensor, $h_{ab}(t)$, and the curved space where this particle moves in the space of symmetric Riemannian $3$-metrics. That is, the evolution of the universe can be seen as the \emph{trajectory} in the superspace\footnote{Spacetime becomes a 'trajectory of spaces', cfr. p 107, Ref. \cite{Kiefer2007}.} (see, Fig. \ref{figure102}) .

The momenta conjugated to the metric tensor components, $h_{ab}$, are given by \cite{Kiefer2007}
\be
p^{ab} \equiv \frac{\partial \mathcal L_{EH}}{\partial \dot h_{ab}} =  G^{abcd} K_{cd} = \sqrt{h} \left( K^{ab} - K h^{ab} \right) .
\ee
Thus, in terms of the momenta, the total action (the gravitational action plus the action of the matter field) can be written as
\be\label{ACT06}
S = S_{EH} + S_{\rm matter} = \int dt d^3x \left( p^{ab} \dot h_{ab} +p_\varphi \dot\varphi - {N} \mathcal{H} - N^a \mathcal{H}_a \right) ,
\ee
{where} the lapse and {the} shift functions act as Lagrange multipliers, {with} \cite{Kiefer2007}
\beq\label{HC00}
\mathcal H &=&  G_{abcd} p^{ab} p^{cd} - \sqrt{h} \left( ^{3}R - 2 \Lambda \right) + \mathcal{H}_{\rm matter} , \\ \label{MC00}
\mathcal H_a &=& - 2 D_b p_a^{b} + \sqrt{h} J_a ,
\eeq
where \cite{Kiefer2007}, $J_a \equiv h_a^\mu T_{\mu \nu} n^\nu$, and $G_{abcd}$ is the inverse of the DeWitt metric \eqref{DWM02}, 
\be\label{DWM01}
G_{abcd} = \frac{1}{2 \sqrt h}  \left( h_{ac} h_{bd} + h_{ad} h_{bc} - h_{ab} h_{cd} \right) ,
\ee
with \cite{DeWitt1967},
\be
G^{abcd} G_{cd ef} = \frac{1}{2} \left( \delta^a_e \delta^b_f  + \delta^a_f \delta^b_e  \right)  .
\ee
Therefore, variation of the action \eqref{ACT06} with respect to the lapse and the shift functions yield{s} the classical Hamiltonian and momentum constraints, respectively, i.e.,
\be\label{CC01}
\mathcal H = 0 \,\,\, , \,\,\, \mathcal H_a = 0 .
\ee
Let us now focus for a moment on the gravitational part of these constraints. The Hamiltonian constraint, $H = 0$, can be seen as the analogue of the momentum constraint of a particle that propagates in the spacetime,
\be
g^{\mu \nu } p_\mu p_\nu + m^2 = 0 ,
\ee
in which the mass is substituted by a non constant potential (this analogy will be further exploited in Sec. \ref{sec03}). On the other hand, the function $\mathcal H_a$ generates infinitesimal diffeomorphisms (change of coordinates) in the spatial hypersurfaces, $\Sigma$. Thus, the second constraint in \eqref{CC01}, $\mathcal H_a = 0$, means that the Einstein-Hilbert action \eqref{EHA02} is invariant under such diffeomorphisms, which turns out to be like a gauge freedom \cite{Kiefer2007}. Thus, the real configuration space is the quotient space of all Riemannian $3$-metrics, $M \equiv {\rm Riem}(\Sigma)$, in which all three metrics related by diffeomorphisms correspond to the same class, i.e.
\be
\mathcal S(\Sigma) = \frac{M }{{\rm Diff}(\Sigma) } ,
\ee
which is called the \emph{superspace} \cite{DeWitt1967, Wheeler1968, Wiltshire2003, Kiefer2007}.

Following Dirac \cite{Dirac1964}, the canonical procedure of quantisation consists in assuming the quantum version of the classical constraints \eqref{CC01} by promoting the classical variables and their conjugate momenta into quantum operators and apply them to a wave function, $\phi$, that is defined in the configuration space, 
\be\label{QC01}
\hat{ \mathcal H} \phi = 0 \,\,\, , \,\,\, \hat{\mathcal H}_a \phi = 0 .
\ee
The quantum version of the momentum constraint in (\ref{CC01}), $\hat{\mathcal H}_a\phi=0$, assures that the wave function $\phi$ is invariant under spatial diffeomorphisms in the $3$-dimensional slices $\Sigma_t$ \cite{Kiefer2007}. For our purposes, it is much more interesting the constraint, $\hat{\mathcal H} \phi = 0$, which under canonical quantisation becomes the so-called Wheeler--DeWitt equation \cite{DeWitt1967, Wiltshire2003, Kiefer2007},
\be\label{WDW00}
\left( - \hbar^2 \nabla \cdot \vec{\nabla} +\sqrt{h} \left(  -\, ^{(3)}R + 2 \Lambda +  \hat T^{00} \right) \right) \phi(h_{ab}, \varphi) = 0 ,
\ee
where, $\nabla \cdot$ and $\vec{\nabla}$, are the divergence and the gradient, respectively, defined in the space of $3$-dimensional Riemannian metrics, $M$,  and for a scalar field $\hat T^{00}$ reads
\be
\hat T^{00} = \frac{-\hbar^2}{2 h} \frac{\delta^2}{\delta \varphi^2} + \frac{1}{2} h^{ij} \varphi_{,i} \varphi_{,j} + V(\varphi) .
\ee
The Wheeler-DeWitt equation \eqref{WDW00} is the keystone of the canonical formalism of quantum cosmology. The solution, $\phi(h_{ab}, \varphi)$, is usually called the \emph{wave function of the universe} \cite{Hartle1983} because it represents the quantum state of the spacetime and the matter fields that propagate therein. This is usually applied to the case of a single universe. However, as we shall see in Sect. \ref{sec03}, the whole spacetime manifold can present a more complicated structure and represent something more than what is typically called a universe. In that sense, the name can be misleading. However, for historical reasons we shall retain sometimes the name \emph{wave function of the universe} even in the cases where it may represent the state of many different universes.

The Wheeler-DeWitt equation \eqref{WDW00} can be seen as a Schr\"odinger like equation with no time variable, which is a consequence of the invariance of the quantum state of the universe with respect to the time variable. In that sense, there is no preferred time in the quantum description of the universe (as well as there are no preferred spatial coordinates because the constraint, $\mathcal H_a = 0$).  It is then sometimes stated that there is no time evolution of the quantum state of the universe. However, this is not true or at least it is not accurate. As we have already pointed out, the evolution of the universe can be seen as the trajectory in the superspace. The spacetime coordinates are the parameters that parametrise the trajectory, which is therefore invariant under reparametrisations but that does not mean that there is no evolution. It is similar to the description of the path followed by a particle in the spacetime, which is independent of the parametrisation of the path but that does not mean that the particle does not move.

\begin{figure}
\centering
\includegraphics[width=15 cm]{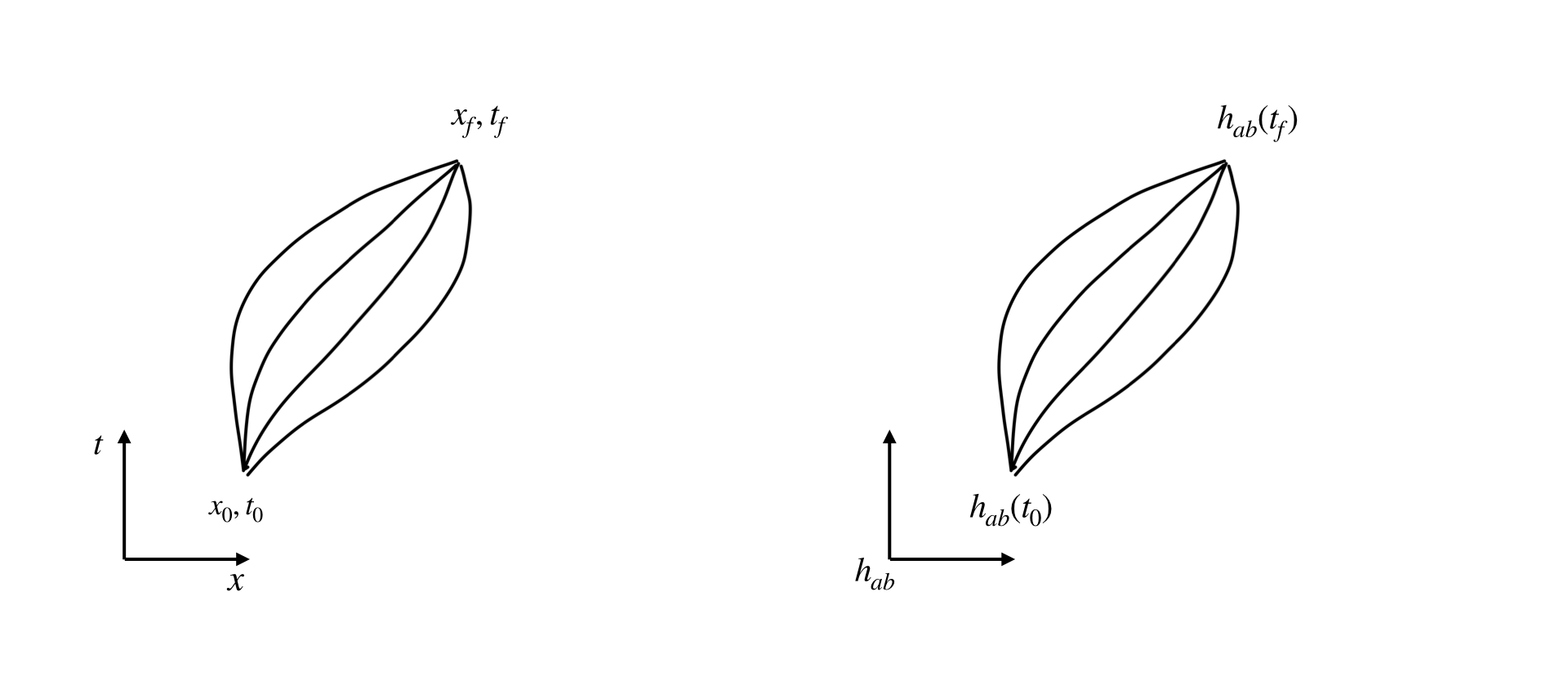}
\caption{Left: path integral in spacetime; Right: path integral in quantum gravity.}
\label{figure103}
\end{figure}

There is also a path integral approach to quantum gravity and quantum cosmology \cite{Hartle1983, Kiefer2007, Wiltshire2003}. It is a generalisation of Feynman's idea that the amplitude for a particle to go from one to another point  point  is given by a functional integral that weights all the paths that start from the point ${\bf x}_0$ at time $t_0$ and end in the point ${\bf x}_1$ at $t_1$ (see Fig. \ref{figure103}).  Following a parallel reasoning, Hartle and Hawking propose \cite{Hartle1983} that the amplitude for the universe to change from the hypersurface $\Sigma$, in which the spatial geometry and the field configuration are given by $h_{ab}$ and $\varphi$, respectively, to the hypersurface $\Sigma'$, where they are given by the values $h'_{ab}$ and $\varphi'$, is given by \cite{Hartle1983} 
\be
\langle h'_{ab}, \varphi' | h_{ab}, \varphi \rangle = \int \delta g \, \delta \varphi \; e^{i S[g, \varphi]} ,
\ee
where the integral must be performed over all $4$-geometries and field configurations that match the given values on the two spacelike hypersurfaces \cite{Hartle1983}, $\Sigma$ and $\Sigma'$. Following that approach, the wave function of the universe is then given by,
\be\label{PIF01}
\phi(h_{ab}, \varphi) = \sum \int_C \delta g \delta\varphi \, e^{i S(g,\varphi)} ,
\ee
where $C$ denotes the class of spacetimes and matter configurations that fulfil the boundary requirements on the hypersurfaces $\Sigma$, and the sum is performed over all kind of topologies. In order to make well defined the path integral in (\ref{PIF01}) one has to rotate to Euclidean time. However, that does not remove all the technical problems, which we are not going to deal with here. In practice, as it happens with the Wheeler-DeWitt equation, the path integral can only be performed for spacetimes and matter field configurations with a high degree of symmetry. Moreover, both formulations become equivalent because the requirement of invariance of the wave function $\phi(h_{ab}, \varphi)$ under reparametrizations of the time variable implies the constraint \cite{Hartle1983}, $\frac{\delta S}{\delta N}=0$, whose quantised version is the Wheeler-DeWitt equation.

Anyway, the path integral formulation has two interesting points. First, the analogy with the Feynman's path integral formulation of the trajectory of a particle in spacetime makes it very intuitive. As we have seen, the classical evolution of the universe can be seen as a path in the superspace. Applying Feynman's idea to gravity means that the quantum state of the universe is given by a quantum superposition of all the paths that go from one to another configuration of the spatial hypersurfaces. As it happens in the spacetime, the \emph{classical path} (i.e. the classical evolution) emerges in some specific limit because the constructive interference between the paths of the quantum superposition, and in the same limit the non-classical paths suffer from a destructive interference or \emph{decoherence} \cite{Halliwell1989, Kiefer1992, Joos2003, Schlosshauer2007}. The second interesting point of the path integral approach is the following: let us assume that we already know how to construct the quantum amplitude for the universe to go from one to another given configuration of the spacelike section. Then, what is it the amplitude for the birth of the universe? What are the boundary conditions that one must impose on the state of the universe to obtain the appropriate probability amplitude for the universe to be created?

\subsection{Boundary conditions}\label{sec0202}

In classical mechanics, as well as in quantum mechanics, we usually work with some given conditions that we know or assume for certain at some initial time. Then,  knowing the law of evolution, the initial conditions determine the state of the system at any later time. In the universe the thing is a bit different. What we only know for certain is the state of the observable universe, say from the inflationary period\footnote{Let us assume that the period of inflation is fully supported by the current observational data.} to the current stage of accelerated expansion, and we have to make some guess about the initial conditions that give rise a universe like that.  But this is all classical cosmology. The question in quantum cosmology is, what are the conditions at the quantum level that give rise a specific initial boundary, $\Sigma_0$, that is propitious to inflate? The probability for the creation of such universe would be given by the modulus squared of the wave function of the universe, when this is evaluated at the initial hypersurface\footnote{Let us notice however that the creation of this initial boundary hypersurface $\Sigma_0$ is not a process occurring in time but it corresponds to the creation of the spacetime itself \cite{Kiefer2007}, actually.}. From that point of view, one can say that the initial state of the (classical) universe is the final state of the amplitude for the universe to be created ... \emph{from what}? And that is actually the issue behind the question of the boundary condition of the universe.

\begin{figure}
\centering
\includegraphics[width=13 cm]{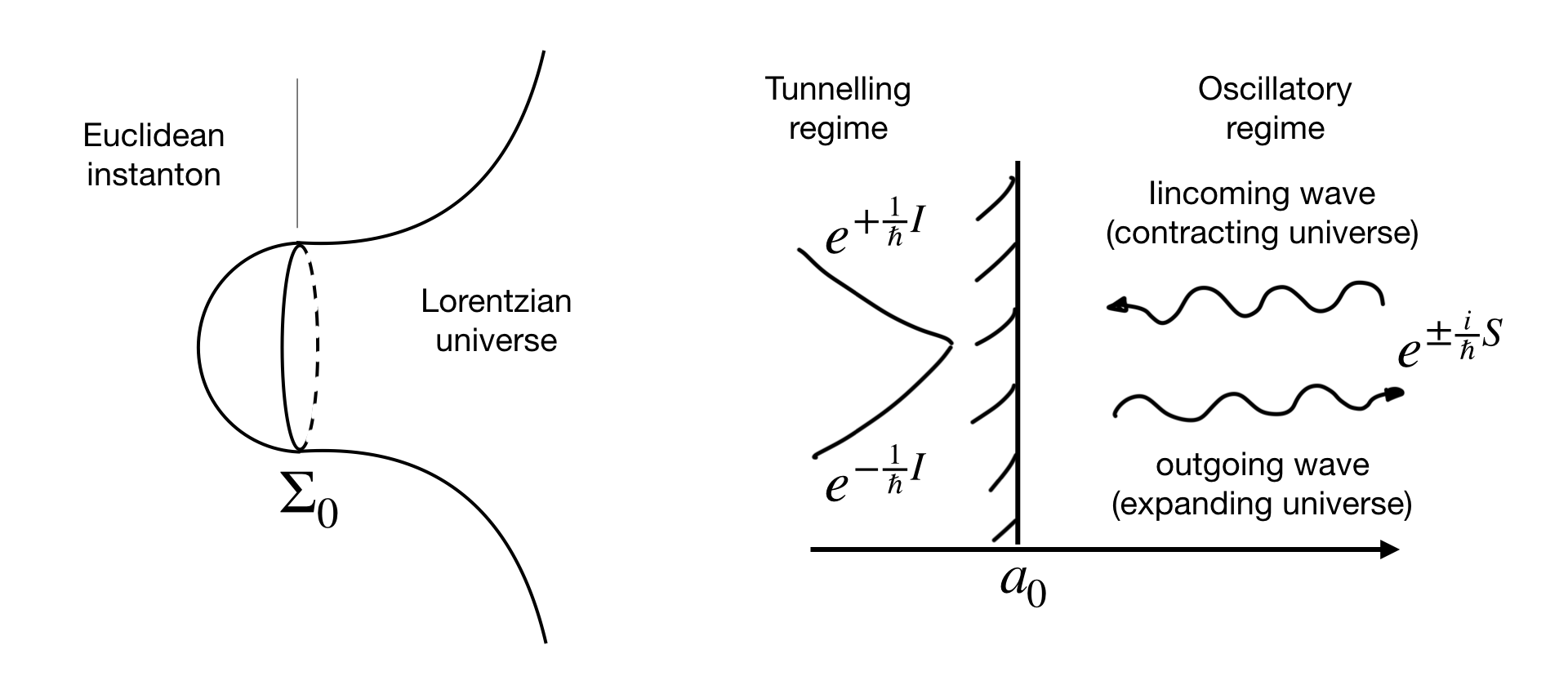}
\caption{Left: the class $C$ over which the path integral has to be performed is the class of compact geometries that have $\Sigma_0$ as their only boundary \cite{Hartle1983}; Right: the tunnelling proposal states that the only modes that survive the quantum barrier are the 'outgoing' modes that represent expanding universes.}
\label{figure104}
\end{figure}

Using the path integral approach, Hartle and Hawking propose \cite{Hartle1983} that the class $C$ over which the integral (\ref{PIF01}) has to be performed is the class of compact geometries (in principle of all topologies) that have $\Sigma_0$ as their only boundary \cite{Hartle1983, Kiefer2007} (see Fig. \ref{figure104}), and matter fields that are regular on those geometries. It means that the \emph{boundary of the universe is that the universe has no boundary}, or equivalently, that the boundary $\Sigma_0$ is created from \emph{nothing}. We shall see later on that for the case of an inflationary spacetime the quantum state that results from the no-boundary condition is \cite{Hartle1983, Halliwell1990, Kiefer2007} 
\be\label{NBWF}
\phi_{NB} \propto {\rm exp} \left( \frac{1}{3 V(\varphi)} \right) {\rm cos}\left( \frac{(a^2 V(\varphi) - 1)^\frac{3}{2}}{3 V(\varphi)} - \frac{\pi}{4} \right) ,
\ee
where $V(\varphi)$ is the potential of the inflaton field and $a$ is the scale factor, which goes from the initial value, $a_0 = V(\varphi)^{-1/2}$, to infinity. It is important to notice that the wave function (\ref{NBWF}) can be written as
\be\label{NBWF2}
\phi_{NB} \propto e^\frac{1}{3 V(\varphi)} \left(  e^{i S} + e^{-i S} \right) ,
\ee
where, 
\be\label{HHA01}
S=  \frac{ (a^2 V(\varphi) - 1)^\frac{3}{2}}{3 V(\varphi)} - \frac{\pi}{4} .
\ee
We shall see in the next section that in terms of the same time variable one of the two terms in \eqref{NBWF2}, say the \emph{branch} with $e^{-iS}$, describes an expanding universe and the branch with $e^{i S}$ describes a contracting universe. It means that the result of imposing the no-boundary condition on the quantum state of the universe is that it is given by the linear combination of two states: one representing an expanding universe and one representing a contracting universe. Typically, one considers the expanding branch of the universe as representing our universe and disregard the contracting branch for being unphysical. However, we shall see that there is another interpretation. In terms of the \emph{physical} time variable of each universe the two universes can both be seen as expanding universes but with their matter fields being CP conjugated. They can be interpreted then as an expanding universe-antiuniverse pair (see, Sec. \ref{sec020503} and Sec. \ref{sec03}). From that point of view, the no-boundary proposal would yield the creation of universes in entangled universe-antiuniverse pairs.

Vilenkin's tunneling proposal is quite different. Perhaps more based on practical grounds, he proposes that the only mode that survives the tunnelling from  the Euclidean region (the region located at, $a < V(\varphi)^{-1/2}$)  is the one that represents an expanding universe. Imposing the tunnelling boundary condition, the resulting wave function of the universe is 
\be\label{TWF}
\phi_{T} \propto {\rm exp} \left( -\frac{1}{3 V(\varphi)} \right) \, {\rm exp} \left(\frac{ -i }{3 V(\varphi)} (a^2 V(\varphi) - 1)^\frac{3}{2} \right) .
\ee
The main difference with respect to the Hartle-Hawking wave function \eqref{NBWF} is the negative sign in the exponent of the exponential pre-factor, which may have important consequences. Let us notice that the probability for the universe to be created from nothing is, $P \propto |\phi|^2$, so in the case of the no-boundary condition we have
\be\label{HHP}
P_{HH} \propto e^{\frac{2}{3 V(\varphi)}} ,
\ee
while in the case of the Vilenkin's tunnelling condition, we have
\be\label{TP}
P_{T} \propto e^{-\frac{2}{3 V(\varphi)}} .
\ee
One immediate consequence is that the Vilenkin's condition seems to favour the creation of a universe with a big value of the potential, which is a necessary condition for the initial hypersurface $\Sigma_0$ to inflate. That would in principle reject the Hartle-Hawking proposal because, on the contrary, the no-boundary proposal seems to favour the creation of a universe with a small (or zero) value of the potential ($P_{HH}\rightarrow \infty$, as $V\rightarrow \infty$). However, this result changes when high order corrections are taken into account so the result is not conclusive \cite{Kiefer2007}. 

From a purely theoretical point of view, it seems that the Hartle-Hawking proposal is more fundamental in the sense that these authors put the focus on the \emph{natural} condition that one should impose on the Euclidean region of the spacetime. As a consequence, they obtain that the universe is represented by two branches, one corresponding to an expanding universe and other describing a contracting branches. These two branches can be considered independently once they suffer a process of decoherence, so in practice they may represent two different universes. Vilenkin's proposal seems to be more practical (although it is also based on a parallelism with some processes of quantum mechanics).

\subsection{Semiclassical quantum gravity}\label{sec0203}

The first thing that the quantum state of the spacetime must provide us with is a consistent explanation of how the classical background can emerge from the full quantum state of the spacetime. Fortunately, not only this can be done in a beautiful manner but it is in fact one of the greatest achievements of quantum cosmology.

Again, the path integral formulation of the spacetime supplies a clear picture of how it can be obtained. At some appropriate limit, which is generally a large length or mass scale compared with their corresponding Planck values, the contribution of most of the paths in the integral vanishes because their destructive interference. The only paths that survive the interference are those that are \emph{in phase}, i.e. those for which $\delta S \approx 0$. These are actually the trajectories of the superspace given by the classical constraints. Therefore, much in a similar way as the classical trajectory of a particle emerges from the constructive interference in the path integral approach of the quantum mechanics of a particle, the classical background spacetime emerges as the constructive interference among the paths in the superspace. In the quantum mechanics of a particle we can then compute the quantum corrections to the trajectory of the particle in terms of \emph{quantum uncertainties}. We shall see in this section that the quantum corrections to the classical background of the universe are caused by the matter fields.

Following the customary approach \cite{Halliwell1990, Hartle1990, Wiltshire2003, Kiefer2007}, let us consider the following semiclassical wave function \cite{Hartle1990},
\be\label{SCWF01}
\phi(h_{ab}, \varphi) = \Delta(h_{ab})e^{\pm \frac{i}{\hbar} S_0(h_{ab})} \psi(h_{ab}, \varphi) ,
\ee
where $\Delta$ and $\psi$ are slowly varying functions of the metric tensor $h_{ab}$, and $S_0(h_{ab})$ is the classical action for gravity alone, given by (\ref{EHA01}). Essentially, the wave function \eqref{SCWF01} contains two parts: one that only depends on the geometric variables of the spacetime and another part that contains all the matter degrees of freedom in the wave function $\psi$. The basic idea is that we expect that the quantum fluctuations of the spacetime will weaken more rapidly than the quantum fluctuations in the state of the matter fields. In that case, we shall find a regime where the spacetime behaves nearly classically with quantum matter fields propagating therein. That is exactly what we need to describe the universe we observe. Also notice the presence of the Planck constant in \eqref{SCWF01}. It means that the classical behaviour of the spacetime will be present whenever the gravitational action is large with respect to the Planck constant.

Now, insert the semiclassical wave function \eqref{SCWF01} into the Wheeler-DeWitt equation \eqref{WDW00} and solve it order by order in $\hbar$ in the geometrical degrees of freedom. At order $\hbar^0$, one finds
\be\label{CC00101}
 G_{abcd} \frac{\delta S_0}{\delta h_{ab}} \frac{\delta S_0}{\delta h_{cd}} -  \sqrt{h} \left( {}^3R - 2\Lambda \right) = 0 ,
\ee
which is the gravitational part of the Hamiltonian constraint (\ref{HC00}) if we make the identification,
\be
p^{ab} = \frac{\delta S_0}{\delta h_{ab}} .
\ee 
At first order in $\hbar$, neglecting second derivatives of the slowly varying terms with respect to the $3$-metric, it is obtained two equations. The first equation \cite{Hartle1990}
\be\label{DFunc}
G_{abcd} \frac{\delta }{\delta h_{ab}} \left( \Delta^2 p^{cd} \right) = 0 ,
\ee
is actually the condition for the function $\Delta(h_{ab})$ to be a \emph{slowly varying} function. In other words, whenever \eqref{DFunc} is satisfied (or to the extent it is satisfied) the wave function \eqref{SCWF01} can be a good candidate to describe the observable universe. Eq. \eqref{DFunc} is also the equation of the conservation of the probability current, $\Delta^2 p^{cd}$. The other equation that is obtained at order $\hbar$ is
\be\label{PSIf}
\mp 2 i G_{abcd} \frac{\delta S_0}{\delta h_{ab}} \frac{\delta \psi}{\delta h_{cd}} + \sqrt h \hat \hat T^{00} = 0 ,
\ee
where the $\mp$ signs correspond to the $\pm$ signs of the exponent of \eqref{SCWF01}. It suggests the identification of a time variable $t$, given by
\be
\frac{\partial }{\partial t} = \pm 2 G_{abcd} \frac{\delta S_0}{\delta h_{ab}} \frac{\delta }{\delta h_{cd}} .
\ee
In that case, \eqref{PSIf} becomes the Schr\"odinger like equation of the matter fields
\be\label{Sch01}
i \frac{\partial \psi}{\partial t} = \sqrt h \, \hat T^{00}(\varphi, -i\partial/\partial\varphi) \psi .
\ee
Therefore, in the semiclassical regime we obtain at order $\hbar^0$ the classical behaviour of the spacetime  and, at first order in $\hbar$,  the quantum evolution of the matter fields. Thus, the semiclassical wave function \eqref{SCWF01} contains all the physical information of the observable universe. One could say that  recovering the classical equations for the spacetime degrees of freedom and the Schr\"odinger equation for the matter fields does not add anything to what we already knew before the quantisation of the universe. That is true, these two features are nothing more than a test of consistency for quantum cosmology. After all, the recovering of the classical spacetime must not be surprising. We started the process of quantisation from the classical action of the spacetime and the matter fields, and the quantisation procedure consists basically in promoting the classical variables to non commuting quantum operators, $[\hat h_{ab}, \hat p^{ab}] \propto \hbar$. Then, it should not be surprising that in the limit $\hbar \rightarrow 0 $ we recover the classical behaviour (this is the essence of the correspondence principle). And something similar for the quantum behaviour of the matter fields.

Even though, the canonical quantisation of the whole universe that we have seen in this  section is interesting for several reasons. First, it suggests that the quantisation processes that leads to the wave function of the universe is consistent and, in that case, one can assume that the wave function of the universe, $\phi(h_{ab}, \varphi)$, would contain in principle all the physical (classical and quantum) information of all the degrees of freedom of the universe. Any physical process should be describable within the formalism of  quantum cosmology. Of course, this reductionist point of view is not practical at all but from the conceptual point of view it results appealing. A more interesting feature is that it allows us to analyse higher order corrections to the semiclassical universe, and this should give novel features that cannot be foreseen in the classical scenario. It might help us to go beyond the quantum description of matter fields in a classical spacetime background. In particular, it can help us to find some exclusive features of the quantum regime of the spacetime, i.e. small deviations from the known behaviour caused by the high order corrections of quantum gravity \cite{Kiefer1991, Kiefer2012, Brizuela2016a, Brizuela2016b}.

Another interesting feature of quantum cosmology is the appearance of time.  From the point of view of the superspace, time is just the parameter that parametrises the curve that describes a particular trajectory. In the picture given by the path integral, the quantum state of the universe is given by the set of all paths that join together the initial and final states. If the quantum wave-packet is spread no definite time variable can be chosen mainly because there is no definite curve that describes the evolution of the universe. It is only when the wave-packet is peaked around a particular solution (ideally becoming a delta function) when we have a definite curve, the classical evolution of the universe, that can be therefore parametrised in terms of a parameter that we can call time\footnote{A classical path of the superspace is invariant under reparametrisations so by \emph{time} we mean any time variable.}. It therefore  appears as the result of a decoherence process between the different \emph{histories} of the universe \cite{Halliwell1989, Hartle1990, Kiefer1992, Hartle1993}.

Furthermore, quantum cosmology relates the two concepts of time of contemporary physics: the one of the theory of relativity and the one of quantum mechanics. This is a very subtle point. Both the theory of relativity and the quantum theory work with a time variable, say $t_r$ and $t_q$, respectively. We usually assume that both time variables are the same, $t_r = t_q = t$, but this is an assumption that is  not guaranteed from the beginning. Of course, they (must) coincide in the Newtonian limit of both theories, but in general, they only coincide if the time variable of theory of relativity would be measured with an actual clock, which is made of matter fields. However, the theory of relativity deals with 'ideal clocks' and the consideration of an actual clock may entail some problems\footnote{For instance, the march of a material clock would be given by the matter fields that are solutions of Einstein's equations, in a circular argument.}.

\subsection{Minisuperspace model}\label{sec0204}

Despite its conceptual importance, it is not hard to see that the Wheeler-DeWitt equation found in the previous chapter is very difficult if not impossible to solve for a general configuration of the spacetime and the matter field(s). In order to make computations one generally has to assume some symmetries in the underlying spacetime of the universe. This process, called \emph{symmetry reduction} \cite{Kiefer2007}, reduces the number of variables of the superspace and the so reduced superspace is called \emph{minisuperspace}\footnote{Sometimes it is distinguished between \emph{minisuperspace} and \emph{midisuperspace} models depending on the number of variables of the reduced superspace.}.

Furthermore,  a minisuperspace model is not necessarily an unrealistic model. On the contrary, the observational data indicates that most of the history of the universe the spacetime presents a high degree of symmetry. Even more, if the initial hypersurface $\Sigma_0$ from which the universe starts evolving is large enough compared with the Planck length, a minisuperspace model could describe the whole history of the universe\footnote{In some cosmological scenarios \cite{Bezrukov2008, Bellido2009, Garay2014}, the length scale of the initial hypersurface can be some orders of magnitude greater than the Planck length, enough for the fluctuations of the spacetime to be subdominant.}. In other cases, it can be taken as a \emph{toy model} from which we can obtain relevant information about some (classical and quantum) aspects of the universe like, for instance, the creation of the initial hypersurface $\Sigma_0$ or the \emph{quantum-to-classical} transition and the appearance of time, among others.

They can also allow us to study the effect of small deviations from the symmetric picture. For instance, we shall consider later on  small departures from the homogeneous spacetime and the matter field in the form of gravitons and matter particles, respectively. These are local perturbations of the otherwise homogeneous and isotropic background. The picture becomes then quite realistic and allows us to analysed potentially observable effects like the  kind of  correlations between the modes of the matter fields in different regions of the spacetime or the quantum gravitational corrections to the Schr\"odinger equation of the matter fields. In all those cases, the study of the minisuperspace model turns out to be justified.

Therefore, let us consider the minisuperspace that is obtained from the foliation of a $4$-dimensional spacetime with closed homogeneous and isotropic spatial sections. The geometry of the spacetime is then characterised by a Freedman-Robertson-Walker (FRW) metric
\be\label{G01}
ds^2 = - N^2(t) dt^2 + a^2(t) d\Omega^2_3  ,
\ee
where $d\Omega_3^2$ is the line element on the unit three sphere. The foliation of the spacetime into space and time is in that case parametrised by just two functions, the scale factor $a(t)$ and the lapse function $N(t)$. The geometry of the spatial sections, $h_{ab}$, is then fully characterised by the scale factor $a(t)$ that parametrises the variation in the distance between two fixed points of the space along the evolution of the universe. It parametrises therefore the expansion or the contraction of the universe. On the other hand, the lapse function $N(t)$ determines the time parametrisation of the foliation. Different values of $N(t)$ entail different time variables (i.e. different time parametrisations), with some special cases. For instance, if $N=1$ the time variable $t$ is called cosmic time and  if $N=a(t)$, $t$ is customary renamed with the Greek letter $\eta$ and is called conformal time because in terms of $\eta$ the metric becomes conformal to the metric of a closed static spacetime. Anyway, we know that the evolution of the universe is invariant under the choice of time reparametrisation so, at the end of the process, we can take the preferable time variable\footnote{Unless otherwise indicated we shall always use cosmic time, $t$, for which $N=1$.}.

The line element of the spacetime is then fully determined by these two functions, $a(t)$ and $N(t)$. The the total action, i.e. the Einstein-Hilbert action (\ref{EHA01}) plus the action of the scalar field (\ref{SMA01}), can be written as \cite{Kiefer2007}
\be\label{AMin201}
S = S_{EH} + S_m =  \frac{1}{2} \int dt N \left( - \frac{a \dot a^2}{N^2} + a - \frac{\Lambda a^3}{3} + \frac{a^3 \dot\varphi^2}{N^2} - 2 a^3 V(\varphi) \right) ,
\ee
where an integration over the spatial variables has been performed and absorbed with a definition of units in which, $2G/3\pi=1$, and the rescalings, $\varphi \rightarrow \varphi/\sqrt{2}\pi$ and $V \rightarrow V/2\pi^2$. The total action has been simplified considerably. The only dynamical degrees of freedom are the scale factor, $a(t)$, and the scalar field, $\varphi(t)$, which according to the homogeneity condition must only depend on the time variable. The superspace has then been reduced to a two dimensional space. Now, we can proceed as described in the preceding sections. The momenta conjugated to the configuration variables are,
\be\label{MMin01}
p_a \equiv \frac{\delta L}{\delta \dot a} = -\frac{a \dot a}{N} \ , \ p_\varphi \equiv \frac{\delta L}{\delta \dot \varphi} =  \frac{a^3\dot\varphi}{N} ,
\ee
and the Hamiltonian then reads
\be\label{HMin01}
H = N \mathcal H = \frac{N}{2} \left( -\frac{1}{a} p_a^2 + \frac{1}{a^3} p_\varphi^2 - a + \frac{\Lambda a^3}{3} + 2 a^3 V(\varphi) \right) .
\ee
The momentum constraint is automatically satisfied by the symmetries of the spacetime and the Hamiltonian constraint, $\frac{\delta H}{\delta N} = 0$, becomes then, $\mathcal H = 0$. Promoting the momenta into quantum operators  and applying  the quantum version of the Hamiltonian constraint, ${\mathcal H} = 0$, to the wave function of the universe, $\phi(a,\varphi)$, which depends now on the two variables of the minisuperspace, $a$ and $\varphi$, we obtain the Wheeler-DeWitt equation, $\hat{\mathcal H} \phi(a,\varphi)= 0$.

\begin{figure}
\centering
\includegraphics[width=13 cm]{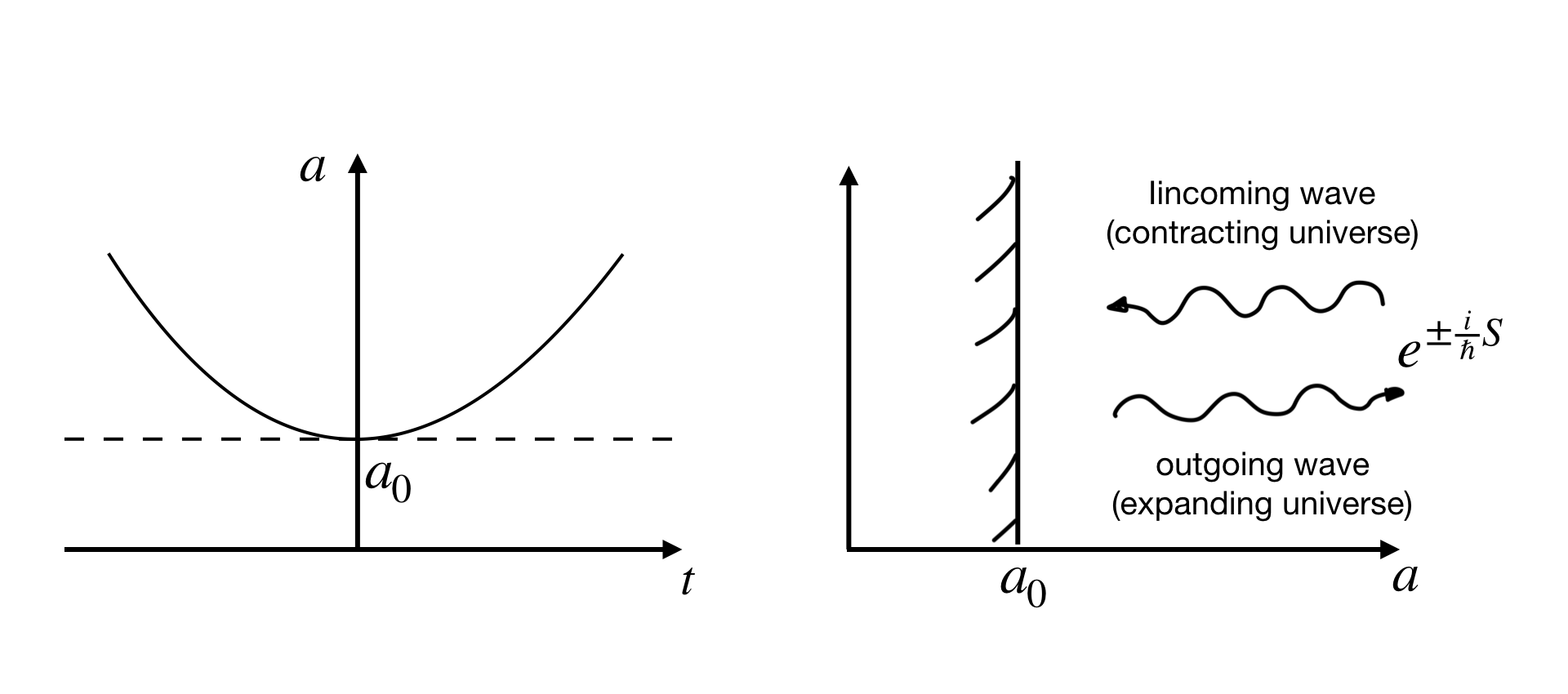}
\caption{Left: the closed DeSitter spacetime describes a universe that starts shrinking from an infinite volume, bounces at the minimum value, $a_0 = H_0^{-1}$, and ends up in an eternal expansion. Right: the value $a_0$ can be seen as a barrier where the incoming waves are reflected and converted into outgoing waves.}
\label{figure201}
\end{figure}

\subsubsection{Inflationary universe}\label{sec020401}

Let us first analyse the initial stage of the universe where the scalar field is assumed to be approximately constant on the small time scale in which the universe rapidly undergoes an inflationary expansion \cite{Linde1993}, $\dot \varphi \approx 0$ and $\varphi_0\gg 1$. In that case, the kinetic term of the scalar field in \eqref{AMin201} can be neglected and the potential  turns out to be approximately  constant, $V(\varphi_0)$. From \eqref{AMin201} it can be seen that a constant value of the potential is equivalent to a cosmological constant term, so the universe effectively behaves like a DeSitter spacetime. Then, we can write \eqref{AMin201} as
\be
S = \frac{1}{2} \int dt N \left( - \frac{a \dot a^2}{N^2} + a - H_0^2 a^3 \right) , 
\ee
where, $H_0^2 = 2 V(\varphi_0)$, in the case of the inflationary universe or, $H_0^2 = \Lambda/3$, in the case of a 'pure' DeSitter spacetime, or the sum of both in a general case. The corresponding Hamiltonian constraint turns out to be
\be\label{HC021}
\mathcal H = - \frac{1}{a} p_a^2 - a + H_0^2 a^3 = 0 ,
\ee
which is nothing more than the Friedmann equation expressed in terms of the momentum conjugated to the scale factor, $p_a$. In terms of the time derivative of the scale factor, using  $p_a =  a \dot a$ (in cosmic time\footnote{Recall that after doing the variation with respect to $N$ we can fix any particular value.}, with $N=1$ in \eqref{MMin01}), the Hamiltonian constraint (\ref{HC021}) can be written as
\be\label{FE001}
\dot a = \sqrt{H_0^2 a^2 - 1 } ,
\ee
whose solutions can easily be obtained,
\be\label{CSF01}
a(t) = \frac{1}{H_0} \cosh H_0 t .
\ee
If $t \in (-\infty, \infty)$, the scale factor \eqref{CSF01} describes a universe that starts shrinking from an infinite volume, bounces at the minimum value, $a_0 = H_0^{-1}$, and ends up in an eternal expansion (see, Fig. \ref{figure201} Left). However, it does not seem quite plausible that the universe is created with an infinite volume, so the most reasonable possibility consist in restricting ourselves to the domain, $t \in (0, \infty)$, which describes a 'bubble' of spacetime that is created with radius $a_0=H_0^{-1}$ at $t=0$ (the origin of time), and starts expanding exponentially in an inflationary like expansion. If the length scale of the initial 'bubble' is some orders of magnitude greater than the Planck scale, i.e. $H_0^{-1} >> l_P$, then, the quantum fluctuations of the spacetime would be small and the homogeneous and isotropic picture described here could reasonably represent the creation of a universe like ours\footnote{However, one would generally expect that the creation of the universe comes from a quantum fluctuation of the spacetime, and therefore be of order of the Planck scale, $H^{-1} \sim l_P$. In that case, new elements should be incorporated although the picture described here and in the next section would still be instructive.}.

Let us now analyse the solutions of the corresponding Wheeler-DeWitt equation. Making the substitution, $p_a \rightarrow -i\hbar \partial_a$, and leaving aside the ambiguity of the factor ordering, the Hamiltonian constraint \eqref{HC021} can conveniently be written as
\be\label{WDW212}
\hbar^2  \frac{d^2\phi(a)}{d a^2} + \omega^2(a) \phi(a) = 0 ,
\ee
where,
\be\label{OME211}
\omega(a) =  \sqrt{H_0^2 a^4 - a^2} .
\ee
Written in this way, the Wheeler-DeWitt equation \eqref{WDW212} resembles the equation of a harmonic oscillator with time dependent frequency (such parallelism  will be further exploited in Sec. \ref{sec03}). Equation  \eqref{WDW212} is not exactly solvable. However, far from the turning point, $a_0=H_0^{-1}$, we can approximate the solutions by the WKB wave functions
\be\label{WF211}
\phi^\pm(a) = \frac{N^\pm}{\sqrt{\omega(a)}} e^{\pm\frac{i}{\hbar} S(a) } ,
\ee
where $N^\pm$ is a normalisation constant and $S(a)$ is given by
\be\label{A211}
S(a) = \int da \, \omega(a) = \frac{\left( H_0^2 a^2 - 1 \right)^\frac{3}{2}}{3 H_0^2 }.
\ee
The two wave functions $\phi^\pm$ in \eqref{WF211} represent incoming and outgoing wave functions. Let us see it by inserting them into the Wheeler-DeWitt equation \eqref{WDW212}.  In that case, at order $\hbar^0$ it is obtained the Hamilton-Jacobi equation
\be
\left( \frac{\partial S}{\partial a} \right)^2 + a^2 - H_0^2a^4 = 0 ,
\ee
which corresponds to the Hamiltonian constraint \eqref{HC021} if one makes the identification
\be
p_a = \pm \frac{\partial S}{\partial a} .
\ee
In fact, let us note that at leading order in $\hbar$ it is satisfied
\be\label{CEQ211}
- a \dot a \equiv p_a \approx \langle \phi^\pm |\hat p_a | \phi^\pm \rangle = \pm \frac{\partial S}{\partial a} + \mathcal O(\hbar^1).
\ee
Therefore, it is obtained
\be\label{FE002}
\dot a = \mp \frac{\omega(a)}{a} = \mp \sqrt{H_0^2 a^2 - 1} ,
\ee
which is the Friedmann equation \eqref{FE001}, with two signs: the $-$ sign that corresponds to $\phi^+$ and the $+$ sign that corresponds to $\phi^-$. Thus, $\phi^-$ describes an outgoing wave, i.e. a wave that travels towards greater values of the scale factor -- it thus represents an expanding universe -- and $\phi^+$ an incoming wave, i.e. a wave that travels towards smaller values of the scale factor, which corresponds therefore to a contracting universe. Let us notice however that the interpretation of $\phi^\pm$ in terms of incoming and outgoing wave must be taken carefully \cite{Rubakov1999}. The Friedmann equation \eqref{HC021} is invariant under the time reversal change, $t\rightarrow - t$, and in terms of $- t$ ($\dot a \rightarrow -\dot a$), $\phi^+$ would represent the expanding universe (i.e. an outgoing wave) and $\phi^-$ the contracting universe (i.e. an incoming wave). Clearly, it depends on the (time) parametrisation. The important thing is that the general quantum state of the universe (far from the turning point) can be written as
\be
\phi(a) = A_0 \, \phi^+(a) + B_0 \, \phi^-(a) ,
\ee
which represents a quantum superposition of incoming and outgoing waves, irrespective of the particular wave function that represents each one. Another important feature that is worth noticing is that  the minimum value $a_0$ constitutes a classical barrier below which the universe cannot go through. Let us notice that for a value $a < a_0$ there is no real solution of the Friedmann equation \eqref{FE002} so an incoming wave function representing a contracting universe is classically reflected (bounced) into an outgoing wave (see, Fig. \ref{figure201} Right).

\subsubsection{Small perturbations and backreaction}\label{sec020402}

Let us now consider a more realistic scenario by introducing two important changes. First, we shall  not neglect the kinetic term of the scalar field in the action \eqref{SMA01}, and we shall consider a general form (i.e. not necessarily a constant) for the potential $V(\varphi)$. Contrary to what may be thought, that will not introduce qualitative changes. In return, it will allow us to represent other stages of the evolution of our universe as well as many other types of universes. The second change that we are going to make is the introduction of small perturbations around the homogeneous and isotropic background spacetime. This will help us to analyse several phenomena like the behaviour of the matter fields in the semiclassical universe or the appearance of a physical time variable.

Regarding the last question, we can assume that except for the very beginning the inhomogeneities of the universe are relatively small\footnote{This is specially clear in the case of the Cosmic Microwave Background radiation (CMB), in which the relative scale of the energy fluctuations in the last scattering surface are of order $10^{-5}$.}. Therefore, to a good order of approximation the universe can be represented by a homogeneous and isotropic background with relatively small inhomogeneities propagating in the background. In that case, it seems reasonable to expand the variables of the spacetime and the matter fields around the homogeneous and isotropic values and study the inhomogeneities as perturbations of the homogeneous and isotropic background.  We can still consider the $3+1$ splitting of the spacetime given in \eqref{STfol}. However, the idea is now to expand the configuration variables ($h_{ab}, N, N_a$ and $\varphi$)  around their homogeneous and isotropic values and retain just the first order terms. Then, let us consider the following expansions \cite{Halliwell1985, Kiefer1987}
\beq\label{hdecomp}
h_{ab}(t,\textbf{x}) &=&  a^2(t) \Omega_{ab} + a^2(t)  \sum_\textbf{n} 2 d_\textbf{n}(t) G^\textbf{n}_{ab}(\textbf{x}) + \ldots  , \\ \label{fidecomp}
\varphi(t,\textbf{x}) &=&  \frac{1}{\sqrt{2\pi}} \varphi(t) + \sum_\textbf{n} f_\textbf{n}(t) Q^\textbf{n}(\textbf{x}) ,
\eeq
where $\Omega_{ab}$ is the metric on the unit three-sphere, $\varphi(t)$ is the homogeneous mode of the scalar field, $Q^\textbf{n}(\textbf{x})$ are the scalar harmonics on the three-sphere and $G^\textbf{n}_{ij}(\textbf{x})$ the transverse traceless tensor harmonics \cite{Halliwell1985}, with, $\textbf{n}\equiv(n,l,m)$. More harmonics can be present in \eqref{hdecomp}. However, we shall only focus on the tensor modes, $d_\textbf{n}$, as the representative of the perturbation of the spacetime. Eventually, these modes will represent gravitons propagating in the background spacetime and, analogously, the perturbation modes $f_n$ will represent the particles of the scalar field(s). The lapse and shift functions must also be expanded in terms of the spherical harmonics. Then, all these perturbed functions are inserted in the action \eqref{ACT06}. The configuration variables are now the scale factor $a(t)$, the homogeneous mode of the scalar field $\varphi(t)$, and the infinite number of modes $d_\textbf{n}(t)$ and $f_\textbf{n}(t)$, denoted generically by $x_\textbf{n}(t)$. The minisuperspace is then the infinite dimensional space spanned by the variables $(a, \varphi, x_\textbf{n})$, and the time evolution of the universe is represented by a parametrised trajectory in that space, $(a(t), \varphi(t), x_\textbf{n}(t))$. We should have included the modes of the perturbed lapse and shift functions. However, as it happens with their homogeneous counterparts they are not dynamical variables but Lagrange multipliers that generate a set of constraints that can be used to simplify the equations \cite{Halliwell1985}. After a cumbersome computation, one arrives at the  Hamiltonian constraint
\be\label{THAM211}
\mathcal H = \mathcal H_0 + \mathcal H_{m} = 0 ,
\ee
where the Hamiltonian $\mathcal H_0$ contains only degrees of freedom of the homogeneous and isotropic background, and $\mathcal H_m$ is the Hamiltonian density that contains also the degrees of freedom of the perturbation modes.

Let us first consider the Hamiltonian of the background spacetime, which is given by the Hamiltonian density of \eqref{HMin01},
\be\label{HC041}
\mathcal H_0 = \frac{1}{a} \left( - p_a^2 - a^2  + \frac{1}{a^2} p_\varphi^2 + 2 a^4 V(\varphi) \right)= 0 .
\ee
After canonical quantisation, the Wheeler-DeWitt equation reads
\be\label{WDW041}
 \left( \hbar^2 \frac{\partial^2}{\partial a^2}  - \frac{\hbar^2}{a^2} \frac{\partial^2}{\partial \varphi^2} + a^4 H^2(\varphi)  - a^2  \right)  \phi_0(a, \varphi) = 0 ,
\ee
where, $H^2(\varphi)  \equiv 2 V(\varphi)$, is not necessarily constant now. Let us consider the semiclassical solutions,
\be\label{SCWF043}
\phi_0^\pm(a,\varphi) = \Delta(a,\varphi) e^{\pm\frac{i}{\hbar} S(a,\varphi)} ,
\ee
where $\Delta(a,\varphi)$ is a slow varying field of the scale factor. Inserting the wave function (\ref{SCWF043}) into the Wheeler-DeWitt equation  (\ref{WDW041}), as we did in the previous section, and disregarding second order derivatives with respect to the background variables, one obtains  at order $\hbar^0$ the Hamilton-Jacobi equation \cite{Kiefer1987}
\be\label{HJ041}
-\left( \frac{\partial S}{\partial a} \right)^2 +\frac{1}{a^2} \left( \frac{\partial S}{\partial \varphi} \right)^2 + a^4 H^2(\varphi) - a^2 = 0 .
\ee
Following the semiclassical development of the previous sections, let us define a WKB-time parameter $t_w$ as \cite{Kiefer1987} 
\be\label{WKBt041}
\frac{\partial}{\partial t^\pm_{w}}  \equiv \pm \left( -\frac{1}{a} \frac{\partial S}{\partial a}\frac{\partial }{\partial a} +\frac{1}{a^3} \frac{\partial S}{\partial \varphi}\frac{\partial }{\partial \varphi} \right) ,
\ee
in terms of which,
\be\label{MOM041}
\dot{a}^2 = \frac{1}{a^2} \left( \frac{\partial S}{\partial a}\right)^2 \ , \ \dot{\varphi}^2 = \frac{1}{a^6} \left( \frac{\partial S}{\partial \varphi} \right)^2 ,
\ee
and the Hamilton-Jacobi equation (\ref{HJ041}) turns out to be
\be\label{FE041}
\dot{a}^2 + 1 - a^2 \left( \dot{\varphi}^2 + 2 V(\varphi) \right) = 0 ,
\ee
which is  the Friedmann equation of the background spacetime. The WKB wave functions $\phi_0^\pm$ describe universes with a background spacetime that evolves according to the Friedmann equation \eqref{FE041}. The wave function $\phi_0$ may thus represent the quantum state of a large variety of classical models of the universe. The two signs in \eqref{WKBt041} are irrelevant at the classical level because the Friedman equation \eqref{FE041} is invariant under the reversal change of the time variable. However, they will play an important role at the semiclassical level.

Let us now consider the complete Hamiltonian constraint \eqref{THAM211}.  The corresponding Wheeler-DeWitt equation is
\be\label{WDW051}
\left( \hat{\mathcal H}_0 + \hat{\mathcal H}_{m} \right) \phi(a,\varphi;x_\textbf{n}) = 0 ,
\ee
where $ \hat{\mathcal H}_0$ is the Hamiltonian of the background that we have already seen (\ref{HC041}-\ref{WDW041}), and $\hat{\mathcal H}_{m}$ is the  Hamiltonian of the perturbation modes, which for the moment we do not need to specify. The wave function of the universe can be separated in two factors, the wave function $\phi_0(a,\varphi)$ of the background \eqref{SCWF043} and another wave function that contains the matter degrees of freedom\footnote{By 'matter degrees of freedom' we mean the perturbation modes that can represent matter, radiation or even fluctuations of the gravitational field (gravitons).}. Therefore, the semiclassical wave function can be written now as 
\be\label{SCWF053}
\phi^\pm(a,\varphi;x_\textbf{n}) = \Delta(a,\varphi) e^{\pm\frac{i}{\hbar} S(a,\varphi)}  \psi_\pm(a,\varphi; x_\textbf{n}) ,
\ee
If we insert the wave function \eqref{SCWF053} into the Wheeler-DeWitt equation \eqref{WDW051} and we solve it order by order in $\hbar$, it is obtained at order $\hbar^0$ the Hamilton-Jacobi equation \eqref{HJ041}. Therefore, the wave function \eqref{SCWF053} still describes the background spacetime that evolves according to \eqref{FE041}. On the other hand, at order $\hbar^1$ in $\mathcal H_{0}$, one obtains
\be\label{SCH00}
\pm i \hbar \left( -\frac{1}{a} \frac{\partial S}{\partial a}\frac{\partial }{\partial a} +\frac{1}{a^3} \frac{\partial S}{\partial \varphi}\frac{\partial }{\partial \varphi} \right) \psi_\pm = \hat{\mathcal H}_m \psi_\pm .
\ee
Here comes a subtle but crucial point of the semiclassical regime. In terms of the initial proper time, $t$, the two branches of the universe represent an expanding and a contracting universe because, from \eqref{MMin01},
\be\label{ECW401}
- a \frac{d a}{d t} = p_a \approx \langle  \phi^\pm_0 | \hat p_a | \phi^\pm_0 \rangle \sim \pm\frac{\partial S}{\partial a}  ,
\ee
so $\phi_0^-$ describes a universe whose spacetime background is expanding and $\phi_0^+$ a universe whose background spacetime is contracting. In that case, in order for the WKB-time \eqref{WKBt041} to represent the proper time variable $t$ in the two branches, we have to choose for the branch $\phi^-$ the WKB-time variable $t_-$ defined by
\be\label{WKBt041-}
\frac{\partial}{\partial t_-}  \equiv -\left( -\frac{1}{a} \frac{\partial S}{\partial a}\frac{\partial }{\partial a} +\frac{1}{a^3} \frac{\partial S}{\partial \varphi}\frac{\partial }{\partial \varphi} \right) ,
\ee 
in which case, 
\be\label{SF41-}
\frac{\partial a}{\partial t_-}  = \frac{1}{a} \frac{\partial S}{\partial a} ,
\ee
that represents an expanding universe. For the WKB-time variable in the $\phi^+$ branch, we must choose
\be\label{WKBt041+}
\frac{\partial}{\partial t_+}  \equiv  -\frac{1}{a} \frac{\partial S}{\partial a}\frac{\partial }{\partial a} +\frac{1}{a^3} \frac{\partial S}{\partial \varphi}\frac{\partial }{\partial \varphi} ,
\ee 
in terms of which, 
\be\label{SF41+}
\frac{\partial a}{\partial t_+}  = - \frac{1}{a} \frac{\partial S}{\partial a} ,
\ee
describes a contracting universe. It is worth noticing that this assignation is somehow arbitrary \cite{Rubakov1988, Rubakov1999}. If we would have started  with a time variable, $t \rightarrow - t$, then, $\phi_0^+$ would represent the contracting universe and $\phi_0^-$ the expanding one and the assignations of the WKB-time variables would have been the other way around. However, in terms of the definitions \eqref{WKBt041+} and \eqref{WKBt041-} the equation \eqref{SCH00} reads,
\be\label{SCH052a}
i \hbar \frac{\partial }{\partial t_\pm} \psi_\pm(t_\pm, \varphi) = \hat{\mathcal H}_m \psi_\pm(t_\pm, \varphi) ,
\ee
where, $\psi_\pm(t_\pm, \varphi) \equiv \psi_\pm[a(t_\pm), \varphi]$, evaluated in the background solutions of \eqref{SF41+} and \eqref{SF41-}. Therefore, we have ended up with two universes, one contracting and another expanding, both filled with matter.

\begin{figure}
\centering
\includegraphics[width=10 cm]{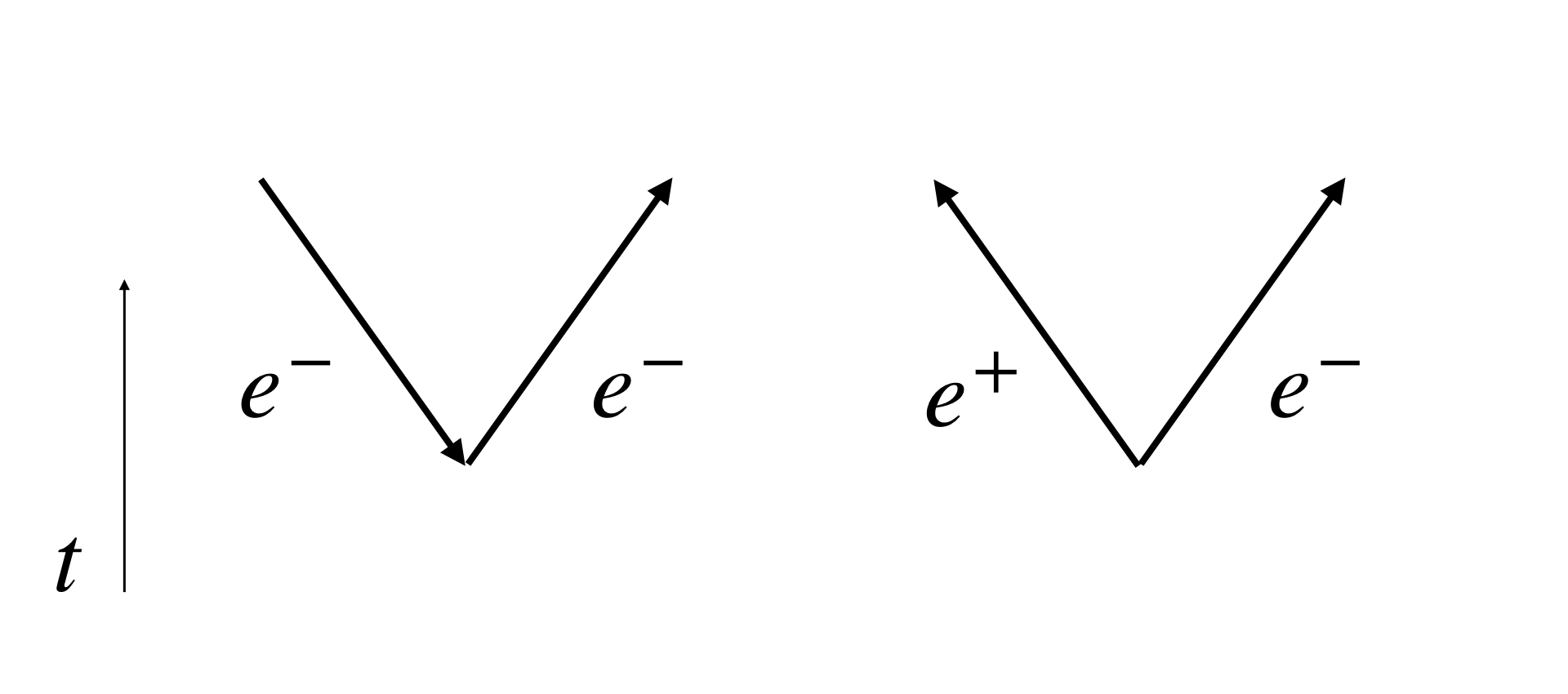}
\caption{An electron-positron pair can equivalently be seen as either an electron propagating backwards in time, bouncing, and then propagating forward in time; or as a particle-antiparticle pair propagating forward in time.}
\label{figure202}
\end{figure}

There is however a different interpretation. One may assume that the physical time variable, i.e. the time variable measured by actual clocks that are made of matter, is the time variable that appears in the Schr\"odinger equation. In that case, it is worth noticing that the physical time variable of the two universes is reversely related, $t_+ = - t_-$. Let us assume that we fix the time variable by fixing the time that a particular observer measures and consider thus $t_-$ as the physical time. Then, in terms of the time variable $t_-$ the evolution of the scale factor is given by \eqref{SF41+} so the two wave functions, $\phi^+$ and $\phi^-$, represent both with a universe with an expanding background spacetime, i.e from the point of view of this hypothetical observer both universes are expanding. The Schrödinger equation \eqref{SCH00} in the observer's universe becomes,
\be\label{SCH052a1}
i \hbar \frac{\partial }{\partial t_-} \psi_-(t_-, x_\textbf{n}) = \hat{\mathcal H}_m \psi_-(t_-, x_\textbf{n}) .
\ee
However, the Schrödinger equation for the fields of the partner universe is
\be\label{SCH052a2}
- i \hbar \frac{\partial }{\partial t_-} \psi_+(t_-, x_\textbf{n}) = \hat{\mathcal H}_m \psi_+(t_-, x_\textbf{n}) ,
\ee
for the wave function $\psi_+$. The 'wrong sign' in \eqref{SCH052a2} is not problematic. It is only indicating that \eqref{SCH052a2} is the Schr\"odinger equation of the complex conjugated wave function $\psi_+^*$ with a $CP$-transformed Hamiltonian \cite{Rubakov1999}. Let us notice that \eqref{SCH052a2} can be written as
\be
i\hbar \frac{\partial}{\partial t_-} \psi_-(t_-, \bar{x}_\textbf{n}) = \hat{\mathcal H}_m(\bar{x}_\textbf{n}) \psi_-(t_-, \bar{x}_\textbf{n}) .
\ee
It is therefore the Schrödinger equation for the antimatter fields of the observer's universe. In this case, we have ended up in the description of two universes, both expanding but one filled with matter and the other filled with antimatter\footnote{In terms of $t_+$ we would have ended up with two contracting universes, one made of matter and the other made up of antimatter. However, that case is not interesting because the two newborn contracting universes would rapidly delve into the spacetime foam from which they came up}. The two interpretations raised here for the linear combination of incoming and outgoing wave look very similar to the two interpretations made in QED of a electron-positron pair (see, Fig. \ref{figure202}). In Sec. \ref{sec030204}, we shall interpret this combination as the creation of a universe-antiuniverse pair.

Let us now specify the Hamiltonian of the perturbation modes. If we restrict to small linear perturbations the different modes do not interact  and $\mathcal{H}_m$ turns out to be the sum of the Hamiltonian of a set of harmonic oscillators \cite{Halliwell1987, Kiefer1992}
\be\label{MH051}
\mathcal  H_m =  \sum_\textbf{n} \mathcal  H_\textbf{n} ,
\ee 
with
\be\label{Hn}
\mathcal{H}_\textbf{n} = \frac{1}{2 M} p_{x_\textbf{n}}^2 + \frac{M \omega_n^2}{2} x_\textbf{n}^2 ,
\ee
where, $M(t) = a^3(t)$, and \cite{Kiefer1987}
\be\label{OMEGAM01}
\omega_n^2 = \frac{n^2-1}{a^2} ,
\ee
for the tensorial modes of the spacetime ($x_\textbf{n} \equiv d_\textbf{n}$), and 
\be\label{OMEGAM02}
\omega_n^2 = \frac{n^2-1}{a^2} + m^2    ,
\ee
for the perturbation modes of the scalar field  ($x_\textbf{n} \equiv f_\textbf{n}$). Thus, for small perturbations the Hamiltonian of the modes turns out to be the Hamiltonian of a set of uncoupled harmonic oscillators with time dependent mass, $M(t)=M[a(t)]$, and frequency given by, $\omega_n(t)=\omega_n[a(t)]$, where $a(t)$ is the solution of the Friedmann equation of the background spacetime (possibly including the backreaction).

The Schr\"odinger equation \eqref{SCH052a} of the perturbation modes is then the Schr\"odinger equation of a set of uncoupled harmonic oscillators, whose general solution can be written as \cite{Kiefer1987, Grishchuk1990, Kiefer1992}
\be\label{CHI01}
\psi_\pm = \prod_\textbf{n} \psi_\textbf{n}(t_\pm, x_\textbf{n}) ,
\ee
where the function $\psi_\textbf{n}(t,x_\textbf{n})$ is the wave function of a harmonic oscillator with time dependent mass and frequency, which can be written in terms of the wave function of a harmonic oscillator with constant mass and frequency \cite{Lewis1969, Leach1983, Kanasugui1995, Sheng1995}. The general solution of $\psi_\textbf{n}(t,x_\textbf{n})$ can then be expanded in the basis of number eigenstates of the invariant representation, $\psi_{N,\textbf{n}}$, as
\be\label{CHI02}
\psi_\textbf{n} = \sum_N c_N \, \psi_{N,\textbf{n}} ,
\ee
where $c_N$ are constants coefficients and the wave function of the invariant number state, $\psi_{N,\textbf{n}}$, is given by \cite{Leach1983, RP2017f}
\be\label{NS01}
\psi_{N,\textbf{n}}(a,\phi; x_\textbf{n}) \equiv \langle a, \phi; x_\textbf{n} | N_\textbf{n} \rangle =  \frac{1}{\sigma^\frac{1}{2}} \exp\left\{ \frac{i M}{2}\frac{\dot{\sigma}}{\sigma} x_\textbf{n}^2 \right\} \bar{\psi}_N  \left(  \frac{x_\textbf{n}}{\sigma}, \tau \right) 
\ee
where $\bar{\psi}_N(q,\tau)$ is the customary wave function of the harmonic oscillator, 
\be\label{HO051}
\bar\psi_N(q,\tau) = \left( \frac{1}{2^N N! \pi^\frac{1}{2}} \right)^\frac{1}{2} e^{-i (N+\frac{1}{2}) \tau} e^{-\frac{q^2}{2}} {\rm H}_N(q)
\ee
with ${\rm H}_N(q)$ the Hermite polynomial of order $N$, $q\equiv \frac{x_\textbf{n}}{\sigma}$,
\be\label{TAU01}
\tau(t) = \int^t \frac{1}{M(t') \sigma^2(t')} dt' ,
\ee
and $\sigma(t)$ is an auxiliary function that satisfies the non-linear equation \cite{Lewis1969, Leach1983}
\be\label{SIGM01}
\ddot{\sigma} + \frac{\dot{M}}{M} \dot{\sigma} + \omega_n^2 \sigma = \frac{1}{M^2 \sigma^3} ,
\ee
plus some boundary condition \cite{RP2017c}.  The interesting property of the invariant representation is that once the \emph{field}\footnote{Here, the field is the Schr\"odinger wave function $\psi_\textbf{n}$.} is in a number state of the invariant representation it remains in the same state along the entire evolution of the field. For instance, let us assume that the perturbation modes are in the vacuum state of the invariant representation, $|0\rangle = \prod_\textbf{n} |0_\textbf{n}\rangle$. In that case, the mean value of the energy of the perturbations reads\footnote{In Ref. \cite{Halliwell1987} it is shown that the distribution of a large number harmonic oscillators becomes highly peaked around its average value.}
\be\label{BR051}
\langle \mathcal H_m \rangle = \sum_\textbf{n} \hbar \omega_n \left( \langle \hat N_\textbf{n} \rangle +\frac{1}{2} \right) = \sum_n \frac{\hbar \omega_n}{2} \rightarrow \frac{\hbar}{2} \int^{n_{\rm max}} dn \; n^2 \; \omega_n ,
\ee
where $\omega_n$ is given by \eqref{OMEGAM02} and \eqref{OMEGAM01} for matter particles or spacetime gravitons, respectively. The sum in \eqref{BR051} diverges and some cut-off $n_{\rm max}$ must be taken. The energy $\langle \mathcal H_m \rangle$ is the backreaction of the perturbations on the homogeneous and isotropic background that would induce a modification of the Friedmann equation \eqref{FE041},
\be\label{BRBG01}
\left( \frac{\dot a}{a} \right)^2 = 2 \rho_\varphi - \frac{1}{a^2} + \frac{\langle \mathcal H_m \rangle}{a^3} ,
\ee
where, $\rho_\varphi = \frac{1}{2}\dot{\varphi}^2 + V(\varphi)$, is the energy of the homogeneous mode of the matter field and, $\langle \mathcal H_m \rangle \propto a^{-1}$, for the massless tensor modes or, $\langle \mathcal H_m \rangle \propto m $, for the perturbations of the scalar field.

\subsection{Paradigms for the creation of the universe in quantum cosmology}\label{sec0205}

In the preceding section we ended up with a Friedman equation that was corrected by the backreaction of the perturbation modes of the spacetime. We shall see in this  section that such term may induce important consequences in the way in which the universes can be created. However, for historical reasons instead of considering the backreaction of the perturbation modes we shall consider the model of a massless scalar field conformally coupled to gravity and a cosmological constant, which eventually raises the same term in the Friedmann equation, so the two models effectively entail similar effects. Later on, we shall briefly comment on the nature of this term. The conformally coupled massless scalar field  is the field used by Hartle and Hawking in Ref. \cite{Hartle1983} to describe the quantum state of the universe and the process of quantum creation. Besides, it will also help us to introduce different paradigms for the creation of the universe.

Therefore, let us consider the following action for the spacetime and the massless, i.e. $V(\varphi)=0$, scalar field,
\be\label{AMin01}
S = S_{EH} + S_m =  \frac{1}{2} \int dt N \left( - \frac{a \dot a^2}{N^2} + a - \frac{\Lambda a^3}{3} + \frac{a^3 \dot\varphi^2}{N^2} - \frac{1}{6} a^3 \ {}^4R \varphi^2  \right) ,
\ee
where the last term represents the conformal coupling of the scalar field, and  ${}^4R$  is the Ricci scalar. Then, with the change, $\chi = a \varphi$, and after an integration by parts the total action \eqref{AMin01} can be written as \cite{Hartle1983}
\be\label{ACT031}
S = S_{EH} + S_m = \frac{1}{2} \int dt N \left( - \frac{a \dot a^2}{N^2} + a - H_0^2 a^3 + \frac{a \dot \chi^2}{N^2} - \frac{\chi^2}{a} \right) ,
\ee
where $H_0^2$ can be a pure cosmological constant, $H_0^2 = \Lambda/3$, or the constant value of the potential of an auxiliary inflaton field (different from the scalar field $\varphi$ that we are considering in \eqref{ACT031}). In any case, it will be assumed that it is a constant. Now, the momenta conjugated to the scale factor and the conformally coupled massless field $\chi$ can be easily obtained and the Hamiltonian constraint associated to the action \eqref{ACT031} reads
\be\label{HAM031}
H = N \mathcal H = \frac{N}{2 a} \left( - p_a^2 - a^2 + H_0^2a^4 + p_\chi^2 + \chi^2 \right) = 0 ,
\ee
which, taking into account that   $p_\chi^2 + \chi^2$ is nothing more than (twice) the energy of the scalar field \cite{RP2017c}, $2E$, can also be written as
\be\label{HAM032}
\frac{1}{2} p_a^2 +  \mathcal U(a)  = E  ,
\ee
with
\be\label{PotU031}
\mathcal U(a) = \frac{1}{2} \left( a^2 - H_0^2 a^4 \right) .
\ee
which is formally similar to the energy equation of a particle that propagates under the action of the potential \eqref{PotU031}. The first term of the l.h.s. of \eqref{HAM032} would be the kinetic energy, the second term would be the potential energy and $E$ would be the total energy\footnote{Let us note however that this is only a formal analogy. In fact, the Hamiltonian constraint \eqref{HAM031} indicates is that the total energy of the universe is zero, i.e. the (negative) energy of the spacetime exactly balances the (positive) energy of the matter fields.}

\begin{figure} 
\centering
\includegraphics[width=12 cm]{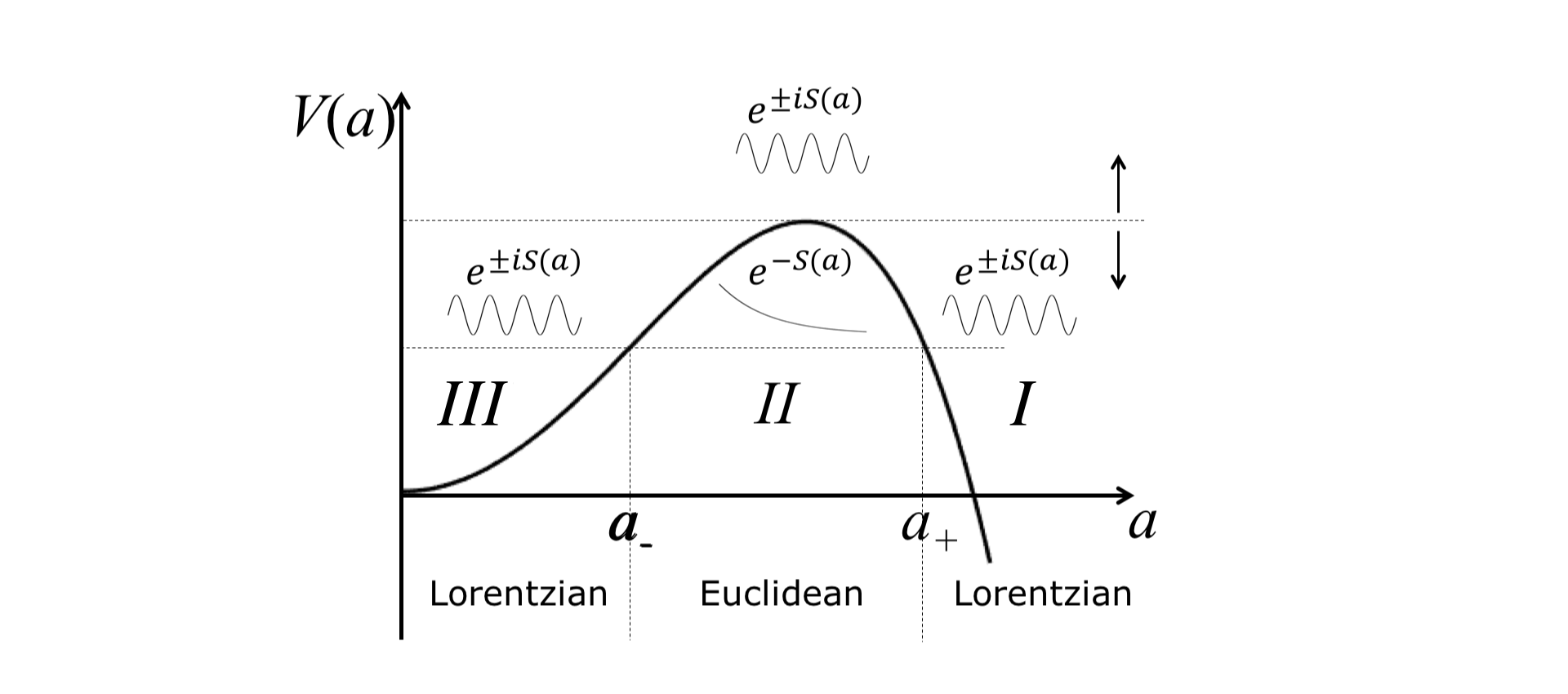}
\caption{Potential $\mathcal U(a)$. For $E>0$ there are three regions, two Lorentzian regions separated by an Euclidean one. Compare it with the case $E=0$ analysed in  Fig. \ref{figure201}.}
\label{figure301}
\end{figure}

Quantum mechanically, the wave function  of the universe, $\phi(a,\chi)$, is the solution of the Wheeler-DeWitt equation associated to the Hamiltonian constraint \eqref{HAM031},
\be\label{WDW301}
 \left( \frac{\partial^2}{\partial a^2} + H_0^2 a^4 - a^2 - \frac{\partial^2}{\partial \chi^2} + \chi^2 \right)  \phi(a,\chi) = 0 ,
\ee
which  can be solved by the method of the separation of variables. Making, $\phi(a,\chi) = \xi(\chi) \psi(a)$, the Wheeler-DeWitt equation \eqref{WDW301} can be split into the two following equations,
\begin{eqnarray}\label{eq1}
 \left( -\frac{d^2}{d\chi^2} + \chi^2 \right) \xi(\chi) = 2 E \xi(\chi) , \\ \label{eq2}
 \frac{d^2 \psi(a)}{da^2}  +  \left( H_0^2 a^4 - a^2 \right) \psi(a) = - 2E \psi(a) .
\end{eqnarray}
The first of these equations is the equation of a quantum harmonic oscillator with unit mass and frequency. It can be solved in terms of Hermite polynomials, ${\rm H}_n(x)$,
\be\label{HO401}
\xi(\chi) \equiv \xi_n(\chi) = \frac{1}{\sqrt{2^n n!}} \left( \frac{1}{\pi }\right)^\frac{1}{4} e^{-\frac{\chi^2}{2 }}   {\rm H}_n(\chi) ,
\ee
with a quantised energy given by
\be\label{En411}
E \equiv E_n =  (n + \frac{1}{2}) .
\ee
On the other hand, Eq. (\ref{eq2}) can be written as
\be\label{WDW32}
\frac{1}{2} \frac{d^2 \psi(a)}{da^2}  +  \left( E - \mathcal U(a) \right) \psi(a) = 0 ,
\ee
with $\mathcal U(a)$ given by \eqref{PotU031}. Eq. \eqref{WDW32} is formally similar to the Schr\"odinger equation of a particle of energy $E$ moving under the action of the potential $\mathcal U(a)$, see Fig. \ref{figure301}. For the value, $E \in (0, \mathcal U_{max})$, where $\mathcal U_{max}$ is the maximum value of the potential, we can distinguish three regions, two classically allowed regions (regions $I$ and $III$ in Fig. \ref{figure301}) separated by a classically forbidden region (region $II$). We have already analysed in the preceding section the behaviour of the wave function in the region $I$. There are incoming and outgoing waves that represent a contracting universe that shrinks to the minimum value of the scale factor $a_+$ (see, Fig. \ref{figure301}) and bounces (or it is reflected) becoming an expanding universe. In the other classically allowed region, region $III$, we shall see that there are also incident and reflected waves that represent a universe that is confined to oscillate between $a=0$ and the maximum value $a_-$.

In the case of a particle moving under the action of a similar potential, the particle that is placed in the region $III$ is classically confined to move in that region like the universes in our example. However, we know that quantum mechanically there is a non zero probability for the particle to tunnel though the quantum barrier and appear in the region $I$. Something similar happens with the universe. If $E>0$, the small oscillating universe of region $III$ can tunnel out through the Euclidean barrier appearing in region $I$ as a new born universe. The universe is then said to be created from \emph{something}. In the limiting case, $E=0$, which was the case analysed by Hartle and Hawking \cite{Hawking1982, Hartle1983, Hawking1983, Hawking1984} and by Vilenkin \cite{Vilenkin1982, Vilenkin1984, Vilenkin1986} for the creation of the universe, there is no region $III$ from which to tunnel out to region $I$. The universe appears in region $I$ from a pure tunnelling phenomena (like the creation of particles from the quantum vacuum). In that case, the universe is said to be created from \emph{nothing}\footnote{Let us note however that this process does not violate the conservation of the energy because the total energy, i.e. the gravitational energy plus the energy of the matter fields is, as we have already said, balanced.}. Let us analyse the two cases separately.

\subsubsection{Creation of the universe from \emph{nothing}}\label{sec020501}

\begin{figure} 
\centering
\includegraphics[width=13 cm]{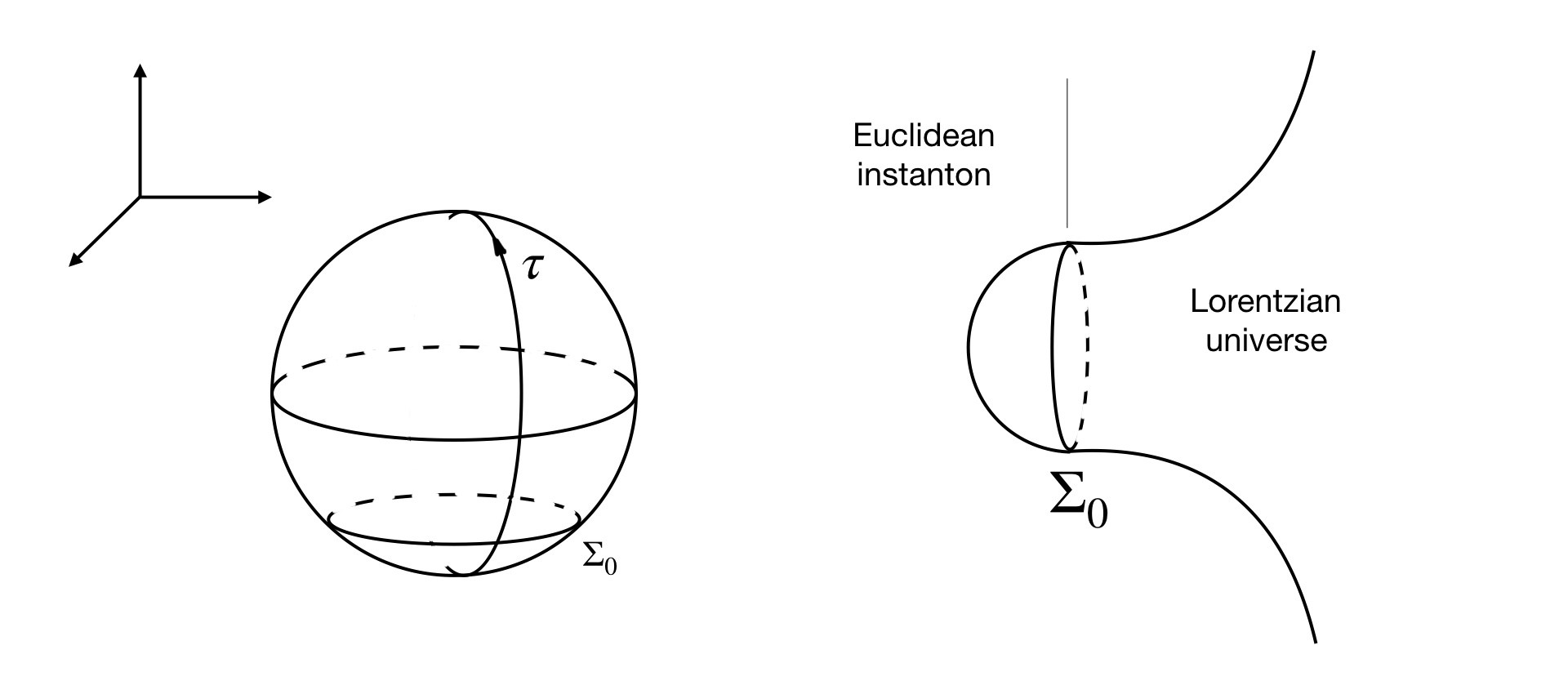}
\caption{Left: a DeSitter instanton can be seen as a $4$-dimensional sphere in the $5$-dimensional Euclidean flat spacetime. Right: the creation of a DeSitter spacetime from a DeSitter instanton of the Euclidean sector of the spacetime.}
\label{figure302}
\end{figure}

Let us first analyse the creation of the universe from nothing, i.e. $E=0$ in \eqref{HAM032}. In that case, in terms of the time derivative of the scale factor, $p_a = - a \dot a$, the Hamiltonian constraint \eqref{HAM032} reduces to the Friedmann equation \eqref{FE001} already studied in Sec. \ref{sec020401},
\be\label{EFE401}
\dot a = \sqrt{H_0^2 a^2 - 1} ,
\ee
which  yields the well-known solution,
\be\label{SF401}
a(t) = a_0 \cosh H_0 t ,
\ee
with, $a_0= H_0^{-1}$. If we restrict ourselves to the 'expanding branch', $t \geq 0$,  the scale factor \eqref{SF401} represents a universe that starts expanding from the initial boundary $\Sigma(a_0)$ until infinity. For values $a<a_0$ there is no real solution and the value $a=a_0$ constitutes a classical barrier for the universe (see, Fig. \ref{figure201}). However, one can perform a Wick rotation into Euclidean time, $t=-i \tau$, in terms of which the Friedmann equation \eqref{EFE401} becomes
\be\label{EFE402}
\frac{d a_E}{d\tau} = \sqrt{1-H_0^2 a_E^2} .
\ee
Now, the Euclidean equation \eqref{EFE402} has the solution
\be\label{ESF401}
a_E(\tau) = a_0 \cos H_0\tau ,
\ee
where, $\tau \in (-\frac{\pi}{2 H_0}, 0)$. Let us notice that in Euclidean time, $- dt^2 \rightarrow + d\tau^2$, the line element turns to be
\be\label{EM311}
ds^2 = d\tau^2 + a_E^2(\tau) d\Omega_3^2 ,
\ee
which is the line element of  a $4$-sphere of radius $1/H_0$ embedded in a $5$-dimensional flat Euclidean spacetime. The 'spatial section' starts expanding from a single point of the sphere (at $\tau = -\pi/2H_0$) until it reaches the value $H_0^{-1}$ at the Euclidean time, $\tau = 0$ (see, Fig. \ref{figure302}). It is called a DeSitter \emph{instanton} and it gives the maximum contribution to the tunnelling wave function (it is the extremal solution of  the Euclidean action \cite{Hartle1983}). At the boundary hypersurface $\Sigma(a_0)$ it appears in the Lorentzian region as a new born DeSitter universe that starts expanding exponentially.

\begin{figure}
\centering
\includegraphics[width=14 cm]{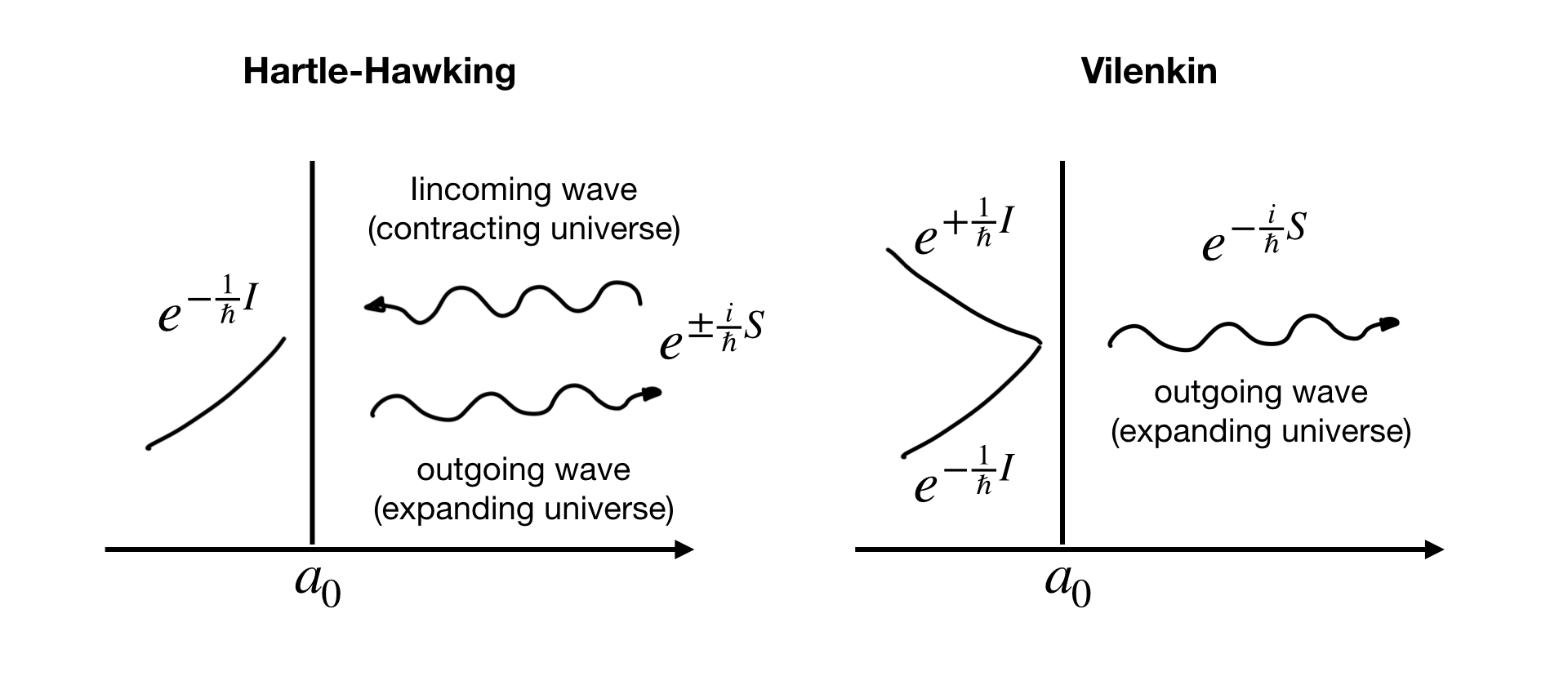}
\caption{Hartle-Hawking 'no-boundary' boundary proposal vs. Vilenkin's tunnelling boundary proposal. The H-H state implies a linear combination of expanding and contracting wave functions in the Lorentzian region, where the tunnelling wave function corresponds only to expanding universes.}
\label{figure303}
\end{figure}

Quantum mechanically the situation is the following. The initial boundary $\Sigma(a_0)$ separates the Lorentzian region of the spacetime (see, Fig. \ref{figure302}), where classical solutions are allowed, from the Euclidean region, where only tunnelling solutions can exist. The wave function that represents a classical spacetime in  the Lorentzian region is given in terms of the oscillatory wave functions $e^{+ i S}$ and $e^{-i S}$, where $S$ is given by \eqref{A211}
\be\label{LAC401}
S(a,\varphi) = \frac{(H^2 a^2 - 1 )^\frac{3}{2}}{3 H^2} .
\ee 
In the Euclidean sector, $a <a_0$, the wave function of the universe can be written in terms of the tunnelling wave functions, $e^I$ and $e^{-I}$, where $I$ is given by the integral \eqref{A211} with $\omega(a)$ being replaced by $|\omega(a)|$, i.e. 
\be
I(a, \varphi) = \frac{(1- H^2 a^2 )^\frac{3}{2}}{3 H^2} .
\ee
The picture is then similar to the problem of a wave-particle tunnelling through a quantum barrier. The oscillatory wave functions $e^{\pm iS}$ can be seen as the  incoming and the reflected waves that may represent a photon or another quantum mechanical particle. Classically, the boundary $\Sigma(a_0)$ acts as a barrier that cannot be crossed (see, Fig. \ref{figure303}). Quantum mechanically, however, there is a non-null probability of penetrate into the barrier although the amplitude is exponentially suppressed in the Euclidean region. Analogously, one can say in cosmology that there is a non-zero probability for the universe to appear \emph{from nothing}, i.e. from the Euclidean barrier of the spacetime. An essential difference is that here the tunnelling is not from another classically allowed region of the spacetime. It is therefore more similar to the creation of virtual particles from the quantum vacuum in a quantum field theory.  In a quantum field theory, the pair of virtual particles can only exist a small amount of time compatible with the Heisenberg's uncertainty relations, otherwise the principle of energy conservation would be violated. In the universe, however, the energy is zero (the negative gravitational energy balances the positive energy of the matter fields) so the creation of the universe from nothing does not violate the conservation of energy.

From the above reasoning it is clear that the name 'creation from nothing' does not refer to the absolute meaning of nothing, i.e. to something to which we can ascribe no properties. As we have seen, the Euclidean region of the spacetime has geometrical properties. In the standard literature (see, for instance, Refs. \cite{Hawking1982, Hawking1983, Vilenkin1982, Wiltshire2003, Kiefer2007}), it usually refers to two meanings. One, perhaps the most consistent, is that the universe is created from a region of the spacetime where nothing real exists; in particular, there is no actual time (i.e. time measured by clocks). In that sense, there is nothing. Another sense with which the term 'nothing' is used in the creation of the universe is that the universe, in the paradigmatic case of the creation of a DeSitter spacetime from the Euclidean instanton \eqref{ESF401}, begins from a single non-singular point of the Euclidean $4$-sphere (which is geometrically equivalent to any other point in the sphere). In that case, the single point is meant to be 'nothing'.

 \paragraph{\bf Vilenkin's vs. Hartle-Hawking's versions}

The general quantum state of the universe is therefore given in the region $I$ by the linear combination
\be\label{WKB421}
\phi_I(a,\varphi) = A_I \frac{1}{\sqrt{\omega(a)}} e^{+\frac{i}{\hbar} S(a)} + B_I \frac{1}{\sqrt{\omega(a)}} e^{-\frac{i}{\hbar} S(a)} ,
\ee 
and in the tunnelling region $II$ by
\be\label{WKB422}
\phi_{II}(a,\varphi) = A_{II} \frac{1}{\sqrt{|\omega(a)|}} e^{+\frac{1}{\hbar} I(a)} + B_{II} \frac{1}{\sqrt{|\omega(a)|}} e^{-\frac{1}{\hbar} I(a)} .
\ee 
The particular combination of WKB wave functions in the Lorentzian and in the Euclidean regions of the spacetime, i.e. the particular values of the constants $A_{I, II}$ and $B_{I,II}$, depend on the boundary condition that we impose on the state of the universe.

Hartle and Hawking \cite{Hawking1982, Hartle1983, Hawking1984} propose as the boundary condition that the path integral must be performed over compact Euclidean geometries that fit with the 'final' values\footnote{These are the final values of the Euclidean regime. From the point of view of the Lorentzian sections, these are the 'initial' values.} $a_0$ and $\varphi$ in the  boundary $\Sigma(a_0)$ (see, Fig. \ref{figure302}). In the homogeneous and isotropic minisuperspace, it is equivalent to the conditions \cite{Halliwell1990}
 \be
 a(\tau_0) = 0 \, , \frac{da}{d\tau}({\tau_0}) = 1 \ , \    \frac{d\varphi}{d\tau}({\tau_0}) = 1 .
 \ee
 The wave function obtained for the Euclidean sections is $e^{- I}$, with
\be\label{EAC403}
I = \int_0^a da \, a (1 - H^2 a^2 )^\frac{1}{2} = \frac{-1}{3H^2} \left( 1 - (1 - H^2 a^2)^\frac{3}{2} \right) .
\ee 
It yields \cite{Hartle1983, Halliwell1990, Kiefer2007}
\be\label{HHE41}
\phi_{HH}^{II}(a,\varphi) = \frac{1}{(1-H^2 a^2)^\frac{1}{4} } \,  {\rm exp}\left( \frac{1}{3H^2} \left( 1 - (1 - H^2 a^2)^\frac{3}{2} \right) \right) .
\ee
From the matching conditions of the WKB method, the Hartle-Hawking wave function \eqref{HHE41} implies a linear combination of oscillatory wave functions in the Lorentzian region ($a>a_0$)  \cite{Hartle1983, Halliwell1990, Kiefer2007},
\be\label{HHos01}
\phi_{HH}^{I}(a,\varphi) = \frac{1}{(H^2 a^2  - 1)^\frac{1}{4} } \, {\rm exp}\left( \frac{1}{3H^2}  \right) \cos\left( \frac{(H^2 a^2 - 1)^\frac{3}{2} }{3H^2} - \frac{\pi}{4} \right) .
\ee

On the other hand, Vilenkin \cite{Vilenkin1982, Vilenkin1984, Vilenkin1986} proposes as the boundary condition, in an analogy with the tunnelling process of quantum mechanics, that the only modes that survive the quantum barrier are the 'outgoing' modes, i.e. those that represent an expanding universe\footnote{As we have seen, it is somehow arbitrary determining which solution describes an expanding universe and accordingly there is an ambiguity in determining which modes are the 'outgoing' modes.}. It means that in the region $I$ the wave function of the universe is given by \cite{Vilenkin1984, Vilenkin1995, Halliwell1990, Kiefer2007}
\be
\phi^{I}_{T}(a,\varphi) = \frac{A_{I}(\varphi)}{(H^2 a^2-1)^\frac{1}{4} } \, {\rm exp}\left(- i \frac{(H^2 a^2 - 1 )^\frac{3}{2}}{3H^2} \right) ,
\ee
where $A_{I}(\varphi)$ is a normalisation 'constant' that can be found by imposing the regularity conditions \cite{Halliwell1990}, $\partial\phi/\partial\varphi \rightarrow 0$ as $a\rightarrow 0$. Then, $A_{I}(\varphi ) = {\rm exp} \left( -1/3H^2\right)$, for which $\phi_T \sim e^{-\frac{1}{2}a^2}$, that is regular at $a\rightarrow 0$ for any value of the scale factor. By following the same WKB procedure of matching conditions, we found in the Euclidean sector \cite{Halliwell1990}
\be\label{TWFE401}
\phi^{II}_{T}(a,\varphi) = \frac{1}{(1-H^2 a^2)^\frac{1}{4} } \, \left( e^{I} -  \frac{i}{2} e^{-I}  \right) \approx \frac{1}{(1-H^2 a^2)^\frac{1}{4} } \,  e^{I} ,
\ee
where $I$ is given by \eqref{EAC403}. Except for the values of the scale factor close to $a_0$, the second term in \eqref{TWFE401} is exponentially smaller than the first, so is usually neglected.

Besides their conceptual meaning the main difference between the wave function of the two proposals is the different sign in the exponent of the pre-factor, ${\rm exp} \left( -1/3H^2\right)$, in the case of the tunnelling wave function, and ${\rm exp} \left( 1/3H^2\right)$, in the case of the no-boundary wave function. The probability measures are different in both cases. Because the similarity of the Wheeler-DeWitt equation with the Klein-Gordon equation, Vilenkin proposes to use the probability current\footnote{In Sec. \ref{sec0303} we shall define more concretely the operator $\nabla$ in the minisuperspace.} \cite{Vilenkin1989, Vilenkin1995}
\be\label{TPM401}
J = \frac{i}{2} \left( \phi^* \nabla\phi - \phi \nabla \phi^* \right) ,
\ee
which is conserved because in virtue of the Wheeler-DeWitt equation it satisfies, $\nabla \cdot J = 0$. The Hartle-Hawking wave function is real so the probability measure \eqref{TPM401} would yield zero. These authors propose instead to use the customary probability measure of quantum mechanics,
\be\label{NBPM401}
J = |\phi|^2 .
\ee
With these two choices the probability for the creation of the universe reads \cite{Halliwell1990}
\be
P = J \cdot d\Sigma \approx {\rm exp } \left( \pm \frac{2}{3 H^2(\varphi)}\right) d\varphi ,
\ee
where the $+$ sign is for the no-boundary wave function and the $-$ sign for the tunnelling wave function. Thus, as we have already notice in Sec. \ref{sec0202}, the no-boundary proposal seems to favour small values of the potential ($P_{HH} \rightarrow \infty$ for $H(\varphi) \rightarrow 0$) and the tunnelling proposal seems to favour the creation of universe with a large value of the potential ($P_{T} \rightarrow 0$ for $H(\varphi) \rightarrow 0$). Therefore, it is usually stated that Vilenkin's tunnelling condition fits better with the inflationary scenario \cite{Vilenkin1995, Linde1993}.

\subsubsection{Creation of the universe from \emph{something}}\label{sec020502}

\begin{figure} 
\centering
\includegraphics[width=9 cm]{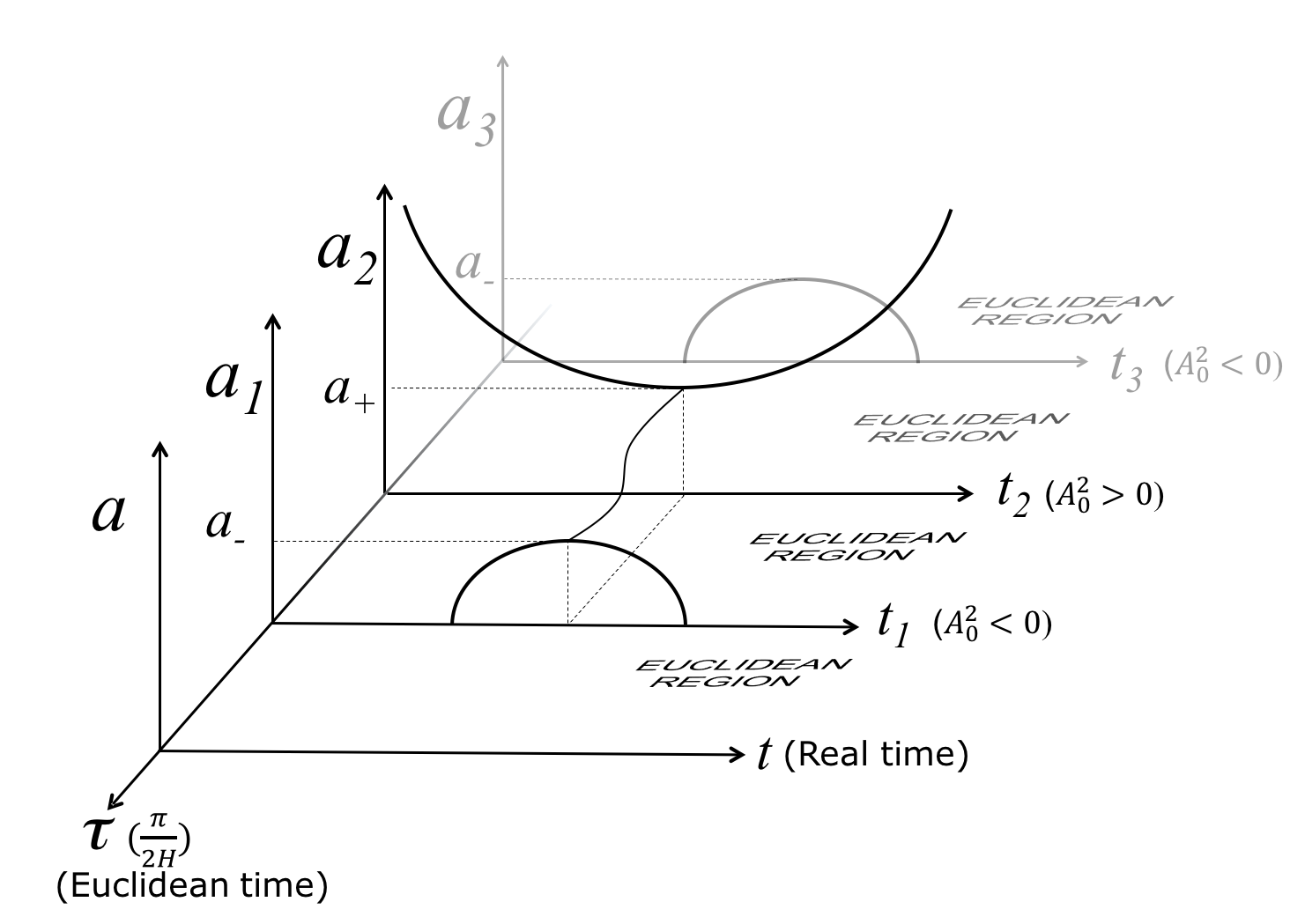}
\caption{The universes of region $III$ are cyclic universes that start from a big-bang singularity and end in a big-crunch one. In region $I$ the universe effectively behaves like a closed DeSitter spacetime.}
\label{figure304}
\end{figure}

 Let us now analyse the case, $E \in (0, \mathcal U_{max})$ in \eqref{HAM032}. The corresponding Friedman equation is obtained by substituting the value, $p_a = - a \dot a$, in \eqref{HAM032}. It yields
 \be\label{FE431}
 \left( \frac{\dot a}{a} \right)^2 = H_0^2 - \frac{1}{a^2} + \frac{2 E}{a^4} .
 \ee
We can see that the conformally massless scalar field can reproduce the effect of a radiation content of the universe ($\rho \sim a^{-4}$, $\langle \mathcal H_m \rangle \propto a^{-1}$ in \eqref{BRBG01}). The Friedman equation \eqref{FE431} can also be written as
\be\label{FE402}
\dot a = \frac{H_0}{a} \sqrt{(a^2-a_+^2)(a^2 - a_-^2)} ,
\ee
with \cite{Gott1998, Rubakov1999}
\be\label{apm}
a_\pm^2 = \frac{1}{2H_0^2} \left( 1\pm \left(1 - 8 E H_0^2 \right)^\frac{1}{2} \right)  .
\ee
For the value, $\frac{1}{8H_0^2} =  \mathcal U_{max} > E > 0$, the two Lorentzian regions are located at, $a > a_+$ and $a<a_-$, respectively. These two sectors represent two separated regions where the universe may exist. In between, there is the Euclidean sector, which is a classically forbidden region. We have therefore two different types of universes separated by a quantum barrier (see Fig. \ref{figure301}). In region $III$, the solution of \eqref{FE402} can be written as \cite{RP2017c}
\be\label{aIV}
a(t) = \left( a_-^2 \cosh^2H_0\Delta t - a_+^2 \sinh^2H_0\Delta t \right)^\frac{1}{2} ,
\ee
where, $\Delta t =t-t_0$, with
\be
\tilde{t}_0 = \frac{1}{H_0} \arctanh\frac{a_-}{a_+} .
\ee
It represents a small universe that starts in a big-bang like singularity at $t = 0$, expands to the maximum value $a_-$, at $t = \tilde{t}_0$, and then it re-collapses to a big-crunch like singularity, at $t = 2 \tilde{t}_0$. For a value $H_0 \ll 1$, the evolution of this type of universes is like the evolution of a  radiation dominated universe (see, Fig. \ref{figure304}). This kind of universes are called \emph{baby} universes \cite{Strominger1990}, which are typically associated with quantum fluctuations of the spacetime.

On the other hand, the solution of \eqref{FE431} in  region $I$ can be written as \cite{RP2017c}
\be\label{aIII}
a(t) = \left( a_+^2 \cosh^2H_0\Delta t - a_-^2 \sinh^2H_0\Delta t \right)^\frac{1}{2} ,
\ee
with, $\Delta t \in (-\infty,\infty)$. It represents a universe that contracts from infinity to the minimum value $a_+$, reached at $\Delta t = 0$, and then, it expands again to infinity (see, Fig. \ref{figure304}). This solution is essentially very similar to the closed DeSitter universe. In fact, it can continuously be  transformed into the customary solution of the closed DeSitter spacetime in the limit $E \rightarrow 0$, for which $a_-\rightarrow 0$ and $a_+\rightarrow 1/H_0$.

\begin{figure} 
\centering
\includegraphics[width=15 cm]{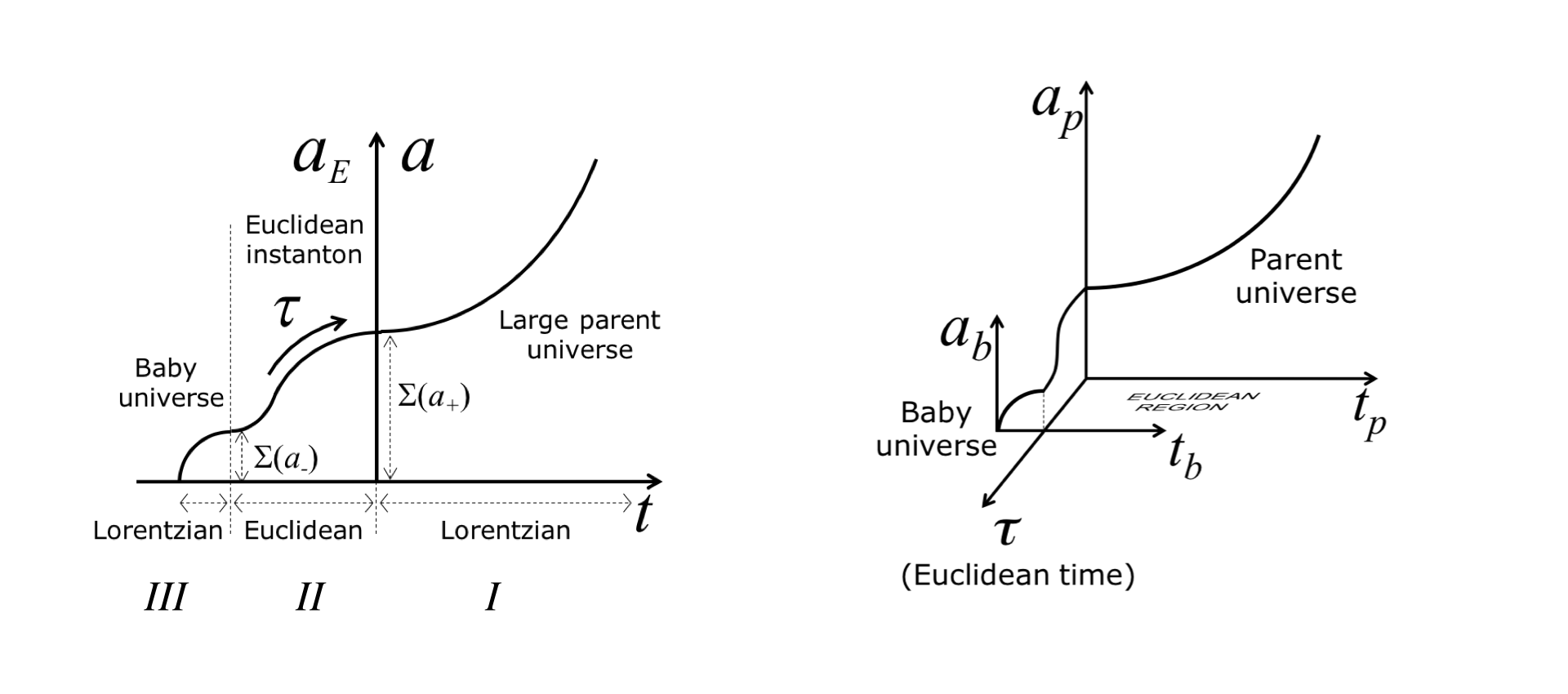}
\caption{The creation of a large parent universe from a baby universe.}
\label{figure305}
\end{figure}

In between there is a tunnelling region, where the solution of the Euclidean version of the Friedman equation is the Euclidean instanton \eqref{EM311} with scale factor given by 
\be\label{aE}
a_E(\tau) = \left( a_+^2 \sin^2H_0\Delta\tau + a_-^2 \cos^2H\Delta\tau \right)^\frac{1}{2} ,
\ee
where $a_\pm$ is given by (\ref{apm}), $a \in (a_-, a_+)$, and 
\be
\Delta\tau = \tau - \tau_{0} \in (0, \frac{\pi}{2 H_0}) .
\ee
The Euclidean instanton with scale factor \eqref{aE} connects the maximum expansion hypersurface $\Sigma(a_-)$ of the baby universes of region $I$ with the initial hypersurface $\Sigma(a_+)$ of the large parent universe in region $I$ (see, Fig. \ref{figure305}). It therefore connects the two regions and it also provides the first order contribution to the probability of crossing the quantum barrier and appearing in region $I$ as a new born universe (see, Ref. \cite{RP2017c}). The universe is then said to be created \emph{from something}.

\paragraph{\bf Gott and Li: the universe is its own mother}

The question of whether the universe is created from \emph{nothing} or from \emph{something} seems to be rooted in the value of the energy $E$ in the Friedman equation \eqref{HAM032}. If $E=0$ the universe must be created from nothing and if $E>0$ the universe must be created from something. However, the value $E=0$ is controversial  because from \eqref{En411}, $E= n + 1/2$, so the value $E=0$ would violate the uncertainty principle of quantum mechanics \cite{Gott1998}. In its original paper \cite{Hartle1983}, Hartle and Hawking suggested that this term might be cancelled by some renormalisation procedure. However, Gott and Li \cite{Gott1998} argue that there is no expectation for such an exact cancellation and in fact, Barvinsky and Kamenshchik \cite{Barvinsky2006, Barvinsky2007a, Barvinsky2007b} compute the renormalisation corrections and not only the energy term is not cancelled but new similar terms appear. Even more, we have seen that the backreaction of the perturbation modes of the spacetime produces a similar term in the Friedmann equation (their energy density is given by, $\langle \mathcal H_m \rangle/ V \propto a^{-4}$, where $\langle \mathcal H_m \rangle \propto a^-1$ in \eqref{BR051} for the perturbations of the spacetime). Therefore, a term with $E> 0$ seems to be unavoidable. One may argue that it can be effectively small and thus it could be neglected. However, regardless the small it can be we have seen that the consequences for the creation of the universe are very important.

A conceptual problem arises however  if the universe is created from 'something' because then, we need the existence of that 'something' prior to the creation of the universe,  i.e. if the universe is created from the tunnelling of a quantum fluctuation of a pre-existing spacetime, then, one should explain how has it been created  the \emph{first} spacetime from which the rest of universes have subsequently been generated. Gott and Li give an apparently exotic although quite interesting explanation. They argue and show \cite{Gott1998} that in region $III$ there can exist closed temporal curves (CTC's). In that case, the spacetime fluctuations of a large parent spacetime can travel through a CTC and become the baby universe that 'tunnelled out' through the Euclidean barrier to give rise to the parent universe in an atemporal  process in which  terms like 'after' or 'before' becomes meaningless. They are only meaningful within the large parent regions of the spacetime where CTC's do not exist. Thus, according to these authors, the universe could be its own mother!.

\begin{figure} 
\centering
\includegraphics[width=13 cm]{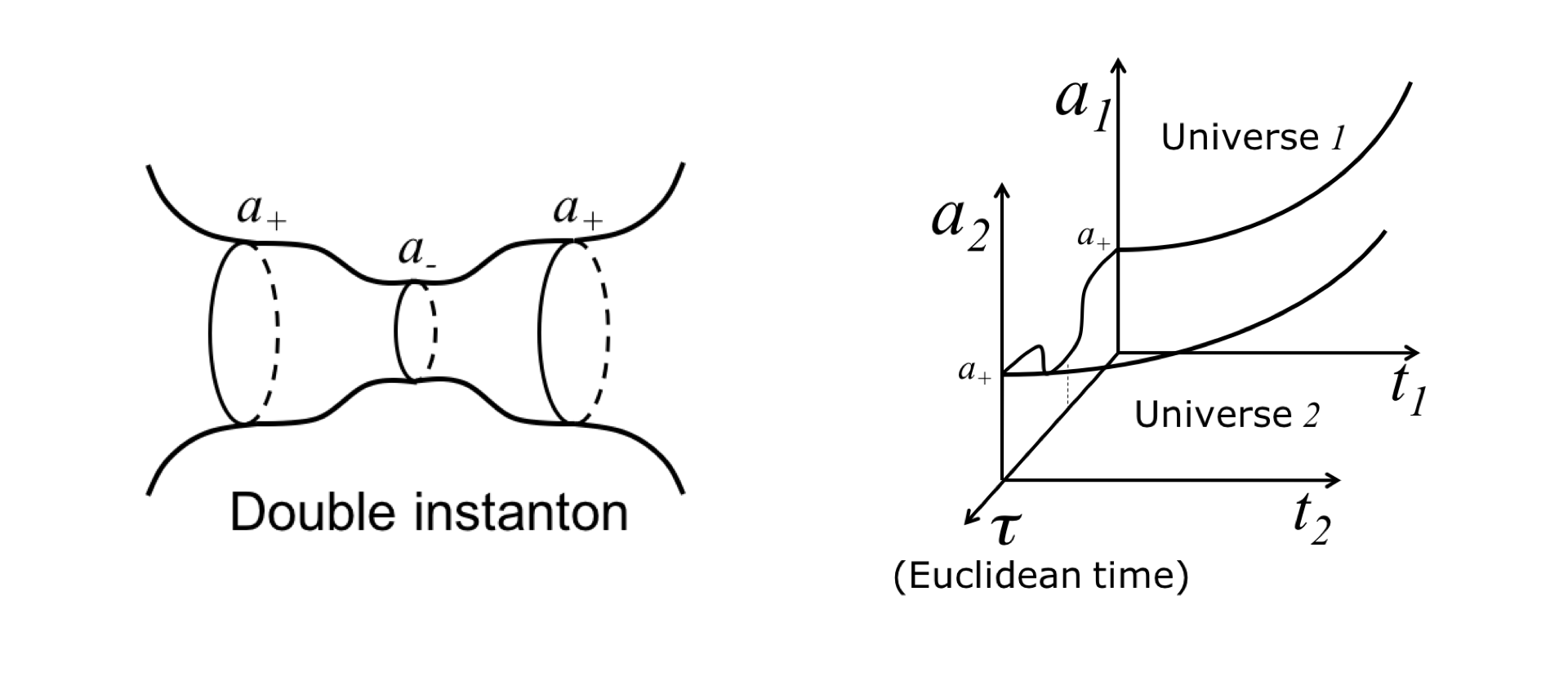}
\caption{Left) a double Euclidean instanton can be formed by matching two single Euclidean instantons. Right) The creation of a pair of entangled universes from \emph{nothing}, i.e. from a double Euclidean instanton.}
\label{figure306}
\end{figure}

\subsubsection{Creation of universes in pairs}\label{sec020503}

There is a way-out to this paradoxic explanation. Even with the value $E>0$ in \eqref{FE431}, there is still room for the universes to be created from \emph{nothing}, i.e. from the Euclidean region without the need of a pre-existing spacetime. However, as we have seen, the process cannot be the one studied in Sect. \ref{sec020501} for the creation of a single universe from the single Euclidean instanton \eqref{ESF401}. Instead, one has to consider more elaborated instantons. For instance, one can consider the double Euclidean instanton that is formed by joining together two single Euclidean instantons through their minimal hypersurfaces $\Sigma_i(a_-)$ (see, Refs.  \cite{Barvinsky2006, Barvinsky2007a} and Fig. \ref{figure306}). The result is the creation \emph{from nothing} of a pair of entangled universes in the region $I$ \cite{RP2011b, RP2014, Chen2017} (see, Fig. \ref{figure306}).

Let us notice that in terms of the same time variable one of the universes of the entangle pair is a contracting universe and the other is an expanding universe so the situation is very similar to the case of coexisting incoming and outgoing waves. The wave function $\phi(a,\chi)$ can therefore be written as
\be\label{3QWF01}
\phi(a,\chi) = \phi^+(a,\chi)  + \phi^-(a,\chi) ,
\ee
with $\phi^\pm(a,\chi)$ given by \cite{RP2017c}
\be\label{WFI301_}
\phi^\pm(a,\chi) = \frac{N}{ \sqrt{\omega_{DS}}} e^{\pm\frac{i}{\hbar} \int \omega_{DS}(a) da } \; \xi_\pm(\eta_\pm, \chi) ,
\ee
where $\omega_{DS}$ is the potential of the Wheeler-DeWitt equation for the DeSitter spacetime, 
\be\label{OMEDS_0101}
\omega_{DS} = \sqrt{H^2 a^4 - a^2} ,
\ee
and
\be\label{SM301_}
\xi_\pm(\eta_\pm, \chi) \equiv \xi(a = a(\eta), \chi) = \sum_n c_n e^{-\frac{i}{\hbar} (n+\frac{1}{2}) \eta_\pm} \xi_n(\chi) ,
\ee
where $\xi_n(\chi)$ are the eigenfunctions of the harmonic oscillator\footnote{In the superposition \eqref{SM301_} it should appear $\xi_n^\pm$, with $(\xi_n^+)^*=\xi_n^-$. However, the eigenfunctions of the harmonic oscillator are real functions so,  $\xi_n^+=\xi_n^-\equiv \xi_n$.}. As we have already seen in the Sec. \ref{sec020402}, the two newborn universes can also be interpreted as two expanding universes filled with matter and antimatter, respectively. Let us notice that if one inserts the wave functions \eqref{WFI301_} into the Wheeler-DeWitt equation \eqref{WDW301}, it is obtained at order $\hbar^1$
\be
\pm 2 i \hbar \omega_{DS} \frac{\partial{\xi}}{\partial a} - \hbar^2 \frac{\partial^2\xi}{\partial \chi^2} + \chi^2 \xi = 0,
\ee
which is the time dependent Schr\"{o}dinger equation
\be\label{SCH01}
i \hbar \frac{\partial }{\partial \eta_\pm} \xi(\eta_\pm,\chi) = \frac{1}{2} \left( -\hbar^2 \frac{\partial^2}{\partial \chi^2} + \chi^2 \right) \xi(\eta_\pm,\chi) ,
\ee
provided that  one identifies the (conformal) time variable of the background spacetime of the two universes, $\phi^\pm$, by 
\be
\frac{\partial }{\partial \eta_\pm} = \mp \ \omega_{DS} \frac{\partial}{\partial a } \ \ \Rightarrow \ \ \eta_\pm = \mp \int \frac{da}{\omega_{DS}}   = \mp \int \frac{dt}{a} .
\ee
If we assume that the physical time variable is the variable measured by real clocks, which are made up of matter, and thus that it is the time that appears in the Schr\"odinger equation, then, in terms of the physical time of an observer in one of the universes the Schrödinger equation for the fields in the partner universe turns out to be the Schrödinger equation of a field $\bar \varphi$ that is CP conjugated with respect the field in the observer's universe (see, Sec. \ref{sec03}). The two universes form then a universe-antiuniverse pair. The process can be compared with the creation of a electron-positron pair (see, Fig. 1 of Ref. \cite{Vilenkin1982}, and Fig. \ref{figure202}), which can be seen as the creation of an electron moving backward in time and an electron moving forward in time. Here, the 'time' variable is the scale factor so 'moving forward in time' means an expanding universe and 'moving backward in time' means a contracting universe, so the creation of a contracting-expanding pair of universes can be paralleled  with the creation  of a electron-positron pair.

%%%%%%%%%%%%%%%%%%%%%%%%%%%%%%%%%%%%%%%%%%
%%%%%%%%%%%%%%%%%%%%%%%%%%%%%%%%%%%%%%%%%%
%%%%%%%%%%%%%%%%%%%%%%%%%%%%%%%%%%%%%%%%%%
%%%%%%%%%%%%%%%%%%%%%%%%%%%%%%%%%%%%%%%%%%

\section{Third quantisation formalism}\label{sec03}

\subsection{Historial review}\label{sec0301}

There was a great excitation in the 80's with the formulation of the third quantisation formalism and the associated description of topology change in quantum gravity \cite{Caderni1984, Coleman1988, Coleman1988b, McGuigan1988, Rubakov1988, McGuigan1989, McGuigan1990, Strominger1990}. At that time the cosmological paradigm was a universe in a decelerating expansion with a zero value of the cosmological constant. The third quantisation formalism, which was initially proposed \cite{Caderni1984} as an analogue to the second quantisation formalism of a quantum field theory, fitted well with the description of the quantum fluctuations of the spacetime. As a consequence of the interaction with the fluctuations of the spacetime, it turns out that  the coupling constants become dynamical functions and this was seen as a possible  explanation for the expected vanishing value of the cosmological constant.

\begin{figure} 
\centering
\includegraphics[width=13 cm]{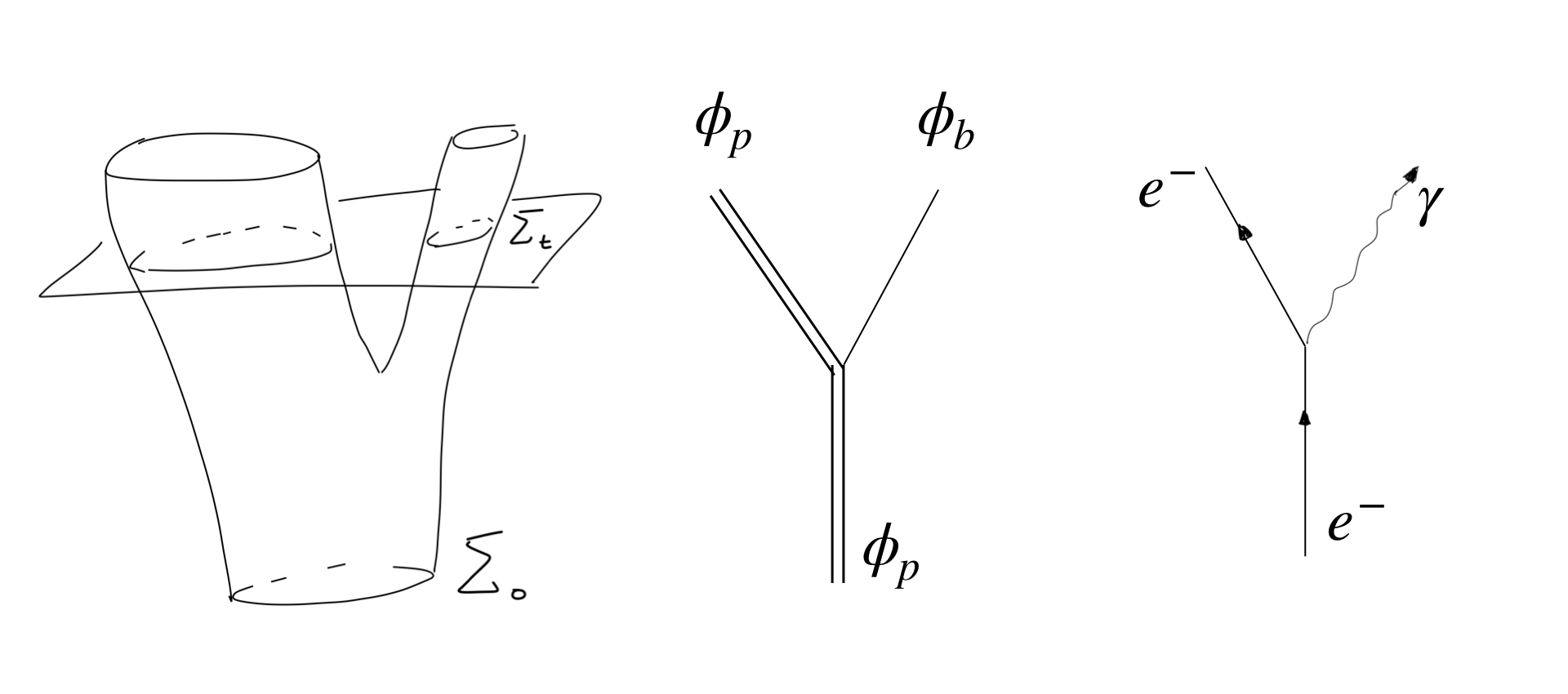}
\caption{Left) the foliation of a given spacetime can give rise, at some value $t_0$, two disconnected spatial sections. Middle: schematic representation; Right: creation of a photon from the scattering of an electron.}
\label{figure501}
\end{figure}

On the other hand, general relativity is a (local) geometrical theory and therefore it does not account for  the global topology of the spacetime manifold. However, it may perfectly happen that the foliation of the spacetime gives rise, at some given value $t$, a collection of simply connected spatial sections (see Fig. \ref{figure501} Left). From the point of view of the evolution of the universe the topology change represented in Fig. \ref{figure501} (Middle) can be seen as the creation of a universe (universe $B$) from the pre-existing one (universe $A$). In fact, the process is formally similar to the QED process depicted in Fig. \ref{figure501} (Right), which represents the creation of a photon from the propagation of an electron. As we know from QED this kind of processes are better explained in the formalism of quantum field theory where we can define creation and annihilation operators of particles. Therefore, it seems reasonable to develop a  field theoretical approach too to describe the creation (and annihilation) of the universe(s).

As an introductory example, let us consider the Wheeler-DeWitt equation of a closed DeSitter spacetime [see \eqref{WDW212}], which can be written as
\be\label{HO01}
\ddot{\phi} + \omega^2(a) \phi = 0 ,
\ee
where, $\phi=\phi(a)$, $\dot \phi \equiv \frac{d\phi}{da}$, and
\be
\omega^2(a) = \frac{H_0^2 a^4 - a^2}{\hbar^2} ,
\ee
with, $H^2 = \Lambda/3$. Clearly, \eqref{HO01} is the equation of a harmonic oscillator with the scale factor $a$ playing the role of the time variable. One can assume then that \eqref{HO01} is the result of the variational principle of the action of a harmonic oscillator with time dependent frequency,
\be
S_3 = \int da \, \left( \dot\phi^2 - \omega^2 \phi^2 \right) ,
\ee
from which one can obtain the conjugate momentum, $P_\phi = \dot \phi$, and construct a the corresponding Hamiltonian,
\be
H_3 = \frac{1}{2} P_\phi^2 + \frac{\omega^2}{2} \phi^2 .
\ee
The third quantisation procedure consists in promoting the variables $\phi$ and $P_\phi$ to quantum operators in the usual way, $\hat\phi \rightarrow \phi$  and $\hat P_\phi \rightarrow -i \hbar \partial_\phi$, and describe the quantum state of the whole spacetime manifold by the use of a new wave function, $\Psi(\phi)$, constructed as
\be\label{3PSI01}
\Psi(\phi) = \int \delta \phi e^{\frac{i}{\hbar} S_3} ,
\ee
in the path integral approach, or via the Schrödinger equation,
\be\label{3H01}
H_3 |\Psi \rangle = i \hbar \frac{\partial |\Psi \rangle}{\partial a} .
\ee
One can also define the \emph{ladder} operators, $\hat b$ and $\hat b^\dag$, of this particular harmonic oscillator in terms of the operators $\hat \phi$ and $\hat P_\phi$, as usual, and construct the state of the whole spacetime manifold, whatever the topology it has\footnote{\label{fn05}In general, a non simply connected manifold can be divided into $N$ simply connected parts \cite{Hawking1990} and this $N$ parts can be seen as $N$ classically independent universes.}, in terms of the eigenstates of the number operator, $|N, a \rangle$, so
\be\label{3PSI02}
| \Psi \rangle = \sum_N C_N |N, a \rangle  .
\ee
The number states, $|N, a \rangle$, would represent  $N$ universes with a scale factor $a$ and the state of the whole spacetime manifold, $|\Psi\rangle$, would be a quantum superposition of different number states (see, fn. \ref{fn05}). Classically, these $N$ universes are disconnected and therefore one should just consider one of them as representing our universe and disregard the rest as being physically irrelevant. However, from the quantum mechanical standpoint new phenomena may appear as quantum correlations and other collective behaviour so it seems interesting, at least in principle, to analyse the quantum description of the whole many-universe state.

\begin{figure} 
\centering
\includegraphics[width=13 cm]{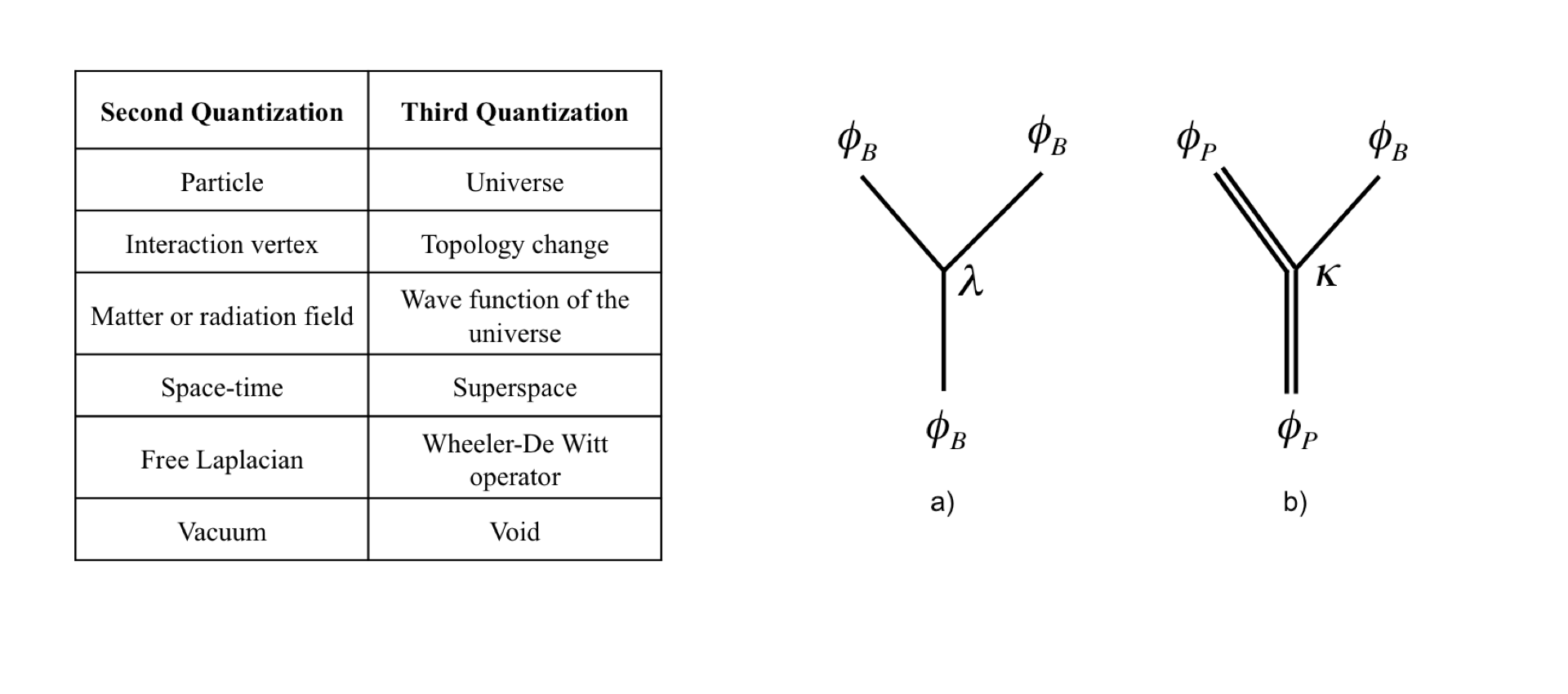}
\caption{Analogy between the second and the third quantisation  \cite{Strominger1990, RP2010}.}
\label{figure502}
\end{figure}

 \paragraph{\it Parent and baby universes: the hybrid action}

The third quantisation formalism was mainly applied in the 80-90's to the description of the quantum fluctuations of the spacetime. Let us notice that the quantum gravity of multiply connected spacetime manifolds  can be applied to two well distinguished cases: one that accounts for the properties of large regions of the spacetime, called \emph{parent universes}\footnote{Typically, large regions of order of the Hubble length of our universe.}, and another that focus on a local region of the spacetime where small pieces of length of the order of the Planck length can branch off and disconnect from the parent spacetime and become small \emph{baby universes} \cite{Strominger1990}. In both cases, the spacetime manifold under study turns out to be non simply connected. However, in the 80-90's the idea of a multiverse was not seriously considered and the main problem at that time was to explain the supposedly zero value of the cosmological constant\footnote{In the 80-90's, the paradigm was a universe in a non accelerated expansion.}.

Then, from the point of view of the third quantisation formalism, our universe can then be seen as a large parent universe propagating in a plasma of baby universes. The effects of the baby universes could be measured  by their influence on the observable properties of the parent universe, and the most representative picture of the baby-parent universe interaction becomes the so-called \emph{hybrid action}\cite{Strominger1990}, in which the parent universe is described by a second quantised wave function, $\phi_p(q^a)$,  and the baby universes are described by the third quantised wave functions, $\hat \phi_b$, i.e. the behaviour of the spacetime is assumed classical and its fluctuations are seen as small particles propagating in the parent spacetime. The total action is then given by
\be
S_T = S_0(a,\varphi) + S_b(\hat \phi_b) + S_I(a,\varphi; \hat\phi_b) , 
\ee
where $S_0(a,\varphi)$ is the Einstein-Hilbert action of the homogenous and isotropic parent spacetime with scale factor $a$ and matter field $\varphi$, $S_b(\hat \phi_b)$ is the third quantised action of the baby universes, and $S_{I}$ is the action of interaction,
\begin{equation}\label{SI501}
S_I(a,\varphi; \hat\phi_b)  = \int dt \mathcal{N} \sum_i \mathcal{L}_i(t, \vec{x}) \hat \phi_b^i  = \int dt \mathcal{N} \sum_i \mathcal{L}_i(t, \vec{x}) (\hat b_i^+ + \hat b^-_i) ,
\end{equation}
where the index $i$ labels the different modes of the baby universe field (i.e. it labels different species of baby universes), and $\mathcal{L}_i(t, \vec{x})$ is called the insertion operator at the nucleation event \cite{Strominger1990}. It defines the space-time points of the parent universe in which the baby universes effectively nucleate.

Two main problems were addressed in the 80's using the third quantisation formalism: the dynamical value of the coupling constants and the loss of quantum coherence (decoherence) produced by the plasma of baby universes.

As an example of the former, let us consider a universe with a matter field $\varphi(t)$ that is coupled to the spacetime through the interaction with the baby universes. The parent universe two point function becomes \cite{Strominger1990}
 \beq\nn
 G(\varphi_f, \varphi_i) &\propto& \langle \hat \phi_b, 0 | \phi_p(\varphi_f) \phi_p(\varphi_i) | \hat \phi_b, 0 \rangle \\  \label{Gf501} &=&\langle \hat \phi_b, 0 | \int_{\varphi_i}^{\varphi_f} d\varphi(t) \int_0^\infty dN \, e^{i  S_p+ iS_I}  | \hat \phi_b, 0 \rangle ,
 \eeq
 where $S_I$ is given by \eqref{SI501} (for simplicity, let us assume just one specie of baby universes). Now, suppose the baby universes are in a wave function eigenstate $|\alpha\rangle$ , with $\hat \phi_b |\alpha \rangle = \alpha |\alpha\rangle$, where $\alpha$ satisfies
 \be\label{a501}
 \left( \nabla^2_q + m_b^2 \right) \alpha = 0 .
 \ee
 In that case, the function \eqref{Gf501} becomes
 \be
 G(\varphi_f, \varphi_i) \propto \int_{\varphi_i}^{\varphi_f} d\varphi(t) \int_0^\infty dN \, e^{i \tilde S_p} ,
 \ee
 where
 \be
 \tilde S_p = \int dt N a^3 \left( \frac{1}{2 N^2} \dot\varphi^2 - V(\varphi) - \kappa \alpha \right) .
 \ee
The effects of the baby universes is thus encoded in an addition of an ordinary potential to the second quantised action. The new term is dynamical in the sense that it must satisfy the dynamical equation \eqref{a501}. That was used as an argument for a possible mechanism for the vanishing value of the cosmological constant, which was the expected value at that time.

The other question addressed with the third quantisation formalism was the loss of quantum coherence of the matter fields caused by their propagation in the plasma of baby universes. Basically, the argument was the following \cite{Coleman1988}. Let us suppose the composite state between a matter field $\varphi$ and the baby universes. Let $|\varphi, n\rangle$ be the state in which the matter field is in the state $|\varphi\rangle$ and there are $n$ baby universes. Then, the initial state is
\be\label{in501}
|{\rm in}\rangle = | \varphi^{\rm in}, 0\rangle , 
\ee
where $|\varphi^{\rm in}\rangle$ is the initial state of the matter field. If we do not measure the state of the baby universes and we therefore integrate out their quantum state from the composite state \eqref{in501}, then, the initial state is described by the density matrix \cite{Coleman1988}
\be
\rho^{\rm in} = |\varphi^{\rm in}\rangle \langle \varphi^{\rm in} | .
\ee
After the interaction with the baby universes, the final state becomes a linear combination of the states of the fluctuations and the corresponding states of the matter field that come out from the interaction with the $|n\rangle$ states. For simplicity, let us consider just two $|n\rangle$ states, $|0\rangle$ and $|1\rangle$. The composite state after the interaction would be
\be\label{out501}
|{\rm out} \rangle = |\varphi_0 , 0\rangle + |\varphi_1 , 1\rangle ,
\ee
where $\varphi_0$ and $\varphi_1$ are in general different and the linear combination in \eqref{out501} can be weighted accordingly. Then, the reduced density matrix that describes the state of the matter field alone becomes after the interaction
\be
\rho^{\rm out} = | \varphi_0 \rangle \langle \varphi_0 | + | \varphi_1 \rangle \langle \varphi_1 |  .
\ee
The field turns out to be in a statistical mixture of two states. The initial state was a pure state, i.e. a state of total information with zero entropy, $S=0$. The final state, instead, becomes a mixed state with entropy, $S > 0$, so information (quantum coherence) has been lost. Coleman's argument was that the operators of the baby universes must be independent of the coordinates of the parent spacetime and thus, the coupling with the matter fields is independent of their evolution. In that case, the state of the field does not change along the time evolution because the states of the baby universes do not change \emph{in time}\footnote{It means that the state \eqref{out501} would actually be, $$|{\rm out} \rangle = |\varphi_0 \rangle \left( |0\rangle + | 1\rangle\right).$$ In that case, when we trace out the state of the baby universes the state of the field remains unaffected in the initial state, $\rho^{\rm out} = | \varphi_0 \rangle \langle \varphi_0 |$}.  However, counter arguments were also given for the loss of quantum coherence \cite{PFGD1992a, PFGD1992b}.

\subsection{Quantum field theory in $M \equiv {\rm Riem}(\Sigma)$}\label{sec0302}

\subsubsection{Geometrical structure of $M$}\label{sec030201}

We have seen in Sec. \ref{sec02} that the evolution of the universe can be seen as the time evolution of the $3$-dimensional metric that is induced on the spatial hypersurfaces by the $4$-dimensional metric that is the solution of the Einstein's equations. Therefore, the evolution of the universe is a trajectory in  the space of Riemannian symmetric $3$-metrics with components, $h_{ab}$. Let us call it $M$. With the DeWitt metric \eqref{DWM02}, $M$ becomes a metric space, where we can define the line element as
\be\label{LE101}
ds^2 =  G^{abcd} dh_{ab} dh_{cd} .
\ee
However, not all of the $h_{ab}$ components are independent. A symmetric $3$-metric has only $6$ independent components so it turns out that $M$ is isomorphic to $\R^6$. Thus, we can make the following choice for the coordinates\footnote{We have followed the normalisation applied in Ref. \cite{DeWitt1967}.} in $M$, 
\be
q^A = \{h_{11}, h_{22}, h_{33}, \sqrt{2}h_{23}, \sqrt{2} h_{13}, \sqrt{2} h_{12} \} ,
\ee
in terms of which the line element \eqref{LE101} can be written
\be\label{LE531}
ds^2 = G_{AB} dq^A dq^B ,
\ee
where $G_{AB}$ is a $6$-dimensional metric tensor that is related to the components of DeWitt's metric, $G^{abcd}$. The signature of $M$ is $(-,+,+,+,+,+)$, whichis easy to checked for the case of the flat metric, $h_{ab} = \delta_{ab}$, and because the signature remains invariant under a change of coordinates, it holds then for the general case too. Thus, DeWitt showed \cite{DeWitt1967} that the $6$-dimensional space $M$ is indeed a $5+1$ dimensional space with a $1$ \emph{time-like} dimension and an orthogonal $5$ dimensional \emph{space-like} subspace. As the coordinate of the time-like subspace it is appropriate to take the coordinate $\tau$ defined by \cite{DeWitt1967, Higuchi1995, Kiefer2007}
\be
\tau = \left( \frac{32}{3} \right)^\frac{1}{2} h^{1/4} ,
\ee
where, $h={\rm det} h_{ij}$, which essentially represents the volume of an infinitesimal volume element of the spatial sections of the spacetime ($V \propto \int dx^3 \sqrt{h}$). The hypersurfaces of constant $\tau$ are the {space-like} sections of $M$, labelled by  $\bar M$ \cite{DeWitt1967}. Then, in terms of the variables, $q^\mu = \{\tau, \bar q^A\}$, where $\bar q^A$, with $A=1,\ldots ,5$, are the five coordinates in $\bar M$, the line element \eqref{LE531} becomes
\be\label{SME01}
d s^2 = - d\tau^2 + h_0^2 \tau^2 d\bar s^2 = - d\tau^2 + h_0^2 \tau^2 \bar G_{AB} d\bar q^A d\bar q^B ,
\ee
where, $h_0^2 = 3/32$, and $d\bar s$ is the line element in $\bar M$, with \cite{DeWitt1967}
\be\label{Gbar01}
\bar G_{AB} = {\rm tr}\left( h^{-1} h_{,A} h^{-1} h_{,B} \right) \equiv h^{ij} \frac{\partial h_{jk}}{\partial \bar q^A} h^{k l} \frac{\partial h_{l i}}{\partial \bar q^B}  .
\ee
The metric \eqref{Gbar01} is invariant under a conformal transformation of the metric and in particular it is invariant under the change, $h_{ab} \rightarrow \xi(h) h_{ab}$, so it is convenient labelling the points of $\bar M$ with the five independent components of the transformed metric,
\be\label{BMET01}
\bar h_{ab} = h^{-\frac{1}{3}} h_{ab} ,
\ee
which has unit determinant.  Furthermore, $\bar M$ is noncompact and diffeomorphic to Euclidean 5-space \cite{DeWitt1967}, with Ricci tensor $\bar R_{AB}$ given by\footnote{This result is corrected from the one given in Ref. \cite{DeWitt1967} by a factor $\frac{1}{2}$, which is already noted in Ref. \cite{Higuchi1995}.}
\be
\bar R_{AB} = -\frac{3}{4} \bar G_{AB} ,
\ee
and scalar curvature
\be
\bar R = -\frac{15}{4} \equiv \frac{k}{a^2} ,
\ee
with, $k=-1$. $\bar M$ is then an 'Einstein space' of constant negative curvature. In a homogeneous universe the time-like variable $\tau$ represents the volume of the spatial sections of the universe and the five coordinates, $\bar h_{ab}$, represent the \emph{shape} of a unit volume. A fixed point in $\bar M$ represents therefore the evolution of a universe that scales the volume of the spatial sections without changing their shape, and lines of constant $\tau$ represent different shapes of the spatial sections of the universe with a fixed given volume. Thus, the line element \eqref{SME01} reveals the space $M$ as  \emph{a set of "nested" 5-dimensional submanifolds, all having the same intrinsic shape} \cite{DeWitt1967}. From that point of view, the $5$-dimensional submanifold $\bar M$ can be seen as a proper realisation of what is called the 'shape space' \cite{Barbour2011}.

\begin{figure} 
\centering
\includegraphics[width=7 cm]{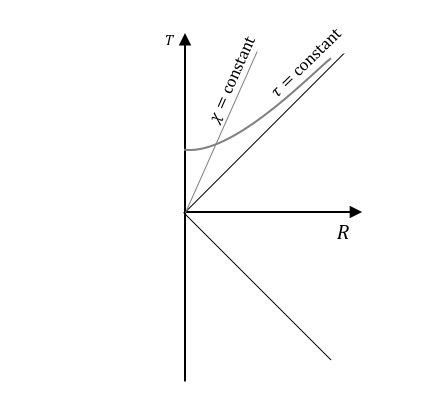}
\caption{The space of three metrics, $M$,  turns out to be a particular coordination of upper (lower) light cone of a $6$-dimensional Minkowki space. Every point in the $T,R$ plane is a four-sphere of unit radius. Lines of constant $\tau$ are lines of constant volume of the spatial sections of the spacetime (with different shapes). Lines of constant $\chi$ correspond to different volumes of the same shape (a scaling universe). Something similar occurs in the lower light cone, which would represent a time reversed copy of the universe.}
\label{figure503}
\end{figure}

On the other hand,  the space $M$ with the metric \eqref{SME01} has the same formal structure of a Friedmann-Robertson-Walker spacetime with the hyperbolic $5$-space $H^5$ as the 'spatial' section. In particular, it has the same formal structure than the Milne spacetime\footnote{For the Milne spacetime, see Ref. \cite{Griffiths2009}}. Therefore, one can find a set of coordinates $(\chi, \theta, \phi, \psi, \zeta)$ in $\bar M$ in terms of which the metric \eqref{SME01} can be written as\footnote{A rescale, $\chi \rightarrow a \chi$, $\theta \rightarrow a \theta$, $\ldots$, has been made to absorb the constant $a$.} 
\be\label{SME102}
d s^2 = - d\tau^2 + \tau^2 \left( d\chi^2 + \sinh^2\chi d\Omega_4^2 \right)  ,
\ee
where, $\chi \in[0,\infty)$, and $d\Omega_4^2$ is the line element on the $4$-sphere of unit radius 
\be
d\Omega_4^2 =  d\theta^2 + \sin^2\theta \left( d\phi^2 + \sin^2\phi (d\psi^2 + \sin^2\psi d\zeta^2) \right) . 
\ee
The Milne spacetime is a particular coordination of part of the Minkowski spacetime. It does not cover the whole Minkowski spacetime but only the interior of the upper (or the lower, with a time reversal change) light cone of the Minkowski spacetime.  Something similar occurs in $M$. Let us introduce the variables 
\be
T = \tau \cosh  \chi \ \ , \ \ R = \tau \sinh  \chi ,
\ee 
in terms of which the line element \eqref{SME102} becomes
\be\label{SME04}
ds^2 = - dT^2 + dR^2 + R^2 d\Omega_4^2 ,
\ee
with, $0 < T < \infty$ and $0 < R < \infty$. The metric \eqref{SME04} is nothing more than the metric of a $6$-dimensional Minkowski space, and the Milne space only cover the upper light cone (see Fig. \ref{figure503}). The interior of the lower light cone is covered by a time reversal change of coordinates, $\tau \rightarrow -\tau$ (let us notice that the metric \eqref{SME01} is invariant under this change). However, although the manifold $\bar M$ is geodesically complete, the manifold $M$ is not. The scalar curvature of $M$,
\be
R = - \frac{20}{\tau^2}
\ee
presents a singular frontier of infinite curvature, located at $\tau = 0$, where all geodesics in $M$ eventually hit \cite{DeWitt1967}. It means that the upper and the lower light cones of the $6$-dimensional Minkowski space must be considered independently. They represent two time reversed copies of the universe, i.e. two universes related by a time reversal change of the time (volume) variable. We shall see that, quantum mechanically, this can be seen as a universe-antiuniverse pair.

\subsubsection{Classical evolution of the universe}\label{sec030202}

Let us now analyse the evolution of the universe in the space, $M\equiv {\rm Riem}(\Sigma)$. From a geometrical point of view, the evolution of the universe is the trajectory that extremizes the Einstein-Hilbert action \eqref{EHA02}, which can conveniently be written as\footnote{Assuming the value, $N^i = 0$.}\cite{Kiefer2007}
\be\label{EHA101}
S_{EH} =   \frac{1}{2} \int_\mathcal{M} dt d^3x N \left( \frac{1}{N^2} G^{abcd} \dot{h}_{ab} \dot{h}_{cd} - m^2(h_{ab}) \right) ,
\ee
where have made the rescale, $G^{abcd} \rightarrow \frac{1}{32 \pi G} G^{abcd}$, with $G$ the Newton's constant, and the potential terms of the Einstein-Hilbert action have been gathered in a \emph{mass term}
\be\label{MASS01}
m^2(h_{ab}) =  \frac{\sqrt{h}}{8 \pi G} \left( 2\Lambda - ^{(3)}R \right) .
\ee
In term of the variables,  $q^A = (\tau, \bar q^A)$, the action \eqref{EHA101} can be written
\be\label{EHA102}
S_{EH} =   \frac{1}{2} \int_\mathcal{M} dt d^3x N \left( \frac{1}{N^2} G_{AB} \dot{q}^{A} \dot{q}^{B} - m^2(q^A) \right) ,
\ee
where, $G_{AB}$, is given by \eqref{SME01} or \eqref{SME102}. The Einstein-Hilbert action has been written in the form of  \eqref{EHA102} to make clear the formal resemblance with respect to the action of a particle that moves in the spacetime,
\be\label{ACT101}
S[x^\mu(\lambda)] = \!\frac{1}{2} \int \left(\!\frac{1}{N^2} g_{\mu\nu} \dot{x}^\mu \dot{x}^\nu - m^2  \right)  N d\lambda ,
\ee
with, $\dot x^\mu =\frac{d x^\mu}{d\lambda}$, for which the trajectory is given by the geodesic equation,
\be\label{GEO01}
\ddot{x}^\mu + \Gamma ^\mu_{\alpha \beta} \dot{x}^\alpha \dot{x}^\beta = 0 ,
\ee
with
\be\label{Chr01}
\Gamma^\mu_{\alpha \beta} = \frac{1}{2} g^{\mu\nu} \left( \frac{\partial g_{\alpha\nu}}{\partial x^\beta} + \frac{\partial g_{\beta\nu}}{\partial x^\alpha} -  \frac{\partial g_{\alpha\beta}}{\partial x^\nu} \right)\, .
\ee
The case of the universe is formally similar. The evolution of the universe is a trajectory in $M$. The only difference is the trajectory is not a geodesic\footnote{The fact that the trajectory is not a geodesic is not really determinant. In fact, using a generalisation of the Maupertuis principle \cite{Biesiada1994, Garay2018}, one can compute the metric where the trajectory of the universe is a geodesic. Let us consider the reparametrisation given by, $d\tilde t = m^2(h_{ab})  dt$ and $G^{abcd}\rightarrow \tilde G^{abcd} = m^2(h_{ab}) G^{abcd}$. In that case, the action \eqref{EHA101} turns out to be
\be
S_{EH} =   \int_\mathcal{M} d\tilde t d^3x N \left( \frac{1}{2 N^2} \tilde G^{abcd} {h}'_{ab} {h}'_{cd} - 1 \right) ,
\ee
where, $h'_{ab}=\frac{d h_{ab}}{d\tilde t}$. In the superspace determined by the supermetric $\tilde G^{abcd}$ the evolution of the universe turns out to be a geodesic.} because the non constant potential, $m^2(h_{ab})$. Instead, it is given by
\be\label{GEQ01}
\ddot q^A + \Gamma^A_{BC} \dot q^B \dot q^C = - G^{AB} \frac{\partial V(q)}{\partial q^B} ,
\ee
where, $\dot q^A=\frac{d q^A}{dt}$ and $2 V = m^2(h_{ab})$, and the Christoffel's symbols are defined analogously in terms of the metric components as
\be
\Gamma^A_{BC} = \frac{1}{2} G^{AD} \left( \frac{\partial G_{DC}}{\partial B} + \frac{\partial G_{BD}}{\partial C} - \frac{\partial G_{BC}}{\partial D} \right) .
\ee
In terms of the variables $(\tau, q^A)$, the equations \eqref{GEQ01} turn out to be
\beq\label{GEQ02}
\ddot{\tau} + h_0^2 \tau \bar G_{AB} \dot{\bar q}^A \dot{\bar q}^B = \frac{\partial V}{\partial \tau}   , \\ \label{GEQ03}
\ddot{\bar q}^A + \frac{2\dot \tau}{\tau} \dot {\bar q}^A + \bar \Gamma_{BC}^A \dot {\bar q}^A \dot {\bar q}^B = - \frac{1}{h_0^2 \tau^2} \bar G^{AB} \frac{\partial V}{\partial {\bar q}^B} ,
\eeq
As we have seen in the preceding section, a geodesic in $M$ eventually hits the singular frontier located at, $\tau = 0$ (the zero volume hypersurface). However,  because the potential term, which may also include the Lagrangian of the matter fields, the universe does not follow a geodesic in $M$ and may thus avoid the singular frontier. The paradigmatic case is the closed DeSitter spacetime. In addition to (\ref{GEQ02}-\ref{GEQ03}), one can use the Hamiltonian constraint \eqref{HC00}, which in terms of the $(\tau, \bar q^A)$ coordinates reads,
\be
\dot \tau^2 = h_0^2 \tau^2 \bar G_{AB} \dot{\bar q}^A \dot{\bar q}^B + m^2 = h_0^2 \tau^2 \dot{\bar s}^2 + m^2 .
\ee
In the case that the right hand side of \eqref{GEQ03} is zero, it can be shown that \cite{DeWitt1967}, $\dot{\bar s} = \alpha/\tau^2$, with $\alpha$ a constant of integration\footnote{For a FRW spacetime, $\alpha=0$ (because, $\dot{\bar s}=0$).}, and then
\be
\dot \tau^2 = \frac{h_0^2 \alpha^2}{\tau^2}  + m^2 .
\ee

\begin{figure} 
\centering
\includegraphics[width=13 cm]{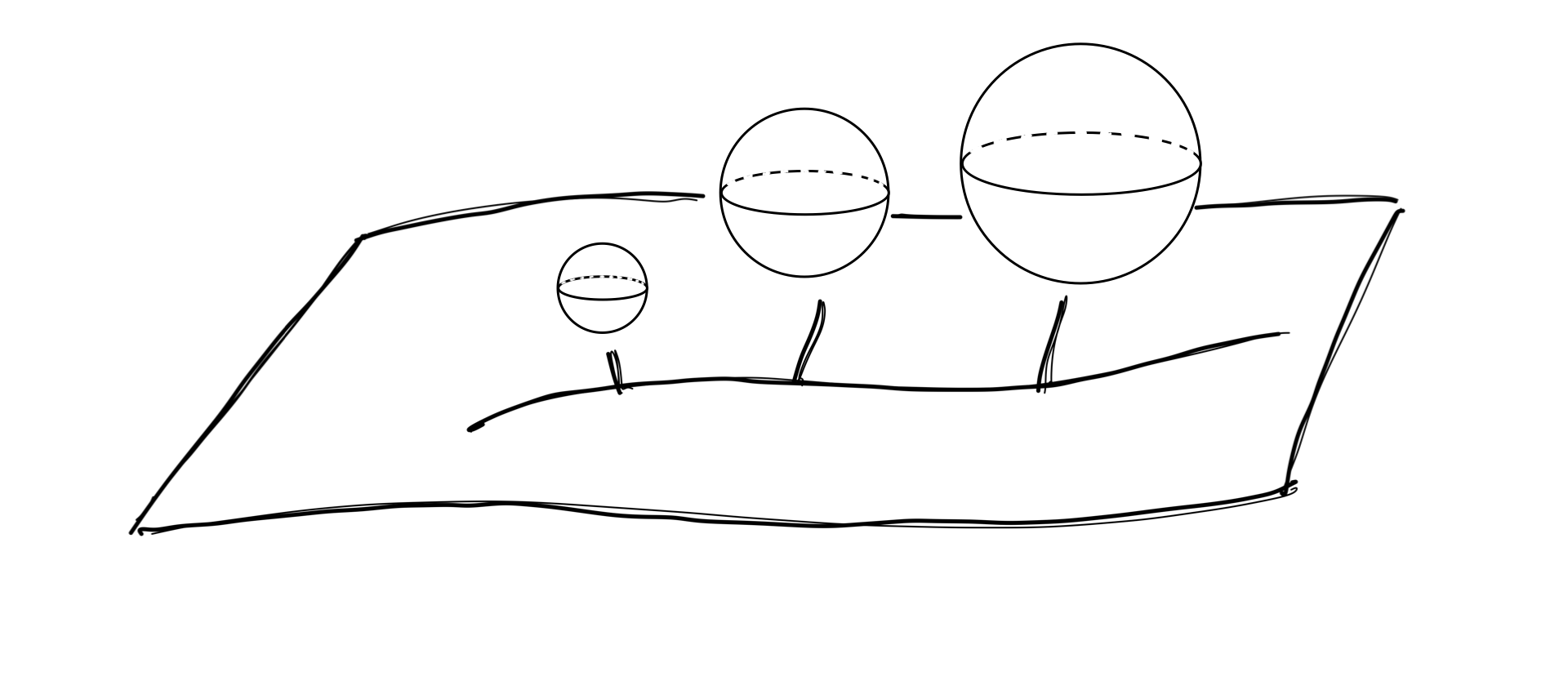}
\caption{The evolution of the universe can be seen as a trajectory in the superspace \cite{Hartle1990}.}
\label{figure504}
\end{figure}

In the case for which, $\Lambda \gg \, ^{3}R$, the potential is proportional to $\tau$, $m^2 = h_0^2 H_0^2 \tau^2$, and
\be
\dot \tau^2 = h_0^2 \left( \frac{\alpha^2}{\tau^2} + H_0^2 \tau^2 \right) ,
\ee
which solutions given by
\be\label{tau01}
\tau^2(t) = \alpha \sinh\left( 2 H_0 \Delta \tilde t - \ln H_0 \alpha \right) ,
\ee
for, $\alpha \neq 0$, and
\be\label{tau02}
\tau(t) \propto e^{H_0 \Delta \tilde t} ,
\ee
for, $\alpha= 0$ (FRW spacetime), with, $\Delta \tilde t = h_0 \Delta t$. From (\ref{tau01}-\ref{tau02}), one can see that, as expected, for a universe stage in which, $\Lambda \gg \, ^{3}R$, the expansion of the volume element is exponential.

\subsubsection{Quantum field theory in $M$}\label{sec030203}

As we have seen above, the third quantisation\footnote{For recent works on the third quantisation, see Refs. \cite{Pimentel2001, Kim2012, Ohkuwa2013, Calgani2012, Faizal2014, Balcerzak2019, Balcerzak2020, Campanelli2020}.} procedure consists in promoting the \emph{field} $\phi(h_{ab})$ and its conjugate momentum to quantum operators. One can then pose another wave function, $\Psi$, as in (\ref{3PSI01}-\ref{3PSI02}), and work with the corresponding Schrödinger equation [see, \eqref{3H01}]. However, it turns out to be much more interesting to develop and study the quantum field theory (QFT) of the field $\phi(h_{ab})$ propagating in the $6$-dimensional space $M$.

Let us first notice that, in terms of the coordinates $q^A = \{\tau, \bar q^A \}$ (see, Eq. \eqref{SME01}), the Hamiltonian constraint \eqref{HC00} can be written as
\be\label{HC01}
\mathcal H = G^{AB} p_A p_B + m^2(q,\varphi) = 0 ,
\ee
where,  $G^{AB}$ is the inverse of \eqref{SME01}, and in $m^2(q, \varphi)$ we have also included the Hamiltonian of the matter fields (which for simplicity have not been considered so far) that can generically be encapsulated in a variable $\varphi$, 
\be\label{MAS01}
m^2(q,\varphi) = m_g^2(q) + 2 \mathcal H_m(q,\varphi) ,
\ee
with, $m_g^2(q)$, given by [see, Eq. \eqref{MASS01}]
\be\label{MAS02}
m_g^2(q) = \frac{h_0^2 \tau^2}{8 \pi G} \left( 2 \Lambda - {}^3R(q) \right) ,
\ee
and
\be\label{HMT01}
\mathcal H_m = \frac{1}{2 h_0^2\tau^2} p_\varphi^2 + \ldots + h_0^2 \tau^2 V(\varphi) ,
\ee
where the dots indicate terms that contain spatial derivatives of the matter fields, which for simplicity we shall consider negligible.

Under canonical quantisation of the momenta in the Hamiltonian constraint  \eqref{HC01} one obtains the Wheeler-DeWitt equation, which with an appropriate choice of factor ordering can also be written as,
\be\label{WDE01}
\left( -\hbar^2 \Box_q + m^2(q,\varphi) \right) \phi(q,\varphi) = 0 ,
\ee
where, $\phi(q,\varphi)$, is the so-called wave function of the universe \cite{Hartle1983}, and
\be\label{BOX01}
\Box_q = \nabla \vec\nabla = \frac{1}{\sqrt{- G}} \frac{\partial }{\partial q^A} \left(\sqrt{-G} \,  G^{AB} \frac{\partial}{\partial q^B} \right) ,
\ee
where, $G = {\rm det}\, G_{AB}$, and we have used the customary definitions of the gradient and the divergence in a curved space,
\be\label{NAB532}
\vec\nabla \phi = G^{AB} \frac{\partial \phi}{\partial q^B} \ , \ {\nabla} \cdot \vec F= \frac{1}{\sqrt{-G}} \frac{\partial}{\partial q^A} \left( \sqrt{-G} F^A \right) .
\ee 
With these definitions, the Wheeler-DeWitt equation \eqref{WDE01} can be obtained from the variational principle of the third quantised action
\be\label{3ACT531}
S_{(3)} = \frac{1}{2} \int dq \sqrt{-G} \left( -\hbar^2 \vec{\nabla}\phi \cdot \vec{\nabla}\phi - m^2(q) \phi^2 \right) ,
\ee
Variation of \eqref{3ACT531} with respect to the wave function $\phi$ gives rise to the wave equation \eqref{WDE01}. Now, following the analogy with a QFT, we can define a conserved current in the superspace,
\be
\vec J =  i \hbar  \left( \phi^* \vec \nabla \phi - (\vec \nabla \phi^*) \phi \right) ,
\ee
from which it can easily be checked that, $\nabla\cdot \vec J = 0$, and the following inner product
\be\label{INPROD531}
(\phi_m, \phi_n) = i \hbar \int d\vec{\Sigma} \left( \phi_m^* \vec \nabla \phi_n - (\vec \nabla \phi_m^*) \phi_n  \right)  ,
\ee
where $d\vec{\Sigma}$ is the future oriented surface element of the $5$-dimensional spacelike subspace.

The procedure of field quantisation consists in promoting the wave function $\phi(q)$ to an operator and expand it into modes that are orthonormal with respect to the inner product \eqref{INPROD531}. Then \cite{McGuigan1988}
\be\label{MEX531}
\hat \phi(q) = \sum_n \phi_n(q) \hat A_n + \phi^*_n(q) \hat A_n^\dag ,
\ee
where $\phi_n(q)$ is a complete set of orthonormal solutions of the Wheeler-DeWitt equation \eqref{WDE01}, the index $n$ symbolises the particular set of quantum number associated to that state \cite{McGuigan1988}, and $\hat A^\dag_n$  and $\hat A_n$ are respectively the creation and annihilation operators of modes $\phi_n$, satisfying the customary commutation relations, 
\be
[\hat A_n, \hat A^\dag_m] = \delta_{nm} \ , \  [\hat A_n, \hat A_m] = 0 = [\hat A_n^\dag, \hat A_m^\dag] .
\ee
In terms of the variables $(\tau, \bar q)$, the label $n$ of the modes $\phi_n$ in \eqref{MEX531} can be associated to the $5$-dimensional spacelike momentum of the \emph{particles} that propagate in the space $M$.  In particular, we have seen that $M$ has the geometrical structure of a $5+1$-dimensional Friedmann-Robertson-Walker universe so we can use this information to develop the quantisation of the field $\phi(q)$.  In terms of the coordinates $q^A =(\tau, \bar q^A)$, the Laplace-Beltrami operator \eqref{BOX01} can be written as,
\be
\Box_q = - \frac{1}{\tau^5} \frac{\partial}{\partial \tau} \left( \tau^5 \frac{\partial}{\partial \tau} \right) + \frac{1}{\tau^2}{\Box}_{\bar q} ,
\ee
where, ${\Box}_{\bar q} $, is the corresponding $5$-dimensional Laplacian (that with $\Box_{\bar q}$ given by \eqref{BOX01} with the $5$-dimensional metric $\bar G_{AB}$ instead of $G_{AB}$ and without the minus sign in the square roots). In \emph{conformal time}, $\lambda = \ln \tau$, and with the rescale, $\phi(q) = e^{-2\lambda} \tilde \phi(\lambda, \bar q)$the wave equation \eqref{WDE01}  (i.e. the Wheeler-DeWitt equation) becomes \cite{RP2021a}
\be\label{WE00101}
\left\{ \frac{\partial^2}{\partial \lambda^2} - \Box_{\bar q} + \left( \frac{m^2}{\hbar^2} e^{2\lambda} - 4 \right) \right\} \tilde \phi(\lambda, \bar q) = 0 .
\ee
However, the 'mass' of the field \eqref{MAS01} is not a constant. Even considering just the geometrical degrees of freedom it continues being a non constant function of the components of the metric tensor $h_{ij}$  (or, equivalently, of the variables $q^A$) through the dependence on the ${}^3R$ curvature (see, \eqref{MAS02}). In that case, the space $M$ turns out to be a dispersive medium for the wave function of the universe. It does not invalidate the formalism but it becomes more complicated from a technical point of view. For that reason let us  focus on the case for which, $ {}^3R \ll 2 \Lambda$, which on the other hand is a very plausible condition for the initial state of the universe\footnote{Let us notice that the condition, $ {}^3R \ll 2 \Lambda$, does not assume that the universe is homogeneous.}, and consider only the geometrical degrees of freedom plus the constant $\Lambda$. We can then assume the value\footnote{The factor $\hbar^2$ has been introduced for later convenience.}
\be\label{MAS03}
m^2_g(q) \approx \frac{ h_0^2 \Lambda  }{4 \pi G} \tau^2 \equiv \hbar^2 m_0^2 e^{2\lambda},
\ee 
in the Wheeler-DeWitt equation \eqref{WDE01}. In that case, the mass \eqref{MAS01} only depends on the \emph{time} variable $\tau$ and we can perform the quantisation of the field $\phi$ in the customary way (see, for instance, Refs. \cite{Birrell1982, Mukhanov2007}). Then, we can decompose the wave function of the universe $\phi(q)$ in normal modes as
\be\label{FD01}
\phi(q)=\int_0^\infty d k \sum_{\vec j}  \, [ a_\textbf{k} u_\textbf{k}(q) + a_\textbf{k}^\dag u^*_\textbf{k}(q) ] ,
\ee
where, $\textbf{k} = (k, \vec j)$,
\be
u_\textbf{k}(q) =e^{-2\lambda}  \chi_{k, J}(\lambda) \mathcal Y_{J, \vec M}(\bar q)  ,
\ee
and, $\mathcal Y_{k, {\vec j}}(\bar q) $, are the  eigenfunctions of the  Laplacian defined on the $5$-dimensional hyperboloid, which satisfy \cite{Bander1966}
\be
\Box_{\bar q}\mathcal Y_{k, {\vec j}}(\bar q) = - (k^2 +4) \mathcal Y_{k, {\vec j}}(\bar q) ,
\ee
with, $ 0 < k < \infty$, and $\vec j$ denotes the $4$ indices that distinguish the four components of the generalisation of the angular momentum on the $4$ sphere\footnote{In the $2$ sphere, $\vec j = \{l , m\}$.}. Thus,  the wave equation \eqref{WDE01}  (i.e. the Wheeler-DeWitt equation) reduces to
\be\label{QO101}
\chi_k'' + \left( m_0^2 e^{2\lambda } + k^2 \right) \chi_k = 0 ,
\ee
where, $\chi' \equiv\frac{d \chi}{d \lambda}$. One interesting thing is that the frequency squared of the oscillator \eqref{QO101}, $\omega_k^2(\lambda) \equiv m_0^2 e^{2\lambda } + k^2$, is never negative. The other interesting thing is that \eqref{QO101} is readily solvable in terms of Bessel functions. With the customary normalisation condition
\be\label{ORN101}
\chi_k \partial_\lambda \chi_k^* - \chi_k^* \partial_\lambda\chi_k = i .
\ee
we easily find two set of orthonormal modes given by \cite{RP2021a}
\beq\label{CHI01}
\bar \chi_k(\tau) &=&  \left( \frac{2}{\pi}\sinh(\pi k) \right)^{-\frac{1}{2}} \mathcal J_{-i k }(m_0 \tau)  , \\ \label{CHI02}
\chi_k(\tau) &=& \frac{\sqrt{\pi}}{2}  e^{\frac{k \pi }{2}} \mathcal H^{(2)}_{ik}(m_0 \tau)  .
\eeq

\subsubsection{Boundary conditions and the creation of the universes in pairs}\label{sec030204}

In order to choose the particular set of modes, we have to impose some boundary condition. For this, we shall consider the multiverse as a  really closed system so no external influence is expected to modify its state. Therefore, it seems appropriate to describe the state of the multiverse in a representation that is invariant under the evolution of the third quantised Hamiltonian, which is an extension of the invariant representation used in quantum mechanics \cite{Lewis1968, Lewis1969, Leach1983, Pedrosa1987, Dantas1992, Kanasugui1995, Song2000, Kim2001, Park2004, RP2017d, Rajeev2018}.

An invariant representation can be given in terms of creation and annihilation operators, $\hat b_\textbf{k}$ and $\hat b^\dag_\textbf{k}$, defined as \cite{Kim2001}
\be\label{INV101}
\hat b_\textbf{k} = \frac{i}{\sqrt{\hbar}} \left( \xi_k^* \, \hat p_\phi - (\xi_{k}^*)' \hat \phi \right) \ \ , \ \ 
\hat b^\dag_\textbf{k} = -\frac{i}{\sqrt{\hbar}} \left( \xi_{k} \, \hat p_\phi - ( \xi_{k})' \hat \phi \right) ,
\ee
where, $\hat \phi$ and $\hat p_\phi$, are the operator version of the wave function and the conjugate momentum, respectively, and $\xi_k$ is a solution of the wave equation \eqref{WE00101}, or equivalently with \eqref{QO101},  the orthonormality condition \eqref{ORN101}, which ensures the usual commutation relations,
\be\label{COM101}
[\hat b_\textbf{k}, \hat b_\textbf{k}^\dag] = 1.
\ee
The operators $\hat b_\textbf{k}$ and $\hat b_\textbf{k}^\dag$ in \eqref{INV101} are \emph{time} dependent operators but the dependence is such that the the eigenstates of the corresponding number operator, $\hat N_\textbf{k} = \hat b_\textbf{k}^\dag \hat b_\textbf{k}$, remains invariant under the action of the third quantised Hamiltonian. It means, for example, that once the multiverse is in the vacuum state of an invariant representation it remains in the  same vacuum state irrespective of the internal histories of the connected pieces of the whole spacetime manifold. From this point of view, the multiverse does not evolve in a proper sense, although the time dependence of the vacuum state makes that the vacuum state at (conformal) time, $\lambda_0$, is functionally different  than the vacuum state at $\lambda_1$, being both however the same vacuum state of the same invariant representation.

The conditions (\ref{INV101}-\ref{ORN101}) do not fix the vacuum state. There is in fact an infinite number of solutions that fit with (\ref{INV101}-\ref{ORN101}), each of which define a particular representation and the associated vacuum state. For instance, the modes (\ref{CHI01}-\ref{CHI02}) define two vacuum states,  $|\bar 0_\textbf k \bar 0_{-\textbf k}\rangle$ and $| 0_\textbf k 0_{-\textbf k}\rangle$, respectively. The modes \eqref{CHI01} can be identify with the Hartle-Hawking no boundary condition. First, because they are regular at the Euclidean origin\footnote{There is here no Euclidean region because we have assumed, $2\Lambda \gg ^{3}R$, in \eqref{MAS03}. Otherwise, the condition $^{3}R > 2\Lambda$ defines an Euclidean region where, in the no-boundary proposal, the universe would be created \emph{from nothing} (see, Sec. \ref{sec02}). }, $\tau \rightarrow i \tau \rightarrow 0$. Second, because that for large values of the variable $\tau$, $\bar \chi_k(\tau) \propto \cos m_0\tau$, which essentially matches with the result \eqref{NBWF}. On the other hand, Vilenkin's tunnelling wave function can be identified with the mode \eqref{CHI02}, because at large values of $\tau$ it reads, $\chi_k(\tau)  \propto e^{-i m_0 \tau}$, which represent, in terms of the time variable $t$, an expanding universe [see, Secs \ref{sec0202} and \ref{sec0203}]. One can also follow the analysis made in Ref. \cite{Birrell1982} to conclude that the state $|\bar 0_\textbf k \bar 0_{-\textbf k}\rangle$ is the conformal vacuum state, and that the state $| 0_\textbf k 0_{-\textbf k}\rangle$ is the vacuum state of the  $6$-dimensional Minkowski space, $M$.

Anyway,  these two set of modes are related by a Bogolyubov transformation, 
\be
\bar \chi_\textbf{k} = \alpha_k \chi_\textbf{k} + \beta_k \chi_\textbf{k}^* ,
\ee
where (see, for instance, Ref. \cite{Birrell1982})
\be
\alpha_k = \left[ \frac{e^{\pi k}}{2 \sinh(\pi k)}\right]^\frac{1}{2} , \beta_k =  \left[ \frac{e^{-\pi k}}{2 \sinh(\pi k)}\right]^\frac{1}{2}  ,
\ee
with, $|\alpha_k|^2-|\beta_k|^2 = 1$. It means that the vacuum state of the $\bar \chi_\textbf{k}$ modes, $|\bar 0_\textbf k \bar 0_{-\textbf k}\rangle$ can be written as \cite{Mukhanov2007, RP2021a}
\be\label{VS01}
|\bar 0_\textbf k \bar 0_{-\textbf k}\rangle = \prod_\textbf{k} \frac{1}{|\alpha_k|^{1/2}} \left( \sum_{n=0}^\infty \left( \frac{\beta_k}{\alpha_k} \right)^n |n_\textbf{k} n_{-\textbf k}\rangle \right) ,
\ee
with a number of universes in the no bar representation given by
\be
N_k = |\beta_k|^2 = \frac{1}{e^{2\pi k} - 1} ,
\ee
which corresponds to a thermal distribution with generalised temperature 
\be
T = \frac{1}{2\pi}.
\ee
Then, one can state that in this case the Hartle-Hawking no-boundary version of the vacuum state is full of (Vilenkin's) universes (and antiuniverses) \cite{RP2021a}. This result is very interesting because it implies that the consideration of universe-antiuniverse pairs seems to be quite unavoidable. It is formally similar to what happens in the quantum field theory of a matter field in an isotropic background spacetime, where the isotropy of the space makes that the particles are created in pairs with opposite values of the field modes, $\textbf{k}$ and $-\textbf{k}$ (see \eqref{VS01}), and in the case of a complex field in particle-antiparticle pairs with opposite momenta. In the third quantisation formalism, the space-like subspace $\bar M$ is homogeneous and isotropic. Therefore, if the potential of the Wheeler-DeWitt equation is also isotropic in $\bar M$, i.e. invariant under rotations in $\bar M$, the universes should be created in pairs with opposite values of the $5$-dimensional $\textbf{k}$ ($\equiv \bar{\textbf{k}}_{ab}$) modes. This is not the most general case, but it is a quite plausible one provided that we assume a high value of the potential of the inflaton field, which can be identified at the initial stage of the universe with $\Lambda$, or equivalently, a small value of the spatial curvature $^{3}R$ of the newborn universe. In both cases, the potential term of the Wheeler-DeWitt equation can be approximated by \eqref{MAS03} and small deviations can be treated as perturbations, which should not significantly violate the isotropy of the space $\bar M$. 

This could be confirmed as well from a more geometrical point of view. Let us notice that the Milne spacetime can separately cover the interior of the upper and the lower light cones of the Minkowski spacetime. These two sections of the full light cone can be seen as the regions of the spacetime where propagate future oriented particles and past oriented particles, or antiparticles, which turn out to be  entangled \cite{JOlson2011}. One would expect something similar in the case of the space $M$, which also covers the upper and lower half light cones of the $6$ dimensional Minkowski space. These two regions would describe expanding and contracting universes (created in pairs as we have seen above), which are equivalent to the future and past oriented particles in Minkowski spacetime. Thus, much in a similar way as particles propagating backwards in time can be interpreted as antiparticles propagating forward in time \cite{Feynman1949}, we have seen in Sec. \ref{sec02}, and we will see it again in the next section, contracting universes can be seen as expanding antiuniverses with a time variable that is reversely related with respect to the time variable of the partner universe. It means that the fields that propagate in one of the two entangled universes appear, from the point of view of an hypothetical observer in the partner universe, as moving backward in time. This is an illusionary effect created by the relative definition of the time variables in the two universes, i.e. internal observers always define the fields that propagate in their universes as matter and the fields that propagate in the partner universe as antimatter. Furthermore,  the value of each mode of the Fourier decomposition in \eqref{FD01} is proportional to the momentum conjugated to the components of the scaled metric tensor. It means that any change that is produced by the momentum associated to $+\bar{ \textbf{k}}_{ab}$ in the shape of the universe with metric $\bar h_{ab}$ is being also produced in the shape of the partner universe with opposite sign, $-\bar{ \textbf{k}}_{ab}$, so it is the parity of the two spatial sections is reversely related too and so it is the relative parity of the fields that propagate in the two universes. In the next section, we shall see that the fields that propagate in the two universes are also charge conjugated as a consequence of the reversely relation of their time variables. It turns out therefore that the field of the two universes is CP conjugated. One can then conclude that, quite generally, the universes of the multiverse are created in symmetric universe-antiuniverse pairs whose composite quantum state is also expected to be entangled \cite{RP2011b, RP2018a, RP2019c}.

\subsubsection{Semiclassical regime}\label{sec030205}

If one takes into account the matter fields, the total Hamiltonian constraint \eqref{HC01} can be written as,
\be\label{HC01}
\hat H_T \phi = \left( \hat H_{G} + \hat H_{SM}\right) \phi = 0 ,
\ee
where $\hat H_{G}$ is the Hamiltonian operator that yields the WDW equation of the spacetime geometry alone \eqref{WDE01} with $\Lambda$ related to the constant part of the potential of the field that drives the inflationary period, $2 \Lambda = 2 V_0 \equiv H_0^2$, and  $\hat H_{SM}$ ($\mathcal H_m$ in \eqref{MAS01}) is the Hamiltonian operator of the matter fields, that essentially are the fields of the Standard Model (SM) with their corresponding potentials and interactions. Following the procedure described in Sec. \ref{sec0203}, the wave function of the universe can be written as the product of two components, a wave function $\phi_0$ that depends only on the gravitational degrees of freedom and the value of the constant $\Lambda$, and a wave function that contains all the dependence on the fields of the SM, collectively denoted by the variable, $\varphi$, i.e
\be\label{PHI00}
\phi^\pm(h_{ij}, \Lambda; \varphi) = \phi_0^\pm(h_{ij}, \Lambda) \psi_\pm(h_{ij}, \Lambda; \varphi) ,
\ee
where the two signs have been introduced for later convenience and, $\phi^+ = \left( \phi^-\right)^*$. The wave function $\phi_0$ is the solution of the WDWE of the geometrical degrees of freedom, computed in the preceding section. In general, it can be written in the semiclassical approach as
\be\la{PHI01}
\phi_0^\pm(h_{ij}, \Lambda) \propto e^{\pm \frac{i}{\hbar} S(h_{ij}, \Lambda)} .
\ee
If one introduces the wave function \eqref{PHI00} into the complete WDW equation and use the classical constraint \eqref{CC00101} one obtains, at order $\hbar^1$, the following equation (see, \eqref{PSIf})
\be\la{SCH01}
\mp 2 i\hbar \vec \nabla S  \cdot \vec \nabla \psi_\pm = H_{SM} \psi_\pm ,
\ee
where $\vec \nabla$ is the gradient in $M$ and the negative and the positive signs correspond, respectively, to $\phi^+$ and $\phi^-$ in (\ref{PHI00}). The Schrödinger equation for the matter fields is then obtained if one defines the (WKB) time parameter $t$ through the condition,
\be\label{WKBt01}
\frac{\partial }{\partial t} = \mp 2 \vec\nabla S \cdot \vec\nabla \equiv \mp 2 G^{\alpha \beta} \frac{\partial S}{\partial q^\alpha} \frac{\partial }{\partial q^\beta} ,
\ee
where, $q^{\alpha} = (\tau, \bar q^A)$, and $\bar q^A$ are the coordinates of $\bar M$ given in \eqref{SME01}. We have now two choices. Typically, it is chosen the positive sign in (\ref{WKBt01}) for the spacetime represented by the wave function $\phi_0^-$ and the negative sign for the spacetime represented by the wave function $\phi_0^+$. With these choice, the Schrödinger equation in the two branches turns out to be 
\be\la{SCH02}
i\hbar \frac{\partial \psi_\pm}{\partial t_\pm} = H_{HSM}(\varphi) \psi_\pm ,
\ee
where it can now be written, $\psi_\pm=\psi_\pm(t_\pm; \varphi)$. From \eqref{WKBt01}, one easily gets
\be
\frac{\partial \tau}{\partial t_\pm} = \pm 2 \frac{\partial S}{\partial \tau} ,
\ee
so the wave functions $\psi_\pm$ represent two universes, one expanding and one contracting (recall that the variable $\tau$ is proportional to the volume of the space), which from \eqref{SCH02} are both filled with matter. An alternative although equivalent interpretation is to choose the positive sign in \eqref{WKBt01} for both universes, i.e. $t \equiv t_+$. In that case, both wave functions represent expanding universes but then the corresponding Schrödinger equations for the internal fields are given by
\beq\label{FI101}
i\hbar \frac{\partial \psi_+}{\partial t} &=& H_{HSM}(\varphi) \psi_+ , \\  
- i\hbar \frac{\partial \psi_-}{\partial t} &=& H_{HSM}(\varphi) \psi_- ,
\eeq
respectively. The last of which can be written as,
\be\label{FI102}
i\hbar \frac{\partial \psi_+}{\partial t} = H_{HSM}(\bar \varphi) \psi_+ ,
\ee
where we have used that, $\psi_-^*(\varphi) = \psi_+(\bar\varphi)$. It is therefore the Schrödinger equation of a field that is  charge conjugated with respect to the field given in \eqref{FI101}. The wave functions  $\phi^+$ and $\phi^-$ represent then two expanding universes but from the point of view of the same time variable one is filled with matter and the other with antimatter, having these two concepts always a relative meaning.

\begin{figure} 
\centering
\includegraphics[width=13 cm]{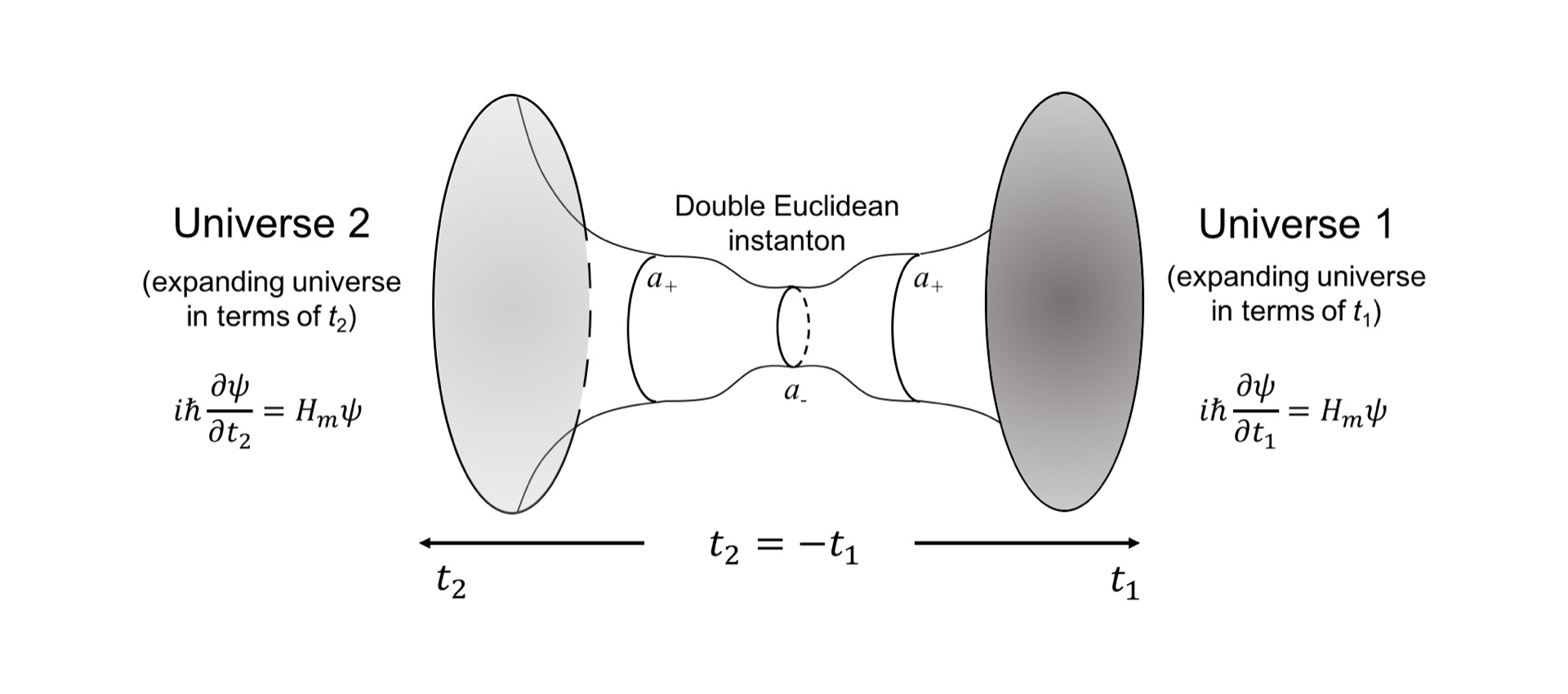}
\caption{The creation of universes in entangled pairs \cite{RP2018a}. In order to obtain the correct value of the Schr\"{o}dinger equation in the two universes, their physical time variables must be reversely related. In that case, particles moving in the symmetric universe look as they were moving backward in time so they are naturally identified with the antiparticles that are left in the observer's universe. The primordial matter-antimatter asymmetry observed in the context of a single universe would thus be restored in the  multiverse. Particles and antiparticles do not collapse at the onset because the Euclidean gap that exists between the two newborn universes \cite{RP2018a, RP2017e}.}
\label{figure704}
\end{figure}

Let us focus on the wave function of the matter fields in one of the universes,  say $\psi_+$. If we consider that the modes of the field are decoupled, then, the Schr\"odinger equation for the scalar field $\varphi_+$, which generically denotes any of the polarisations of the $W^\pm$ and $Z$ bosons, can be written as the product of the wave functions of the modes, i.e.
\be\label{PSI01}
\psi_+(t, \varphi) = \prod_k \psi^{(k)}_{+}(t, \varphi_k) ,
\ee
where $\psi^{(k)}_{+}(t_+, \varphi_k) $ is the solution of the Schr\"odinger equation \eqref{FI101} for each mode, whose general solution can be expressed in the basis of number eigenfunctions of the time dependent harmonic oscillator \cite{Brizuela2019}

\vspace{2cm}

The wave function in the time reversely symmetric universe, $\psi_-(t, \bar \varphi)$, can be obtained from the relation $\psi_-(\bar\varphi) = \psi^*_+(\varphi)$, so the eigenfunctions of the basis for the state of the boson fields in the symmetric universe turns out to be given by (\ref{PSI01}) with the replacements, $t \rightarrow -t$ and $\varphi_k \rightarrow \bar \varphi_k$. Thus, the field $\varphi$ that represents the matter content of one of the universes is the charge conjugated of the field $\bar \varphi$ that represents the matter content of the partner universe. We have seen in the preceding section that their parity is algo reversely related so $\bar \varphi$  turns out to be CP conjugated field of the field $\varphi$. Thus, the matter content of one of the universes is the CP conjugated of the matter in the partner universe and they form thus a universe-antiuniverse pair. It does not necessarily mean that one of the universes is completely made up of matter and the other is made up of antimatter. In fact, the two universes can contain matter as well as antimatter but exactly in the opposite ratio so, from the global point of view, the total amount of matter in the two universes is balanced with the total amount of antimatter.

%%%%%%%%%%%%%%%%%%%%%%%%%%%%%%%%%%
%%%%%%%%%   Minisuperspace model %%%%%%%%%%%%
%%%%%%%%%%%%%%%%%%%%%%%%%%%%%%%%%%

\subsection{Minisuperspace model}\label{sec0303}

\subsubsection{Geometrical structure of the minisuperspace}\label{sec030301}

Let us now apply the third quantisation formalism to the case of the minisuperspace of homogeneous and isotropic metrics with small perturbations that represent the matter content of the universe. The formalism greatly simplifies and one can still obtain a clear picture of the scenario described by the third quantisation formalism. On the other hand, we have seen that the minisuperspace description of the universe, although not complete, is a good approximation for most of the evolution of the universe provided that the universe is  created with a length scale of some orders of magnitude  above from the Planck length. In that case, the small deviations from the homogeneity and the isotropy of the universe can be treated as perturbations described as particles propagating in the homogeneous and isotropic background.

Let us therefore consider the homogeneous and isotropic Friedmann-Robertson-Walker (FRW) metric as the background spacetime \eqref{G01}
\be\nn\label{FRWmetric701}
ds^2 = - N^2(t) dt^2 + a^2(t) d\Omega^2_3 ,
\ee 
where $a(t)$ is the scale factor, and $d\Omega^2_3$ is the line element on the three sphere\footnote{We are considering geometrically closed spatial sections.}. We saw in Sec. \ref{sec02} that the lapse function is not a dynamical variable so the only dynamical variable turns out to be the scale factor, $a(t)$. In this case all the components of the spatial metric are fixed except for the value of the scale factor. $\bar M$ turns out to be then a $0$-dimensional space,  where the spatial sections of the universes are represented by single points and their evolution by (curved) lines in the $1+0$ dimensional space $M$.

This picture can easily be extended by considering as well the homogeneous mode of some matter fields, represented by a set of scalar fields, $\vec \varphi(t)=(\varphi_1(t), \ldots, \varphi_n(t))$, minimally coupled to gravity. We will see that these fields enter as space-like variables in the configuration space. For simplicity, we shall consider only one single scalar field representing the matter of the universe so the configuration space, $M$, will be a $1+1$ dimensional space. In addition, on can also consider the inhomogeneous modes of these fields so the total configuration space would be the $1+n \cdot \infty$ dimensional space spanned by the variables, $(a(t), \varphi_{1,k}(t.x), \ldots, \varphi_{n,k}(t,x))$. For simplicity, we shall only consider\footnote{The inhomogeneities of the spacetime can also be considered as fields propagating in the spacetime (see, Sec. \ref{sec02}).} the homogeneous mode of a single scalar field, $\varphi$, and its inhomogeneities will be treated as a perturbation described by particles propagating in the spacetime. Therefore,  by now let us consider the $1+1$ dimensional configuration space $M$ of coordinates, $q^A \equiv (a, \varphi)$.

The total action, i.e. the Einstein-Hilbert action of gravity plus the action of the scalar field, given by \eqref{AMin201}, can be written as
\be\label{S701}
S = S_g + S_m = \frac{1}{2} \int dt N \left(  G_{AB} \frac{\dot{q}^A \dot q^B}{N^2}  - \mathcal V(q) \right) ,
\ee
where, $q^A = (a,\varphi)$, with the supermetric $G_{abcd}$ in (\ref{DWM01}) given now by \cite{Kiefer2007}
\be\label{MSM701}
G_{AB} = {\rm diag}(-a, a^3) ,
\ee 
from which one can clearly see that the scale factor (i.e. the first component) is a time-like variable and the scalar field (the second component) is a space-like variable. The potential term, $\mathcal V(q)$ in (\ref{S701}), reads
\be\label{V701}
\mathcal V(q) \equiv \mathcal V(a,\varphi) = - a + 2 a^3 V(\varphi)  .
\ee
The first term in (\ref{V701}) comes from the closed geometry of the three space, and $V(\varphi)$ is the potential of the scalar field. The case of a spacetime with a cosmological constant, $\Lambda$, is implicitly included if we consider a constant value of the potential of the scalar field, $V(\varphi)=\Lambda/6$. As we showed in Sec. \ref{sec030202}, the evolution of the universe can be seen as a parametrised trajectory of the superspace with the variable $\tau \propto \sqrt{h}$ formally playing the role of a time variable. In the case of the minisuperspace, $\sqrt{h}=a^3$; but it is interesting to change to \emph{conformal} scale factor, $\alpha = \ln a$, in terms of which the metric $G_{AB}$ turns out to be conformal to the $2$-dimensional Minkowski space,
\be\label{MSM701}
G_{AB} = e^{3\alpha} \eta_{AB}  ,
\ee 
and the action \eqref{S701} can be written as,
\be\label{S701}
S = S_g + S_m = \frac{1}{2} \int dt N e^{3\alpha} \left(  \eta_{AB} \frac{\dot{q}^A \dot q^B}{N^2}  - \left( H^2(\varphi) - e^{-2\alpha}\right)  \right) ,
\ee
where, $e^{3\alpha} = a^3$, is essentially the volume of the spatial sections and, $H^2(\varphi) = 2 V(\varphi)$, is the Hubble function. From the signature of \eqref{MSM701}, it can be seen that the scale factor formally plays the role of the time variable and the matter field(s) the role of the space-like component(s), and the minisupermetric \eqref{MSM701} provides the minisuperspace with a complete metric structure with a line element given by
\be
d s^2 = G_{AB} dq^A dq^B = - a da^2 + a^3 d\varphi^2 = e^{3\alpha} \left( - d\alpha^2 + d\varphi \right).
\ee
It also allows us to define the usual machinery of a geometric manifold. For instance, we can defined the Christoffel symbols associated to the minisupermetric $G_{AB}$, defined as usual by
\be
\Gamma^A_{BC} = \frac{G^{AD}}{2} \left\{ \frac{\partial G_{BD}}{\partial q^C} + \frac{\partial G_{CD}}{\partial q^B} - \frac{\partial G_{BC}}{\partial q^D} \right\} ,
\ee
which in terms of the variables $(a,\varphi)$ the non zero values are
\be\label{CS701}
\Gamma^a_{aa} = \frac{1}{2a} \ , \ \Gamma^a_{\varphi\varphi} = \frac{3a}{2} \ , \ \Gamma^\varphi_{\varphi a} = \Gamma^\varphi_{a \varphi} = \frac{3}{2 a} ,
\ee
or in terms of the variables $(\alpha, \varphi)$,
\be
\Gamma^\alpha_{\alpha\alpha} =  \Gamma^\alpha_{\varphi\varphi} =  \Gamma^\varphi_{\varphi \alpha} = \Gamma^\varphi_{\alpha \varphi} = \frac{3}{2} .
\ee
In any case, we could compute other geometrical properties of the minisuperspace like the corresponding Riemann tensor, the curvature scalar, etc (see, Ref. \cite{DeWitt1967}).

\begin{figure} 
\centering
\includegraphics[width=13 cm]{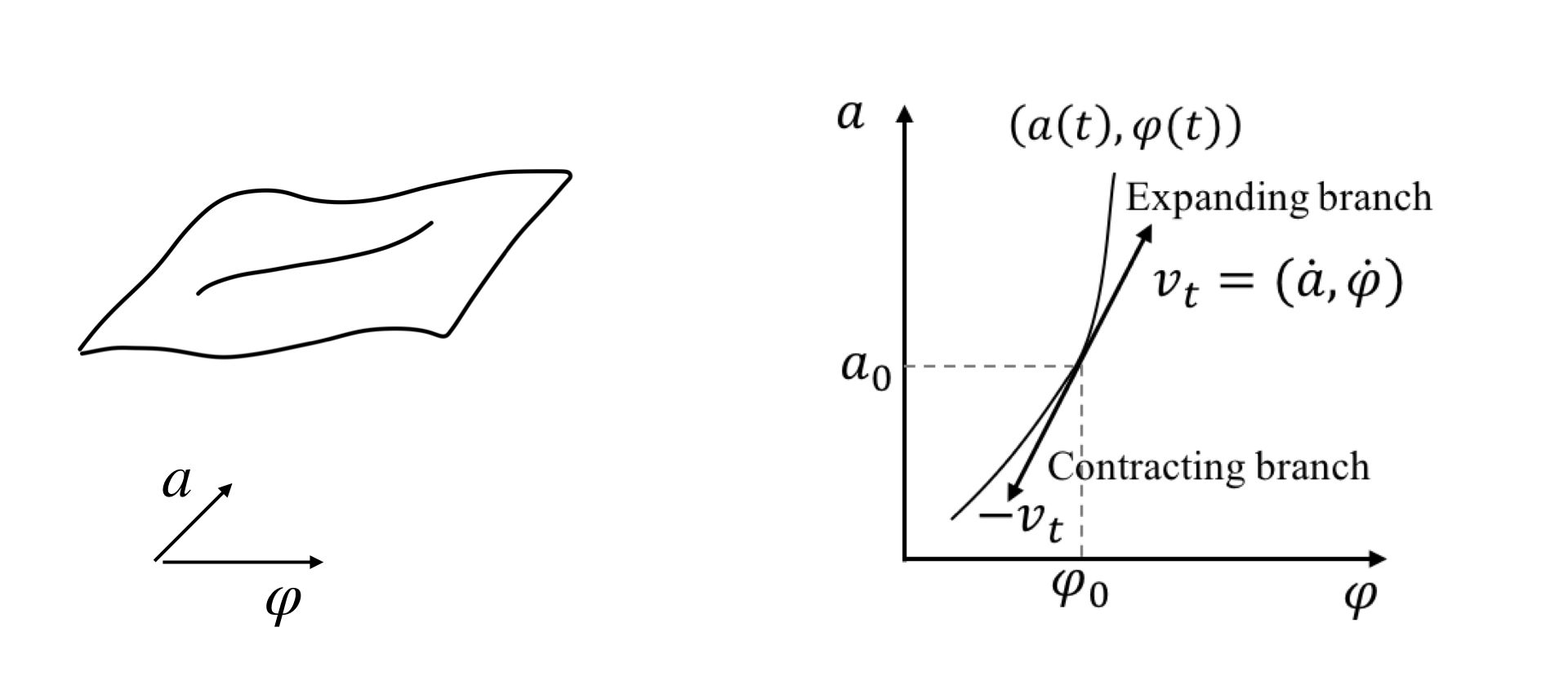}
\caption{Left: the evolution of the universe can be seen as a trajectory in the minisuperspace. Right: a trajectory in the minisuperspace that is positively oriented with respect to the scale factor component describes an expanding universe. Similarly, a negatively oriented trajectory describes a contracting universe.}
\label{figure701}
\end{figure}

From the geometrical point of view the evolution of the universe can be seen as a trajectory in the minisuperspace (see, Fig. \ref{figure701}), with $a(t)$ and $\varphi(t)$ being the parametric coordinates of the universe along the \emph{worldline} of the universe, and the time variable $t$ is the parameter that parametrises the trajectory. From that point of view it is easy to see that the evolution of the universe, i.e. the trajectory of the universe in the minisuperspace, cannot depend on the particular choice of time variable, i.e. the trajectory must be independent of the parametrisation used to describe it. 

However, because the presence of the potential $\mathcal V(a,\varphi)$ in the action \eqref{S701}, the trajectory of the universe along the minisuperspace manifold is not a geodesic. It is instead given by the equation 
\be\label{EM701}
\ddot{q}^A + \Gamma^A_{BC} \dot{q}^B \dot{q}^C = - G^{AB} \frac{\partial \mathcal{V}}{\partial q^B} ,
\ee
which with the help of \eqref{CS701} yields the customary field equations (see, for instance, Refs. \cite{Linde1993, Kiefer2007})
\be
\ddot{a} +\frac{\dot{a}^2}{2 a} + \frac{3a}{2} \dot \varphi^2 = -\frac{1}{2a} + 3 a V(\varphi) \ \ , \ \ 
\ddot \varphi + 3\frac{\dot a}{a} \dot \varphi = -  \frac{\partial V(\varphi)}{\partial \varphi} .
\ee
The fact that the curve $(a(t), \varphi(t))$ is not a geodesic is not a big deal. As we have said, the trajectory of the universe is invariant under reparametrisations of time, so we can make the following change of time variable
\be
d\tilde t =  \mathfrak m^{-2} \mathcal{V}(q)  d t ,
\ee
where $\mathfrak m$ is some constant. Now, if we also perform the following conformal transformation of the minisupermetric
\be\label{CT701}
\tilde G_{AB} =  \mathfrak m^{-2} \mathcal{V}(q) G_{AB} ,
\ee
the action (\ref{S701}) becomes
\be\label{S702}
S = \frac{1}{2}  \int d\tilde t   N \left( \frac{1}{  N^2} \tilde G_{AB} \frac{dq^A}{d\tilde t}  \frac{dq^B}{d\tilde t}  -  \mathfrak m^2 \right)     ,
\ee
which is a similar action but with a constant potential. The new time variable, $\tilde t$, turns out to be the affine parameter of the minisuperspace geometrically described by the metric tensor $\tilde G_{AB}$, and the trajectory of the universe in this minisuperspace is given by the geodesic equation
\be\label{EM702}
\frac{d^2q^A}{d\tilde t^2} + \tilde \Gamma^A_{BC} \frac{dq^B}{d\tilde t} \frac{dq^C}{d\tilde t} = 0 .
\ee
Thus, the classical trajectory of the universe can equivalently be seen as either a geodesic of the minisuperspace geometrically determined by the minisupermetric $\tilde G_{AB}$ or a non geodesic of the minisuperspace geometrically determined by $G_{AB}$.

We can also define the momenta conjugated to the minisuperspace variables
\be
\tilde p_A \equiv \frac{\delta L}{\delta \frac{dq^A}{d\tilde t}} ,
\ee
and the Hamiltonian constraint associated to the action (\ref{S702}) turns out to be
\be\label{HC701}
\tilde G^{AB} \tilde p_A \tilde p_B + \mathfrak m^2 = 0 ,
\ee
or in terms of the metric $G_{AB}$ and the time variable $t$,
\be\label{HC702}
G^{AB}  p_A  p_B + \mathfrak m^2_{\rm ef}(q) = 0 ,
\ee
where for convenience we have written, $\mathfrak m^2_{\rm ef}(q) = \mathcal V(q)$, with $\mathcal V(q)$ given by (\ref{V701}). It is worth noticing that the phase space does not change in the transformation $\{G_{AB}, t\} \rightarrow \{\tilde G_{AB}, \tilde t \}$, because
\be\label{M701}
\tilde p_A   = \tilde G_{AB} \frac{d q^B}{d\tilde t} = G_{AB} \frac{dq^B}{d t} = p_A  ,
\ee
where, $p_A = \{p_a, p_\varphi\}$ and $q^A \equiv \{a, \varphi\}$, and the Hamiltonian constraints \eqref{HC701} and \eqref{HC702} are related by the inverse of the conformal transformation \eqref{CT701},
\be
\tilde G^{AB} = \frac{\mathfrak m^2}{\mathcal V(q)} G^{AB} .
\ee

The field equations given either by \eqref{EM701} or by \eqref{EM702} are invariant under the reversal change in the time variable, $t \rightarrow -t$.  From the geometrical point of view it only changes the direction along which the curved is travelled, i.e. the direction of the tangent vector $\frac{\partial }{\partial t}$. It means that for any given solution $a(t)$ and $\varphi(t)$ one may also consider the symmetric solution, $a(-t)$ and $\varphi(-t)$.

In our case the momenta conjugated to the variables of the minisuperspace, given in \eqref{M701}, turn out to be
\be\label{MOM731}
p_a = -\frac{a \dot a}{N} \,\,\, , \,\,\, p_\varphi = \frac{a^3 \dot \varphi}{N} ,
\ee
in terms of which the Hamiltonian constraint \eqref{HC702} reads
\be\label{HC732}
-\frac{1}{a} p_a^2 + \frac{1}{a^3} p_\varphi^2 + m^2_{\rm eff}(a,\varphi) = 0 ,
\ee
which is the Friedmann equation expressed in terms of the momenta instead of in terms of the time derivatives of the minisuperspace variables. As pointed out before, the geodesic equation and the momentum constraint (\ref{HC732}) are invariant under a reversal change of the time variable. Let us notice however that the momenta \eqref{MOM731} are not invariant but they turn out to be reversely changed, $p_a \rightarrow - p_a$ and $p_\varphi \rightarrow -p_\varphi$. Nevertheless, they appear squared in the Hamiltonian constraint  \eqref{HC732} so it is not affected by the change.

However, by conservation of the momenta one would expect that the cosmological solutions should come in symmetric pairs with opposite values of the associated momenta. From (\ref{MOM731}) and (\ref{HC732}), it is easy to see that in terms of the cosmological time ($N=1$) the two symmetric solutions are given by
\be\label{FE733}
a \frac{da}{dt} = - p_a = \pm \sqrt{\frac{1}{a^2} p_\varphi^2 + a m^2_{\rm eff}(a,\varphi) } .
\ee
It clearly reminds to the solutions of the trajectory of a test particle moving in the spacetime \cite{Garay2018}. For instance, in Minkowski spacetime\footnote{A similar procedure can be followed in a curved spacetime.}, the time component of the geodesics satisfies
\be\label{MST731}
\frac{dt}{d\mu} = - p_t = \pm \sqrt{\vec p^2+m^2} ,
\ee
where $\mu$ is an affine parameter and, $p_t = \pm E$, with $E$ the energy of the test particle. The two solutions are eventually associated to particles and antiparticles in a quantum field theory.

\begin{figure} 
\centering
\includegraphics[width=12 cm]{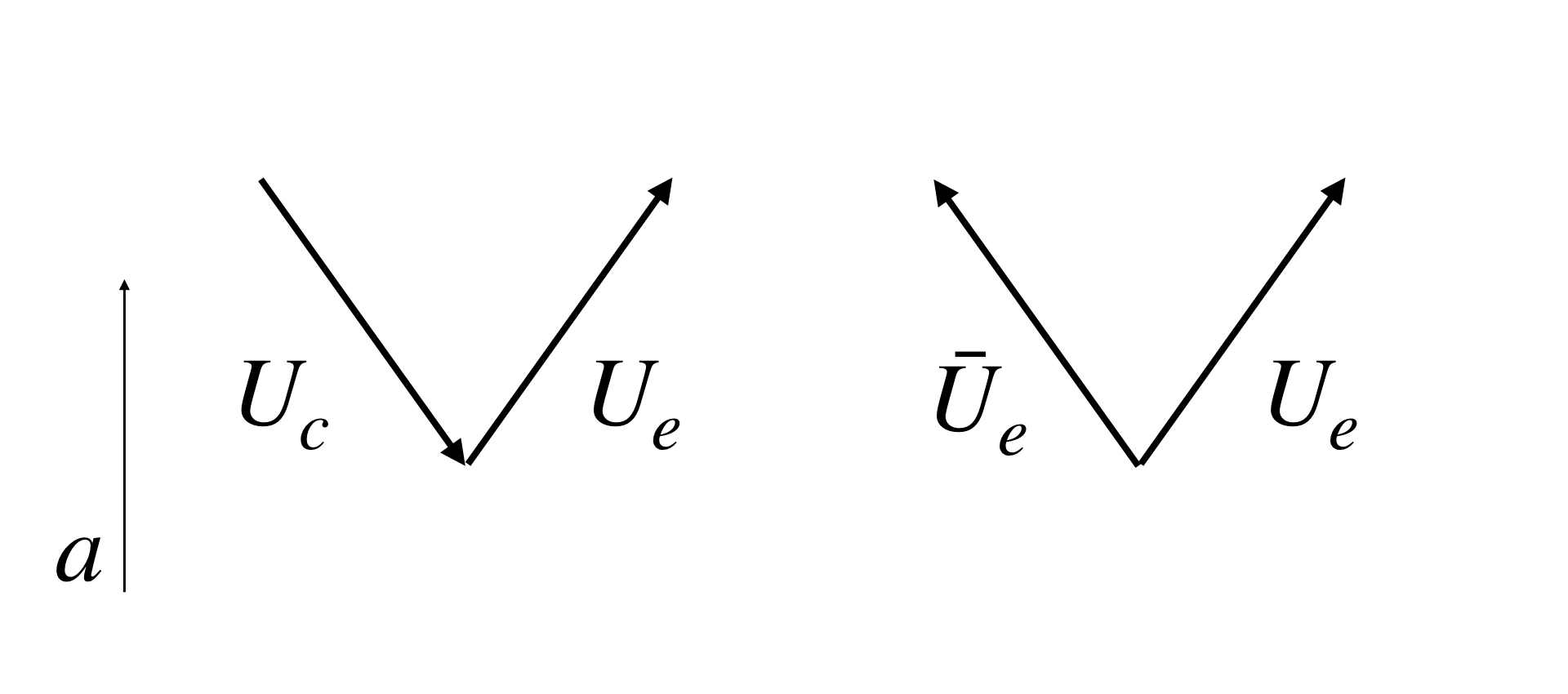}
\caption{A contracting and an expanding universes, both made of matter, can also be seen as a pair of expanding universes, one of them made up of matter and the other made up of antimatter, i.e. they can be seen as a universe-antiuniverse pair.}
\label{figure702}
\end{figure}

In the case of the universe the two solutions given in (\ref{FE733}) also represent two universes: one universe moving forward in the scale factor component and the other moving backward in the scale factor component (see, Fig. \ref{figure702}). In the minisuperspace, however, moving forward in the scale factor component means evolving with an increasing value of the scale factor so  the associated solution represents an expanding universe, and moving backward in the scale factor component means evolving with a decreasing value of the scale factor so the symmetric solution represents a contracting universe. Therefore, the two symmetric solutions form an expanding-contracting pair of universes (see, Fig. \ref{figure702}). However, we  have already showed in previous sections that an expanding-contracting pair filled with matter can also be interpreted as two expanding universes, one of them filled with matter and the other filled with antimatter \cite{Rubakov1999, RP2017e}, i.e. it can be interpreted as a universe-antiuniverse pair \cite{RP2017e, RP2019c}.

\subsubsection{Field quantisation of a FRW spacetime}\label{sec030302}

As we have already seen, the procedure of third quantisation parallels that of a second quantisation in a curved spacetime (see Sec. \ref{sec030203}). Now, the field is the wave function $\phi(a,\varphi)$ that satisfies the corresponding Wheeler-DeWitt equation, which is now seen as a wave equation. With the minisupermetric \eqref{MSM701} in \eqref{WDE01}, the Wheeler-DeWitt equation turns out to be
\be
 a \frac{\partial}{\partial a} \left( a \frac{\partial \phi}{\partial a} \right) - \frac{\partial^2 \phi}{\partial \varphi^2} + a^2 \omega^2(a) \phi = 0 ,
\ee
where,
\be
\omega^2(a,\varphi) = H^2 a^4 - a^2 ,
\ee
with, $H^2= V(\varphi_0)$, evaluated at the moment of the creation of the (inflationary) universe, where it can be approximated by a constant. In that case, following the procedure shown in Sec. \ref{sec030204} we can decompose the wave function $\phi(a,\varphi)$ in Fourier modes,
\be
\phi(a,\varphi) = \int \frac{d K}{2 \pi} e^{i K \varphi} \phi_K(a) ,
\ee
where $\phi_K(a)$ must satisfy
\be\label{WE752}
\ddot{\phi}_K + \frac{1}{a} \dot \phi_K  + \omega_K^2(\alpha) \phi_K = 0 ,
\ee
with \cite{RP2011b, RP2018a}
\be\label{FRE751}
\omega_K^2(a) = H^2 a^4 - a^2 + \frac{K^2}{a^2}  .
\ee
Let us notice that the inner product turns out to be here by \eqref{INPROD531} with \cite{RP2019b}, $d\Sigma^A = n^A d\Sigma$, where $n^A=(a^{-\frac{1}{2}}, 0)$ is a timelike unit vector, and $d\Sigma = d\varphi$, which defines the orthogonal hypersurfaces (one dimensional curves) of constant $a$. It then becomes \cite{RP2011b, RP2019b}
\be
(u_1 , u_2 ) = - i \int_{-\infty}^{+\infty} d\varphi \, a \ \left( u_1(a,\varphi) \overset{\leftrightarrow}{\partial}_a u_2^*(a,\varphi)  \right) .
\ee
We can now define the operator version of the field, $\hat \phi$, and write it as
\be\label{PHI751}
\hat \phi(a,\varphi) = \frac{1}{\sqrt{2}} \int \frac{dK}{2\pi} \left( e^{iK\varphi} v_K^*(a) \hat A_K^- + e^{-iK\varphi} v_K(a) \hat A_K^+ \right)  ,
\ee
where $\hat A_K^+$ and $\hat A_K^-$  are the creation and annihilation operators, respectively, of universes with momentum $K$ conjugated to the scalar field; and the modes are normalised according to the condition
\be\label{NC00101}
\frac{d v_K}{da } v_K^* - v_K \frac{d v_K^*}{da } = \frac{2 i}{a} .
\ee
We can now define the ground state of the invariant representation, $\hat A_K^+$ and $\hat A_K^-$, by
\be
|0\rangle_I = \prod_K |0_K, 0_{-K}\rangle_I ,
\ee
where $| 0_K\rangle_I$ ($| 0_{-K}\rangle_I$) is the state annihilated by the operator $\hat A_K^-$ ($\hat A_{-K}^-$). An excited state, i.e. a state representing different number of universes with momenta $K_1, K_2, \ldots$, is then given by \cite{RP2019b}
\be\label{FS701}
| m_{K_1}, n_{K_2}, \ldots \rangle = \frac{1}{\sqrt{m! n! \ldots}} \left[ \left( \hat A^+_{K_1} \right)^m \left( \hat A^+_{K_2} \right)^n \ldots \right]  | 0 \rangle_I ,
\ee
which represents $m$ universes in the mode $K_1$, $n$ universes in the mode $K_2$, etc. In the case of a field that propagates in a homogeneous and isotropic spacetime the value of the mode $\textbf k$ represents the value of the spatial momentum of the particle. In a homogeneous and isotropic minisuperspace the value of the mode $K$ labels the eigenvalues of the momentum conjugated to the scalar field $\varphi$, which formally plays the role of a spacelike variable in the minisuperspace. In that case, the values $K_1, K_2, \ldots$, in (\ref{FS701}) label the different initial values of the time derivatives of the scalar field in the universes. Thus, the state (\ref{FS701}) represents $m$ universes with a scalar field with $\dot \varphi \propto K_1$, $n$ universes with a scalar field with $\dot \varphi \sim K_2$, etc. They represent different energies of the matter fields, which would correspond to different number of particles in the universes. The general quantum state of the field $\phi$, which represents the quantum state of the spacetime and the matter fields, all together, is then given by 
\be\label{QSM01}
| \phi \rangle = \sum_{m,n,\ldots} C_{mn\ldots} | m_{K_1} n_{K_2} \ldots \rangle_I ,
\ee
which represents therefore the \emph{quantum state of the multiverse} \cite{RP2010} in the model of the minisuperspace that we are considering.

Here it follows the subtle subject of the boundary conditions in quantum cosmology. From a QFT we know that the vacuum state of a given representation may contain a certain number of particles of another representation, so the question is then which representation is the appropriate one. We have already imposed that the representation of the field that represents the state of the multiverse should be an invariant representation because, in that case, once that field is in a given state it will then remain in the same state along the entire evolution of the universe. Furthermore, one would expect that the field would be in the ground state of such an invariant representation provided that we assume that no external \emph{force} is exciting the state of the multiverse. However, that condition does not completely fix the state of the field $\phi$ because there are many different invariant representations. In general, an invariant representation, $\hat A_K^+$ and $\hat A_K^-$, can be defined as \cite{Kim2001}
\beq\label{INV101}
\hat A_K^- &=& \frac{i}{\sqrt{\hbar}} \left( v_K^* \, \hat p_\phi - \dot{v}_K^*  \hat \phi \right)  , \\  \label{INV102}
\hat A_K^+ &=& -\frac{i}{\sqrt{\hbar}} \left( v_K \, \hat p_\phi - \dot{v}_K \hat \phi \right) ,
\eeq
where, $\hat \phi$ and $\hat p_\phi$, are the operator version of the wave function and the conjugate momentum in the Schrödinger picture, respectively, and $v_K$ is a solution of the wave equation \eqref{WE752} satisfying the orthonormality condition \eqref{NC00101}, which ensures the usual commutation relations, $[\hat A_K^-, A_K^+] = 1$. However, there many different solutions of the wave equation \eqref{WE752} satisfying the orthonormality condition \eqref{NC00101} so any of them provides an invariant representation. In fact, we have seen in Sec. \ref{sec030204} that the Hartle-Haking's boundary condition and the Vilenkin's boundary condition provide two sets of solutions.

\begin{figure} 
\centering
\includegraphics[width=13 cm]{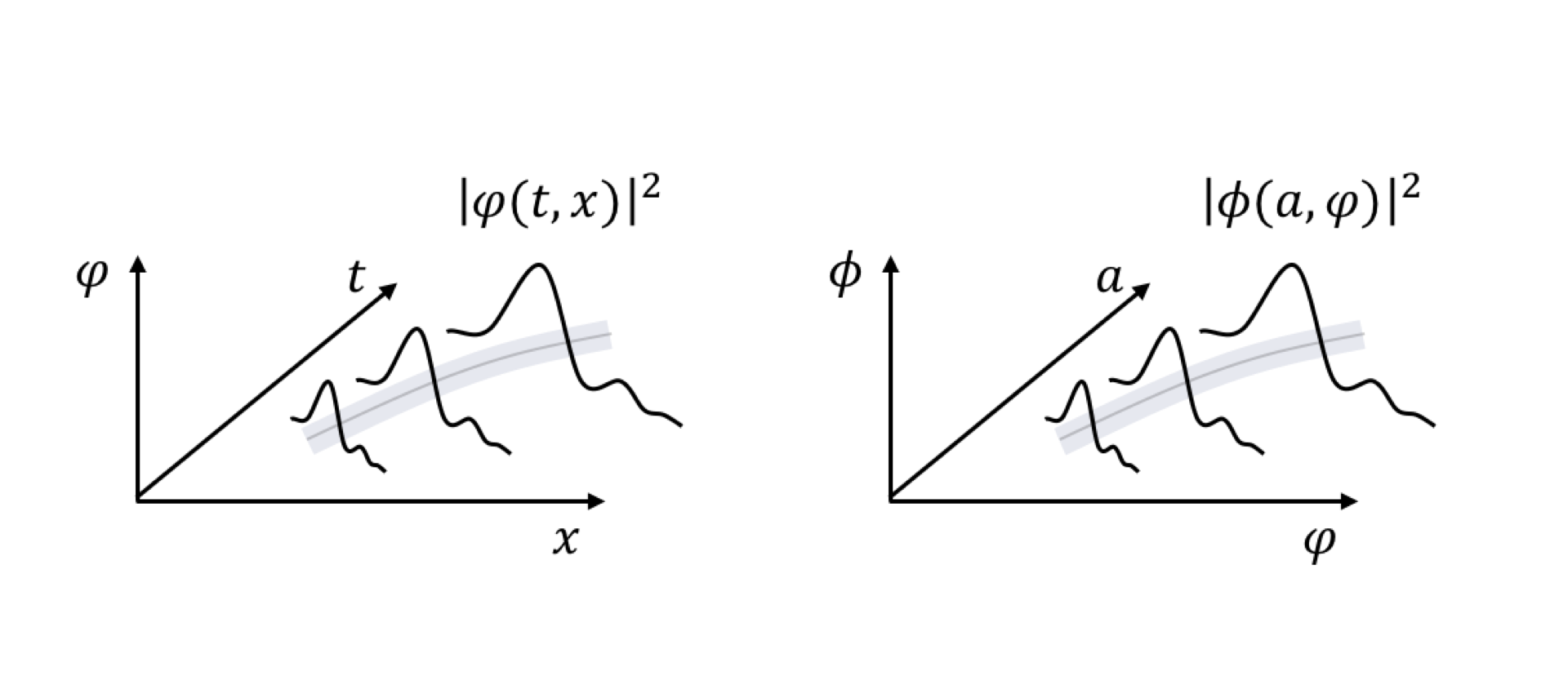}
\caption{Left: in a quantum field theory the field is described in terms of particles that follow with the highest probability the classical trajectories given by the geodesics with however some uncertainties in their positions. Right: the wave function that describes the quantum state of the spacetime and the matter fields, all together, can be seen as a another field, say a super-field, that propagates in the minisuperspace. The universes can then be seen as 'test' particles following classical trajectories in the minisuperspace with quantum uncertainties given by the Schr\"{o}dinger equation of their matter fields.}
\label{figure703}
\end{figure}

One might say that the Hartle-Hawking boundary condition has a more fundamental character because it is rooted on a more ontological reasoning. It essentially rests on the idea that \emph{the boundary conditions of the universe are that it has no boundary} \cite{Hawking1984}, i.e. that the universe, and therefore the multiverse as well, comes from no prior configuration of the space. In that case, it seems consistent to impose or to assume that the multiverse is \emph{always} in the ground state of the Hartle-Hawking invariant representation. However, single universes are better represented by the representation obtained by imposing the Vilenkin's tunnelling condition. In fact, this boundary condition is specifically imposed to assure that it describes single universes created in the Lorentzian region of the (mini)superspace (see Sec. \ref{sec020501}). In that case, as we have seen in Sec. \ref{sec030204}, it turns out that the multiverse is full of Vilenkin's universes  \cite{RP2011b}, which due to the isotropy of the superspace should come in universe-antiuniverse pairs \cite{RP2021a} (see, Sec. \ref{sec030204}, and the next section). Thus, it seems quite unavoidable to assume that our universe has been created in an entangled pair.

\subsubsection{Reheating and the matter-antimatter content of the entangled universe}\label{sec030303}

In Sec. \ref{sec030205} we have seen that the matter fields of the two universes of an entangled pair are CP reversely related and the two universes form thus a unvierse-antiuniverse pair. Let us now apply the same semiclassical formulation to the period after inflation called (p)reheating (see, for instance, Ref. \cite{Mukhanov2008}), where the inflaton field\footnote{We shall use now the variable $\chi$ to represent the inflaton field and leave the variable $\varphi$ to represent collectively the rest of fields of the SM.}, $\chi$, eventually decays into the particles of the Standard Model (SM). In that period, the spacetime can largely be considered homogeneous and isotropic and the inhomogeneities of both the matter fields and the spacetime can be analysed as small perturbations propagating in a homogeneous and isotropic background.

We are not going to repeat the development of Sec. \ref{sec030205} but only to present a particular and detailed example that will show the consequences of the complex conjugated relation between these two wave functions. It is worth noticing that the CP conjugated relation between the matter fields of the universe-antiuniverse pair is based on the fundamental considerations described in Sec. \ref{sec030205} and it is therefore independent of the model chosen for the reheating scenario after inflation, so similar steps can be followed in any other reheating scenario. For concreteness, we shall describe this period in the appealing model of the Higgs-inflaton \cite{Bezrukov2008, Bellido2009}, in which the field that drives the inflationary expansion of the space decays after inflation into the particles of the Standard Model (SM). The idea rest on the form of the potential of the Higgs-inflaton field. At high energy scales, during the first stages of the inflationary period, the functional form of the potential can be approximated by an exponential and it can thus drive inflation. When the field has rolled down the exponential slope of the potential it finds a minimum around which it starts oscillating. Then, inflation ends and the Higgs-inflaton field behaves like the rest of fields of the SM, with interactions that allow the decays of the Higgs-inflaton into the particles of the SM (see below). Finally, in the low energy regime, the functional form of the potential can be approximated by the customary double-well potential of the Higgs that gives the expected masses to the particles of the SM \cite{Bezrukov2008, Bellido2009}.

Therefore, after the inflationary period the potential of the inflaton field, $V(\chi)$, cannot be longer considered a constant. However, we have seen in Sec. \ref{sec020402} that this does not introduce a big qualitative change. The solutions of the Wheeler-DeWitt equation are still expected to come in pairs, $\phi = \phi^+ + \phi^-$, with conjugate complex phases that are the solutions of the Hamilton-Jacobi equation \eqref{HJ041}. In the semiclassical regime, they are given by \eqref{SCWF053}, i.e.
\be\label{SMIN0101}
\phi^\pm(a;\chi,\varphi)=  \Delta(a) e^{\pm\frac{i}{\hbar} S(a)} \psi_\pm(a;\chi,\varphi) ,
\ee
where, $\chi$, is the inflaton field  and $\varphi$ collectively denotes all the fields of the SM. Following the development of Sec. \ref{sec0203}, the complex phase in \eqref{SMIN0101} determines the dynamics of the homogeneous and isotropic background spacetime and the wave functions  of the matter fields in the two universes, $ \psi_\pm( \chi, \varphi)$, satisfy the Schrödinger equation of two sets of CP conjugated fields. They are related by the condition, $\psi_-^*(\chi, \varphi) = \psi_+(\chi, \bar{\varphi})$.

As we have said, at the end of the inflationary period the Higgs-inflaton field $\chi$ has slow rolled down the potential and it then approaches the minimum of the potential located at $\chi_m$, for which $V'(\chi_m)=0$. The expansion rate of the spacetime also slows down and the field starts oscillating around the minimum like a weakly damped harmonic oscillator with mass, $m^2=V''(\chi_m)$. The total Hamiltonian constraint can be written during this period as (\ref{HC01}), with a gravitational part given by 
\be
H_G = -\frac{1}{2M_P^2} p_a^2 - \frac{M_P^2 a^2}{2} + B(a) ,
\ee
where $B(a)$ contains the backreaction of the Higgs-field and eventually the backreaction of the rest of fields of the SM that will contribute to the dynamics of the background spacetime. It may also contain some residual constant term, which is expected to be subdominant at least until the advent of the dark energy period. The Hamiltonian of the Higgs-SM sector, $H_{SM}$ in (\ref{HC01}), can now be written as
\be\label{HC02}
H_{HSM} = H_\chi + H_{SM} ,
\ee
with
\be
H_\chi = \frac{1}{2 a^3} p^2_{\chi} + \frac{1}{2} a^3 M^2 \chi^2 +  \Delta V(\chi) ,
\ee
where \cite{Bellido2009}, $M^2 = \lambda M_P^2/3\xi^2$, with $\xi$ a coupling constant of the theory, $p_\chi$ is the the momentum conjugated to the Higgs-inflaton field, $p_\chi = a^3 \dot \chi$, and $\Delta V(\chi)$ contains high order correction terms that can be neglected in a first approach \cite{Bellido2009}. The interactions between the Higgs and the matter and gauge fields of the SM have been included in the Hamiltonian $H_{SM}$. The Klein-Gordon equation of the Higgs field can then be written,
\be\label{Hig01}
\ddot{\chi} + 3\frac{\dot a}{a} \dot \chi + M^2 \chi = 0 ,
\ee 
where we have assumed that the Higgs is essentially in the zero mode. For instance, for a power-law evolution of the background spacetime, $a(t) \propto t^p$, Eq. (\ref{Hig01}) is a Bessel equation that can be solved analytically. With the appropriate boundary conditions, and assuming $Mt \gg 1$ and $p\approx 2/3$, it can be written as \cite{Bellido2009}
\be\label{CHI01}
\chi(t) = \frac{\chi_{\rm end}}{Mt} \sin (Mt) ,
\ee
where, $\chi(t=0)  = \chi_{\rm end}$, is the value of the Higgs field at the end of the inflationary period, which coincides with the beginning of the appearance of the (p)reheating mechanisms ($t=0$).

Different channels can now be considered for the decaying of the Higgs field into the particles of the SM (see, Ref. \cite{Bellido2009, Koffman1997} for the details). It turns out that the perturbative decay of the Higgs field is only effective when the amplitude of the Higgs is below a critical value that depends on the mass of the final particles. This, together with the time dependence of the decay rate of the Higgs into the particles of the SM makes that the Higgs needs to oscillate a large number of times before decaying into the massive gauge bosons and fermions and much more times to decay into the less massive fermions, so the perturbative decay becomes ineffective during the first oscillations of the Higgs. In that period, the most effective channel turns out to be the parametric resonance \cite{Bellido2009, Koffman1997}. However, this channel is enhanced by the effect of Bose stimulation so the production of fermions through this channel is highly restricted. These will be mainly produced later on through the perturbative channel or through the subsequent decay of the intermediate bosons into fermions.

Therefore, for the purpose of the present analysis, it is enough to focus on the production of the intermediate gauge bosons, $W^\pm$ and $Z$. In the customary SSB mechanism the fields of the SM acquire a constant value of their masses. However, during the reheating period the potential still depends on the value of the Higgs field, $\chi$, and thus the mass acquired from the interaction with the Higgs depends on its value. In that period, it can be approximated by \cite{Bellido2009}
\be\label{MAS01}
m_W^2 \simeq  \frac{ g_2^2 |\chi|}{4 \sqrt{6} \xi } \, , \, m_Z^2 \simeq \frac{m_W^2}{cos^2\theta_W} \, , \,  m_f \simeq \frac{y_f^2 |\chi|}{2\sqrt{6} \xi} ,
\ee
where $g_2$ is the coupling of the intermediate gauge bosons,  $\theta_W$ is the weak mixing angle, and $y_f$ are the Yukawa couplings of the fermion sector \cite{Bellido2009}. Eventually, after the period of reheating, in the low energy regime, the potential takes the customary form of a double well potential and the masses of the particles of the SM become the customary ones \cite{Bellido2009}. Thus, in the low energy limit the Higgs-inflationary scenario is indistinguishable from the Higgs scenario of particle physics, as expected. However, it is in this mid-energy regime, during the reheating period, when the basic components of matter are created in the two universes and the one in which we are mainly interested now.

\begin{figure}
\centering
\includegraphics[width=13 cm]{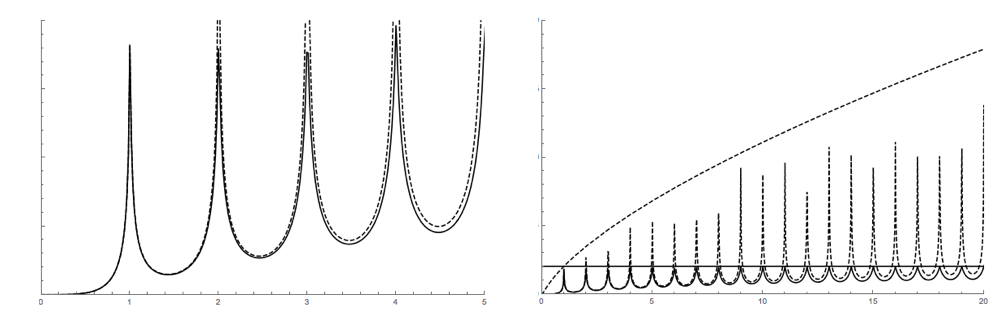}
\caption{Particle production (\ref{N01}) in terms of the number of crossings of zero, $j=\frac{M t}{\pi}$. Dashed line accounts for the expansion of the universe, with $a(t) = t^\frac{2}{3}$, which can be neglected in the first few oscillations. The production of particles is resonant in the points where the Higgs crosses zero (left). If the expansion of the background spacetime is neglected, the production of particles in the peaks rapidly tends to a constant value. However, when the expansion is taken into account, the number of particles in the peaks scale with the scale factor (right).}
\label{figure_705}
\end{figure}

On the other hand, the quantisation of the intermediate gauge bosons $W^\pm$ and $Z$ follows as usual, by decomposing them into normal modes as
\be\label{MEX01}
\hat \varphi(t,\bx) = \int \frac{d^3k}{(2\pi)^{3/2}} \left( e^{-i \tb k \bx}  \varphi_k(t) \hat b_k + e^{i \tb k \bx}  \varphi_k(t) \hat b_k^\dag \right) ,
\ee
where, $\varphi\equiv W^\pm, Z$, and $\hat b_k$ and $\hat b_k^\dag$ are the annihilation and creation operators. As we have seen in Sec. \ref{sec020402}, the inhomogeneous modes of the matter fields can be treated as particles propagating in the background spacetime. The mode amplitude $\varphi_k(t)$ satisfies
\be\label{MEX02}
\ddot \varphi_k + 3\frac{\dot a}{a} \dot \varphi_k + \omega_k^2(t) \varphi_k = 0 ,
\ee
with (see, \eqref{OMEGAM02})
\be\label{FRE01}
\omega_k^2 = \frac{k^2}{a^2} + m_\varphi^2(t) ,
\ee
where $m_\varphi$ is now given by (\ref{MAS01})  with the value of the Higgs given in (\ref{CHI01}). In terms of the of the number of times that it crosses zero, $j=\frac{M t}{\pi}$, the field (\ref{CHI01}) can be written as
\be\label{CHI02}
\chi(j) = \frac{\chi_{\rm end}}{j \pi} \sin(\pi j) ,
\ee
so that the frequency (\ref{FRE01}) turns out to be
\be
\omega_k^2(j) = \frac{k^2}{a^2} +  \frac{\tilde m_0^2 \sin(\pi j)}{\pi j } ,
\ee
where $\tilde m_0$ is the effective mass of the gauge bosons at the beginning of the first oscillation ($j=0$). The time dependence of the frequency entails the production of particles from two different sources. The first one is the expansion of the background spacetime, which can be neglected during the first oscillations of the Higgs. The other one is the time dependence of the Higgs field. The Bogolyubov transformation that relates the creation and annihilation operators of the $j$-crossing, $\hat b_j$ and $\hat b_j^\dag$, with those of the initial oscillation $\hat b_0$ and $\hat b_0^\dag$ is
\beq
\hat b_j &=& \alpha \, \hat b_0 + \beta \, \hat b_0^\dag , \\
\hat b_j^\dag &=& \alpha \, \hat b_0^\dag + \beta \, \hat b_0 ,
\eeq
where
\beq
\alpha &=& \frac{1}{2} \left( a_j^{3/2} \sqrt{\frac{\omega_j}{\omega_0}} + a_j^{-3/2} \sqrt{\frac{\omega_0}{\omega_j}} \right) , \\
\beta &=& \frac{1}{2} \left( a_j^{3/2} \sqrt{\frac{\omega_j}{\omega_0}} - a_j^{-3/2} \sqrt{\frac{\omega_0}{\omega_j}} \right) ,
\eeq 
with, $|\alpha|^2 - |\beta|^2 = 1$, and $a_j = \frac{a(j)}{a(0)}$ is the ratio between the value of the scale factor at the initial oscillation and the value of the scale factor at the oscillation $j$. The number of particles is then
\be\label{N01}
n_j = |\beta|^2 = \frac{1}{4} \left( \frac{\omega_j}{\omega_0} + \frac{\omega_0}{\omega_j} - 2 \right) ,  
\ee
where we have neglected the expansion of the universe during the first part of the reheating period. From Fig. \ref{figure_705} it can be seen that the production of particles is resonant near the points where the Higgs crosses zero. This is why this channel is called narrow resonance \cite{Koffman1997}.

During the first few oscillations and mainly in the adiabatic regime the influence of the expansion of the background spacetime can be neglected and the number of particles at the peaks, which coincide with the value for which the Higgs crosses zero, rapidly tends to the constant value $n_p(a=1)$, given by
\be
n_p(a=1) =  \frac{1}{4} \left( \sqrt{\frac{k^2}{k^2 + \tilde m_0^2}} + \sqrt{\frac{k^2 + \tilde m_0^2}{k^2}} - 2 \right) .
\ee
However, when the expansion of the background spacetime is taken into account, the number of particles created at the peaks scales with the scale factor, as (see, Fig. \ref{figure_705}, Right.)
\be
n_p(t) \approx \frac{ \tilde m_0}{4 k} a(t) - \frac{1}{2} .
\ee
In that case, the contribution to the production of intermediate bosons $W^\pm$ and $Z$ is much more enhanced, and the expansion of the spacetime cannot be neglected.

From the very beginning the intermediate gauge bosons start decaying into the fermions of the SM through their mutual interaction given by the Hamiltonian \cite{Bellido2009}
\be
H_I = -\frac{g_2}{\sqrt{2}} \left( W^+_\mu J^-_\mu + W^-_\mu J^+_\mu \right) - \frac{g_2}{\cos\theta_W} Z_\mu J^\mu_Z ,
\ee
where, $J^-_\mu \equiv \bar d_L \gamma^\mu u_L$ and $J_\mu^+ \equiv \bar u_L \gamma^\mu d_L$, are the charged currents that couple to the boson $W^+$ and to the boson $W^-$, respectively,  and the neutral current
\be
J^\mu_Z \equiv \kappa_1 \bar u_L \gamma^\mu u_L + \kappa_2 \bar d_L \gamma^\mu d_L  ,
\ee
with,
\be
\kappa_1 = \frac{1}{2} - \frac{2 \sin^2\theta_W}{3} , \kappa_2 = \frac{1}{2} - \frac{ \sin^2\theta_W}{3} .
\ee 
These interactions lead to the charged decays
\beq\label{D1}
W^+ \rightarrow  u + \bar d \,\, , \,\, W^- \rightarrow  \bar u +  d ,
\eeq
and the neutral decays
\beq\label{D2}
Z  \rightarrow  u + \bar u \,\, , \,\, Z \rightarrow  d + \bar d ,
\eeq
where $d$ and $u$ stands for the down- and up-type quarks, respectively, and similar decays can also be considered for the rest of quarks.  Analogously, we can consider the following decays in the lepton sector
\beq\label{D3}
W^+ \rightarrow  e^+ + \nu_e \,\, , \,\, W^- \rightarrow  e^- + \bar\nu_e ,
\eeq
as well as the neutral decays
\beq\label{D4}
Z  \rightarrow  e^- + e^+ \,\, , \,\, Z \rightarrow  \nu_e + \bar \nu_e ,
\eeq
all of them with their respective decay widths, $\Gamma_{W^\pm, Z\rightarrow i}$. Let us then notice that an asymmetry in the decay of the Higgs into the intermediate gauge bosons would entail the asymmetry in the production of quarks and leptons and therefore an asymmetry in the creation of primordial matter during the reheating period without the need of any other mechanism\footnote{Although other mechanisms of baryon asymmetry can simultaneously be present.}.

In the scenario of universes created in correlated pairs (see, Sec. \ref{sec03}), the two universes can be seen as expanding universes with the wave functions of their matter fields CP conjugated. Let us focus on one of the two wave functions, say $\psi_+(\varphi)$. If we consider that the modes of the expansion (\ref{MEX01}) are decoupled, then, the Schr\"odinger wave function for the scalar field, $\varphi$, which generically denote any of the polarisations of the $W^\pm$ and $Z$ bosons, can be written as the product of the wave functions of the modes, i.e.
\be
\psi_+(t, \varphi) = \prod_k \psi^{(k)}_{+}(t, \varphi_k) ,
\ee
where $\psi^{(k)}_{+}(t, \varphi_k) $ is the solution of the Schr\"odinger equation
\be\label{SCH03}
i \hbar \frac{\partial \psi^{(k)}_{+} }{\partial t_+} = H_{\varphi_k} \psi^{(k)}_{+} ,
\ee
with 
\be
H_{\varphi_k} = \frac{1}{2 a^3} p^2_{\varphi_k} + \frac{a^3 \omega_k^2}{2} \varphi_k^2 ,
\ee
where $a=a(t)$ is the scale factor of the background spacetime and $\omega_k=\omega_k(t)$ is given by (\ref{FRE01}). The general solution of the Schr\"odinger equation (\ref{SCH03}) can then be expressed in the basis of number eigenfunctions  given by \cite{Brizuela2019}
\be\label{PSI01}
\psi_{+,N_k}(t, \varphi_k) = \frac{e^{-i (N + 1/2) \tau}}{\sqrt{2^N N!}\pi^{1/4}\sqrt{\sigma}} e^{-\frac{\Omega}{2} \varphi_k^2} {\rm H}_N(\varphi_k/\sigma) ,
\ee
where ${\rm H}_N(x)$ is the Hermite polynomial of degree, $N\equiv N_k$, which is  the number occupation of the mode $k$, $\tau = \tau(t)$ is given by
\be
\tau_+(t) = \int^{t} \frac{1}{a^3 \sigma^2} dt ,
\ee
the function $\Omega$ is given by
\be
\Omega = \frac{1}{\sigma^2} - i \frac{a^3 \dot \sigma}{\sigma} ,
\ee
and $\sigma$ is a real function that satisfies the auxiliary equation \cite{Brizuela2019}
\be
\ddot \sigma + 3\frac{\dot a}{a} \dot \sigma + \omega_k^2 \sigma = \frac{k^2}{m^2 \sigma^3} .
\ee
The wave function in the time reversely symmetric universe is given by (see, Sec. \ref{sec030205}), $\psi_+(t, \bar \varphi)$, so the eigenfunctions of the basis for the state of the boson fields in the symmetric universe turns out to be given by (\ref{PSI01}) with the replacement, $\varphi_k \rightarrow \bar \varphi_k$. Therefore, if the scalar field $\varphi$ represents  the boson field $W^-$ in one of the universes, then, $\bar \varphi$ represents the boson field, $\bar W^- = W^+$, in the symmetric universe\footnote{Typically, $\varphi$ would represent a linear combination of the $W^+$ and $W^+$ fields. In that case, $\bar \varphi$ would represent the corresponding conjugated combination.}. The decay of the Higgs into the boson $W^+$ and $W^-$ can then be produced separately in the two symmetric universes.

Then, one can make the hypothesis that the intermediate gauge boson $W^+$ and $W^-$ are created in different universes, or at least at different rates in the two universes, without violating the global matter-antimatter asymmetry, an appealing scenario that is also suggested in \cite{Faizal2014, Boyle2018}. It is not mandatory that the asymmetry is complete but a small asymmetry in the decay of the Higgs into the $W^+$ and $W^-$ bosons in the two universes would eventually derive into an asymmetry in the production of fermions in the two universes due to the different decays of the $W^\pm$ bosons into fermions (see, (\ref{D1}-\ref{D3})). In the universe in which the boson $W^+$ predominates there would be an excess of the up quark with respect to the up antiquark, and accordingly, there would be an excess of protons over antiprotons, and matter would therefore dominate over antimatter. From the global picture of the two correlated universes the total amount of matter is always balanced with the total amount of antimatter so there is no global matter-antimatter asymmetry. It is worth noticing that the creation of a universe-antiuniverse pair does not assure that the content of one of the universes is completely matter and the content of the partner universe is completely antimatter. It is not therefore a mechanism for creating the matter-antimatter asymmetry but a mechanism to restore or explain the apparent asymmetry \cite{RP2017e}. In a multiverse scenario, one may expect a whole range of matter-antimatter distributions along the pairs of universes in the multiverse. In some of them there would be the needed asymmetry to form matter, and therefore galaxies and planets like in our universe, without violating any physical law.

\section{Observable effects of quantum cosmology}\label{sec04}

Testing the predictions of a theory with the observational data is a fundamental keystone of any physical proposal. However, it is not the unique consideration, theoretical consistency must also be taken into account and in fact it may help us to break through new paradigms. Well known examples in contemporary physics are the study of the unobservable black holes from the theoretical consistency of the perturbed motion of an observable companion, or the prediction of the charm quark from symmetry consistencies of the Standard Model of particles physics; not to talk about the unobserved 'dark matter' that is basically supported by consistency arguments. Furthermore, observability and falsifiability are not the same thing, as it is clearly argued\footnote{Tegmark poses the following example: \emph{a theory stating that there are 666 parallel universes, all of which are devoid of oxygen, makes the testable prediction that we should observe no oxygen here, and is therefore ruled out by observation}, cfr. Ref. \cite{Tegmark2007}, p. 105.} in Ref. \cite{Tegmark2007} (see also, Ref. \cite{Alonso2019} for a recent review).

Nevertheless, any theory must eventually be tested. In principle, the effects of quantum gravity are expected to be relevant at a very small length, or equivalent to very large scale of energy, and that makes them to be hardly measurable. For that reason it is quite difficult to propose practical tests in quantum cosmology. Even though,  we may expect some quantum corrections or deviations from the classical behaviour due to quantum cosmological effects that, at least from a theoretical point of view, could be detected. Among these effects, let us here briefly mention two: the pre-inflationary stage induced by the backreaction of the perturbation modes and the corrections due to the high order terms in the WKB approximation \cite{Kiefer1991, Kiefer2012, Brizuela2016a}.

In Sec. \ref{sec020402} we have seen that the vacuum state of the perturbation modes possesses an energy that permeates the whole universe. In principle, that \emph{backreaction} energy is expected to be a small correction to the energy of the unperturbed background spacetime. However, despite being small, it might produce some observable effects. Let us notice that the effective value of the Hamiltonian constraint that is obtained by tracing out from \eqref{THAM211} the degrees of freedom the perturbation modes can is,
\be
\mathcal H = \mathcal H_0 + \langle \mathcal H_m \rangle = 0 ,
\ee
which in terms of the time derivative of the scale factor yields the modified Friedmann equation,
\be\label{FE052}
\frac{\dot a^2}{a^2} = H^2(\varphi) - \frac{1}{a^2} + \frac{\langle \mathcal H_m \rangle}{a^3} ,
\ee
where $H^2(\varphi)$ is here the energy of the homogeneous mode of the scalar field, which for simplicity we shall consider constant, and
\be
\langle \mathcal H_m \rangle \propto  m \, n_{\rm max}^3 ,
\ee
for the particles of the matter fields, and
\be
\langle \mathcal H_m \rangle \propto \frac{ n^4_{\rm max}}{ a} ,
\ee
in the massless case. In the former case, the last term in \eqref{FE052} represents a matter like content in the universe ($\sim a^{-3}$) and in the latter it mimics a radiation energy content ($\sim a^{-4}$). In both cases, it can be shown \cite{Scardigli2011, Bouhmadi2011, Morais2017} that a matter or radiation predominated pre-inflationary period might, under some conditions, leave some observable imprints in the power spectrum of the CMB. However, it is sometimes considered \cite{Halliwell1989, Kiefer1992, Mersini2008d}, $n_{\rm max}\approx Ha$, which means accounting only for the backreaction of the superhorizon modes. In that case, the back reaction becomes equivalent to a cosmological  constant that effectively shifts the value of the potential \cite{RP2018a} (see, also, Refs. \cite{Mersini2008b, Mersini2008c, Mersini2008d}),
\be\label{ESH01}
\varepsilon = \frac{H^4}{8} \left\{ 1 - \frac{m^2}{H^2} \log\frac{b^2}{H^2} + \left( 1+ \frac{m^2}{H^2} \right) \left( 1 - \frac{b^2}{H^2} \right) \right\} ,
\ee
where terms of higher order have been disregarded. The energy shift (\ref{ESH01}) can be seen as a correction to the effective value of the potential of the scalar field, an effect that is expected to produce observable imprints in the properties of the CMB \cite{Mersini2017a, DiValentino2017a, DiValentino2017b}.

A different effect from quantum cosmology can be obtained by considering higher order terms in the WKB wave functions \eqref{SCWF053}. Following \cite{Kiefer1991, Kiefer2012, Brizuela2016a, Brizuela2016b}, let us assume a WKB wave function
\be\label{SCWF053b}
\phi(q;x_\textbf{n}) = C(q) e^{\pm\frac{i}{\hbar} S(q)}  \psi^{(1)}_\pm(q; x_\textbf{n}) ,
\ee
with, 
\beq
S(q) &=& S^{(0)}(q) + \hbar^2 S^{(2)}(q) + \ldots  , \\
 \psi^{(1)}_\pm(q; x_\textbf{n}) &=& \psi^{(0)}_\pm(q; x_\textbf{n})  e^{i\hbar S_\textbf{n}^{(2)}+\ldots} ,
\eeq
where $q$ are the variables of the background and $x_\textbf{n}$ are the perturbation modes. The first order terms give rise the corresponding Hamilton-Jacobi of the background spacetime and a modified Schr\"odinger equation for the corrected wave function of the perturbation modes,  $\psi^{(1)}_\pm$, which can be written as the Schr\"odinger equation for a set of uncoupled harmonic oscillator with a perturbed frequency with respect to the unperturbed frequencies (\ref{OMEGAM01}-\ref{OMEGAM02}) that can be written as
\be\label{SCH251}
\omega_n^2 \rightarrow \tilde{\omega}_n^2 = \omega_n^2 + \mathcal F_n ,
\ee
where, $\mathcal F_n=\mathcal F_n(t)$, is a time dependent function. A term like that is expected to produce a  variation in the power spectrum of the perturbation modes that would be in principle measurable \cite{Kiefer2012, Brizuela2016a, Brizuela2016b, Brizuela2019}. However, the effect is too small to be distinguishable from the statistical uncertainty implied by cosmic variance \cite{Brizuela2016a, Brizuela2016b}.

Perhaps the application to different inflationary models or the expected advances in the field of astronomical detectors and associated space missions, with the detection and analysis of gravitational waves or the 'cosmic neutrino background' (CNB), might make directly testable in the future the deviations from classicality predicted from quantum cosmology. Nevertheless, even though they may be difficult  to be observed these effects may have important conceptual consequences. For instance, we have seen in Sec. \ref{sec0205}  that an energy term like the one produced by the backreaction in the Friedman equation \eqref{FE052} might drastically change the way in which the universes can be created.

On the other hand, the creation of the universe in entangled pairs\footnote{And, in general, the creation of universes in $N$-entangled states, see \cite{RP2017c, AlonsoJL2018}.} may also add new features to be tested in the future. For instance, in an entangled universe the fields of the matter content in the two universe stop being a vacuum state. If one computes the state of the matter field in one single universe of the entangled pair by tracing out from the composite state the degrees of freedom of the matter in the partner universe, then, the resulting state turns out to be a quasi-thermal state with a temperature that depends on the degree of entanglement (which eventually depends on the size of the universe). In Ref. \cite{RP2018a}, it is computed the ratio between the fluctuations of the perturbation modes of a field that is initially in a thermal state and those corresponding to a initial vacuum state, yielding \cite{RP2018a}
 \be\label{QF01}
 \frac{\delta\phi_\textbf{n}^{th}}{\delta\phi_\textbf{n}^{I}} = \sqrt{\frac{1}{2}\left( 1 + \frac{x^2}{(1+x^2 )(1+\frac{m^2}{H^2 x^2})} \right) }  ,
\ee
with,
\be
x \equiv \frac{n }{H a} = \frac{n_\text{ph}}{H} \sim \frac{H^{-1}}{L_\text{ph}} ,
\ee
where, $L_\text{ph}$, is the physical wave length and $H^{-1}$ is the distance to the Hubble horizon.  The  large modes ($x\gg 1$) are in the vacuum state and then, $\delta\phi_\textbf{n}^{th} \approx \delta\phi_\textbf{n}^{I}$. However, the departure is significant for the horizon modes, $x \sim 1$. This would be a distinctive effect of the creation of the universes in entangled pairs and it should leave an observable imprint in the properties of the CMB.

%%%%%%%%%%%%%%%%%%%%%%%%%%%%%%%%%%%%%%%%%%

\section{Conclusions}\label{sec05}

Quantum geometrodynamics provides us with a consistent framework for the quantum description of the universe in terms of a wave function that contains, at least in principle, all the information about both the spacetime and the matter fields that propagate therein. In the semiclassical regime the complex phase of the wave function contains the information about the dynamics of the background spacetime and the wave function of the matter fields satisfy a Schrödinger equation that depends on the geometry of the subjacent spacetime. The dynamics of the spacetime turns out to be invariant under the complex conjugation of the semiclassical wave function. That gives rise two different solutions that have been typically interpreted as representing the expanding and the contracting branches of the universe, both filled with matter. We have seen that a different, more consistent interpretation is that the two solutions represent expanding universes with their matter contents being CP reversely related, so from the point of view of an internal observer of any of the universes the partner universe is always made up of antimatter, having therefore these two terms a relative meaning.

On the other hand,  the Wheeler-DeWitt equation can be seen as the wave equation of a field that propagates in the space of Riemannian $3$-dimensional geometries, $M$, where we can describe the evolution of the universe as a trajectory parametrised by a parameter that we can call \emph{time}.  The quantum mechanical counterpart is a quantum field that, following the customary approach of a quantum field theory, can be expressed in terms of creation and annihilation operators that satisfy the usual commutation relations. These operators represent the creation and the annihilation of modes for the spatial sections of the universe. Thus, the third quantisation formalism allows us to describe the quantum state of a whole spacetime manifold that can in general be a disconnected collection of simply connected manifolds. It is thus an appropriate framework to describe a multiverse scenario, in which the most natural boundary condition turns out to be that the field that represents the whole spacetime manifold remains in the ground state of an invariant representation along the entire history of the universes. This boundary condition implements the idea that the multiverse is the true isolated system and therefore no external interaction may excite its quantum state. However, as we have seen, this invariant boundary condition does not fix completely the quantum representation of the universe and, in fact, the ground state of one invariant representation is in general full of pairs of universes in another invariant representation. For instance, we have seen that the Hartle-Hawking no boundary condition can be seen as more fundamental because it is based on a more ontological argument. In that case, one may assume that the field that represents the  multiverse is in the ground state of the invariant Hartle-Hawking representation. Because the invariance of the invariant representation the field remains then in this ground state irrespective of the evolution of the universes. However, in terms of the invariant representation associated to the Vilenkin's tunnelling boundary condition, which represents the state of single universes, it turns out that the ground state of the Hartle-Hawking no boundary state is full of Vilenkin's pairs of universes. It means that the ground state of the multiverse is full of pairs of universes whose matter contents turn out to be CP conjugated. The charge conjugation comes from the complex conjugation relation between the Schrödinger wave functions of the matter fields in the two universes, and the reversely parity relation comes from the opposite signs of the momentum conjugated to the geometrical variables of the spatial sections of the spacetime. The two universes form thus a universe-antiuniverse pair.

We have seen all these features in a general model but particularly in the model of a homogeneous and isotropic spacetime with particles and matter fields propagating therein, where explicit examples can be analysed. In particular, we have seen that there are three main paradigms for the creation of the universe in quantum cosmology. The universes can be created from \emph{nothing}, i.e. from no preexisting spacetime. However, it requires of a precise fine tuning that seems to be quite unnatural. The other possibility is then that the universe is created from the quantum fluctuation of the spacetime of a preexisting spacetime. However, to be then consistent, one should also give an explanation for the creation of the \emph{first} spacetime. We have seen that the Gott and Li's explanation is that a vacuum fluctuation of our spacetime can travel back along one of the closed temporal curves that are allowed to exist in the vacuum state of the gravitational field to become the seminal spacetime from which our spacetime has been created. From that point of view, the universe would be the mother of itself. There is yet another possibility that avoids this paradoxical conclusion. The universes can be created from nothing, i.e. with no need of any prior spacetime, but they must then be created in pairs, from double Euclidean instantons. That turns out to be the most natural and self-consistent way in which the universes can be quantum mechanically created.

We have also seen in a very specific model that the creation of matter after the period of inflation would be correlated in the two universes of an entangled pair. The decay of the inflaton field into the particles of the Standard Model after inflation in the so-called reheating period is produced in such a way that the matter and antimatter of the two universes is perfectly balanced. The matter-antimatter asymmetry observed in our universe would only be therefore an apparent asymmetry and, in fact, it might be considered as an \emph{evidence} of the existence of an entangled companion of our universe that would contain the amount of antimatter that is left in our universe. However, the observational test of this and other quantum cosmological hypothesis seems to be still far from the current state of observation as the effects of an entangled partner would be in the domain of quantum gravity or at least in a pre-inflationary stage of the universe. Perhaps in the next future the advances in the detection of primordial gravitational waves or a possible cosmic neutrino background may shed some light into these intriguing questions.

\acknowledgments{This review is based on a series of lectures given at the Institute of Physics of the University of Szczecin, in September 2019. I thank Prof. M. Dabrowski for his kind invitation to give these lectures and to the members of the Szczecin Cosmology Group for their hospitality at the University of Szczecin. This work was supported by Comunidad de Madrid (Spain) under the Multiannual Agreement with UC3M in the line of Excellence of University Professors (EPUC3M23), in the context of the 5th. Regional Programme of Research and Technological Innovation (PRICIT).}

%\reftitle{References}


%\externalbibliography{yes}
\bibliography{../bibliography}

\begin{thebibliography}{999}

\bibitem[Wheeler(1957)]{Wheeler1957}
Wheeler, J.A.
\newblock On the nature of quantum geometrodynamics.
\newblock {\em Ann. Phys.} {\bf 1957}, {\em 2},~604--614.

\bibitem[Kiefer(1987)]{Kiefer1987}
Kiefer, C.
\newblock Continuous measurement of mini-superspace variables by higher
  multipoles.
\newblock {\em Class. Quant. Grav.} {\bf 1987}, {\em 4},~1369--1382.

\bibitem[Halliwell(1989)]{Halliwell1989}
Halliwell, J.J.
\newblock Decoherence in quantum cosmology.
\newblock {\em Phys. Rev. D} {\bf 1989}, {\em 39},~2912--2923.

\bibitem[Hartle(1990)]{Hartle1990}
Hartle, J.B.
\newblock The quantum mechanics of cosmology. In {\em Quantum Cosmology and
  Baby Universes}; Coleman, S.; Hartle, J.B.; Piran, T.; Weinberg, S., Eds.;
  World Scientific, London, UK,  1990; Vol.~7.

\bibitem[Gell-Mann and Hartle(1990)]{GellMann1990}
Gell-Mann, M.; Hartle, J.B.
\newblock Quantum mechanics in the light of quantum cosmology. In {\em
  Complexity, Entropy and the physics of Information}; Zurek, W.H., Ed.;
  Addison-Wesley, Reading, USA,  1990.

\bibitem[Kiefer(1992)]{Kiefer1992}
Kiefer, C.
\newblock Decoherence in quantum electrodynamics and quantum gravity.
\newblock {\em Phys. Rev. D} {\bf 1992}, {\em 46},~1658--1670.

\bibitem[Kiefer(2007)]{Kiefer2007}
Kiefer, C.
\newblock {\em Quantum gravity}; Oxford University Press, Oxford, UK,  2007.

\bibitem[Wiltshire(1996)]{Wiltshire2003}
Wiltshire, D.L.
\newblock An Introduction to Quantum Cosmology. In {\em Cosmology: the physics
  of the universe}; Robson, B.; Visvanathan, N.; Woolcock, W., Eds.; World
  Scientific, Singapore,  1996; pp. 473--531,
  \href{http://xxx.lanl.gov/abs/gr-qc/0101003}{{\normalfont [gr-qc/0101003]}}.

\bibitem[Wheeler(1968)]{Wheeler1968}
Wheeler, J.A.
\newblock Superspace and the nature of quantum geometrodynamics. In {\em
  Battelle rencontres}; DeWitt, C.M.; Wheeler, J.A., Eds.; W. A. Benjamin, Inc,
  New York,  1968; chapter~9.

\bibitem[De~Witt(1967)]{DeWitt1967}
De~Witt, B.S.
\newblock Quantum Theory of Gravity. {I}. The Canonical Theory.
\newblock {\em Phys. Rev.} {\bf 1967}, {\em 160},~1113--1148.

\bibitem[Dirac(1964)]{Dirac1964}
Dirac, P.A.M.
\newblock Lectures on Quantum Mechanics. In {\em Belfer Graduate School of
  Science Monographs Series}; Number~2, Belfer Graduate School of Science, NY,
  USA,  1964.

\bibitem[Hartle and Hawking(1983)]{Hartle1983}
Hartle, J.B.; Hawking, S.W.
\newblock Wave function of the Universe.
\newblock {\em Phys. Rev. D} {\bf 1983}, {\em 28},~2960.

\bibitem[Joos \em{et~al.}(2003)Joos et~al.]{Joos2003}
Joos, E.; others.
\newblock {\em Decoherence and the Appearance of a Classical World in Quantum
  Theory}; Springer-Verlag, Berlin, Germany,  2003.

\bibitem[Schlosshauer(2007)]{Schlosshauer2007}
Schlosshauer, M.
\newblock {\em Decoherence and the quantum-to-classical transition}; Springer,
  Berlin, Germany,  2007.

\bibitem[Halliwell(1990)]{Halliwell1990}
Halliwell, J.J.
\newblock Introductory lectures on quantum cosmology. In {\em Quantum Cosmology
  and Baby Universes}; Coleman, S.; Hartle, J.B.; Piran, T.; Weinberg, S.,
  Eds.; World Scientific, London, UK,  1990; Vol.~7.

\bibitem[Kiefer and Singh(1991)]{Kiefer1991}
Kiefer, C.; Singh, T.P.
\newblock Quantum gravitational corrections to the functional Schr{\"o}dinger
  equation.
\newblock {\em Phys. Rev. D} {\bf 1991}, {\em 44},~1067.

\bibitem[Kiefer and Kr{\"a}mer(2012)]{Kiefer2012}
Kiefer, C.; Kr{\"a}mer, M.
\newblock Quantum gravitational contributions to the CMB anisotropy spectrum.
\newblock {\em Phys. Rev. Lett.} {\bf 2012}, {\em 108},~021301.

\bibitem[Brizuela \em{et~al.}(2016{\natexlab{a}})Brizuela, Kiefer, and
  Kr{\"a}mer]{Brizuela2016a}
Brizuela, D.; Kiefer, C.; Kr{\"a}mer, M.
\newblock Quantum-gravitational effects on gauge-invariant scalar and tensor
  perturbations during inflation: The de Sitter case.
\newblock {\em Phys. Rev. D} {\bf 2016}, {\em 93},~104035,
  \href{http://xxx.lanl.gov/abs/arXiv:1511.05545}{{\normalfont
  [arXiv:1511.05545]}}.

\bibitem[Brizuela \em{et~al.}(2016{\natexlab{b}})Brizuela, Kiefer, and
  Kr{\"a}mer]{Brizuela2016b}
Brizuela, D.; Kiefer, C.; Kr{\"a}mer, M.
\newblock Quantum-gravitational effects on gauge-invariant scalar and tensor
  perturbations during inflation: The slow-roll approximation.
\newblock {\em Phys. Rev. D} {\bf 2016}, {\em 94},~123527.

\bibitem[Hartle(1995)]{Hartle1993}
Hartle, J.B.
\newblock Spacetime quantum mechanics and the quantum mechanics of spacetime.
\newblock  Gravitation and Quantization: Proceedings of the 1992 Les Houches
  Summer School; Julia, B.; Zinn-Justin, J., Eds. North Holland, Amsterdam,
  1995,  \href{http://xxx.lanl.gov/abs/gr-qc/9304006}{{\normalfont
  [gr-qc/9304006]}}.

\bibitem[Bezrukov and Shaposhnikov(2008)]{Bezrukov2008}
Bezrukov, F.L.; Shaposhnikov, M.E.
\newblock The {S}tandard {M}odel {H}iggs boson as the inflaton.
\newblock {\em Phys. Lett. B} {\bf 2008}, {\em 659},~703.

\bibitem[Garcia-Bellido \em{et~al.}(2009)Garcia-Bellido, Figueroa, and
  Rubio]{Bellido2009}
Garcia-Bellido, J.; Figueroa, D.G.; Rubio, J.
\newblock Preheating in the {S}tandard {M}odel with the {H}iggs-{I}nflaton
  coupled to gravity.
\newblock {\em Phys. Rev. D} {\bf 2009}, {\em 79},~063531,
  \href{http://xxx.lanl.gov/abs/arXiv:0812.4624}{{\normalfont
  [arXiv:0812.4624]}}.

\bibitem[Garay and Robles-P{\'e}rez(2014)]{Garay2014}
Garay, I.; Robles-P{\'e}rez, S.
\newblock Effects of a scalar field on the thermodynamics of interuniversal
  entanglement.
\newblock {\em Int. J. Mod. Phys. D} {\bf 2014}, {\em 23},~1450043.

\bibitem[Linde(1993)]{Linde1993}
Linde, A.
\newblock {\em Particle physics and inflationary cosmology}; Vol.~5, {\em
  Contemporary concepts in physics, Chur, Switzerland}, Harwood academic
  publishers,  1993.

\bibitem[Rubakov(1999)]{Rubakov1999}
Rubakov, V.A.
\newblock Quantum cosmology.
\newblock {\em Lecture at NATO ASI 'Structure Formation in the Universe',
  Cambridge, 1999} {\bf 1999},
  \href{http://xxx.lanl.gov/abs/arXiv:gr-qc/9910025}{{\normalfont
  [arXiv:gr-qc/9910025]}}.

\bibitem[Halliwell and Hawking(1985)]{Halliwell1985}
Halliwell, J.J.; Hawking, S.W.
\newblock Origin of structure in the Universe.
\newblock {\em Phys. Rev. D} {\bf 1985}, {\em 31},~1777--1791.

\bibitem[Rubakov(1988)]{Rubakov1988}
Rubakov, V.A.
\newblock On third quantization and the cosmological constant.
\newblock {\em Phys. Lett. B} {\bf 1988}, {\em 214},~503--507.

\bibitem[Halliwell(1987)]{Halliwell1987}
Halliwell, J.J.
\newblock Correlations in the wave function of the Universe.
\newblock {\em Phys. Rev. D} {\bf 1987}, {\em 36},~3626--3640.

\bibitem[Grishchuk and Sidorov(1990)]{Grishchuk1990}
Grishchuk, L.P.; Sidorov, Y.V.
\newblock Squeezed quantum states of relic gravitons and primordial density
  fluctuations.
\newblock {\em Phys. Rev. D} {\bf 1990}, {\em 42},~3413--3421.

\bibitem[Lewis and Riesenfeld(1969)]{Lewis1969}
Lewis, H.R.; Riesenfeld, W.B.
\newblock An Exact Quantum THeory of the Time-Dependent Harmonic Oscillator and
  of a Charged Particle in a Time-Dependent Electromagnetic Field.
\newblock {\em J. Math. Phys.} {\bf 1969}, {\em 10},~1458--1473.

\bibitem[Leach(1983)]{Leach1983}
Leach, P.G.L.
\newblock Harmonic oscillator with variable mass.
\newblock {\em J. Phys. A} {\bf 1983}, {\em 16},~3261--3269.

\bibitem[Kanasugui and Okada(1995)]{Kanasugui1995}
Kanasugui, H.; Okada, H.
\newblock Systematic treatments of general time-dependent harmonica oscillator
  in classical and quantum mechanics.
\newblock {\em Prog. Theor. Phys.} {\bf 1995}, {\em 93},~949--960.

\bibitem[Sheng \em{et~al.}(1995)Sheng et~al.]{Sheng1995}
Sheng, D.; others.
\newblock Quantum Harmonic Oscillator with Time-Dependent Mass and Frequency.
\newblock {\em Int. J. Theor. Phys.} {\bf 1995}, {\em 34},~355--368.

\bibitem[Robles-P{\'e}rez(2017)]{RP2017f}
Robles-P{\'e}rez, S.J.
\newblock {\em In preparation} {\bf 2017}.

\bibitem[Robles-P{\'e}rez(2107)]{RP2017c}
Robles-P{\'e}rez, S.
\newblock Quantum cosmology of a conformal multiverse.
\newblock {\em Phys. Rev. D} {\bf 2107}, {\em 96},~063511,
  \href{http://xxx.lanl.gov/abs/arXiv:1706.06023}{{\normalfont
  [arXiv:1706.06023]}}.

\bibitem[Hawking(1982)]{Hawking1982}
Hawking, S.W.
\newblock The boundary conditions of the universe.
\newblock {\em Astrophysical Cosmology, 563-72. Vatican City: Pontificia
  Academiae Scientarium} {\bf 1982}.

\bibitem[Hawking(1984{\natexlab{a}})]{Hawking1983}
Hawking, S.W.
\newblock Quantum cosmology. In {\em Relativity, groups and topology II, Les
  Houches, Session XL, 1983}; De~Witt, B.S.; Stora, R., Eds.; Elsevier Science
  Publishers B. V.,  1984.

\bibitem[Hawking(1984{\natexlab{b}})]{Hawking1984}
Hawking, S.W.
\newblock The quantum state of the universe.
\newblock {\em Nucl. Phys. B} {\bf 1984}, {\em 239},~257--276.

\bibitem[Vilenkin(1982)]{Vilenkin1982}
Vilenkin, A.
\newblock Creation of universes from nothing.
\newblock {\em Phys. Lett. B} {\bf 1982}, {\em 117},~25--28.

\bibitem[Vilenkin(1984)]{Vilenkin1984}
Vilenkin, A.
\newblock Quantum creation of universes.
\newblock {\em Phys. Rev. D} {\bf 1984}, {\em 30},~509--511.

\bibitem[Vilenkin(1986)]{Vilenkin1986}
Vilenkin, A.
\newblock Boundary conditions in quantum cosmology.
\newblock {\em Phys. Rev. D} {\bf 1986}, {\em 33},~3560--3569.

\bibitem[Vilenkin(1995)]{Vilenkin1995}
Vilenkin, A.
\newblock Predictions from Quantum Cosmology.
\newblock {\em Phys. Rev. Lett.} {\bf 1995}, {\em 74},~846--849.

\bibitem[Vilenkin(1989)]{Vilenkin1989}
Vilenkin, A.
\newblock Interpretation of the wave function of the universe.
\newblock {\em Phys. Rev. D} {\bf 1989}, {\em D},~1116.

\bibitem[Gott and Li(1998)]{Gott1998}
Gott, J.R.I.; Li, L.X.
\newblock Can the universe create itself?
\newblock {\em Phys. Rev. D} {\bf 1998}, {\em 58},~023501.

\bibitem[Strominger(1990)]{Strominger1990}
Strominger, A.
\newblock Baby Universes. In {\em Quantum Cosmology and Baby Universes};
  Coleman, S.; Hartle, J.B.; Piran, T.; Weinberg, S., Eds.; World Scientific,
  London, UK,  1990; Vol.~7.

\bibitem[Barvinsky and Kamenshchik(2006)]{Barvinsky2006}
Barvinsky, A.O.; Kamenshchik, A.Y.
\newblock Cosmological Landscape From Nothing: Some Like It Hot.
\newblock {\em JCAP} {\bf 2006}, {\em 0609},~014,
  \href{http://xxx.lanl.gov/abs/hep-th/0605132}{{\normalfont
  [hep-th/0605132]}}.

\bibitem[Barvinsky and Kamenshchik(2007)]{Barvinsky2007a}
Barvinsky, A.O.; Kamenshchik, A.Y.
\newblock Cosmological landscape and Euclidean quantum gravity.
\newblock {\em J. Phys. A} {\bf 2007}, {\em 40},~7043--7048,
  \href{http://xxx.lanl.gov/abs/hep-th/0701201}{{\normalfont
  [hep-th/0701201]}}.

\bibitem[Barvinsky(2007)]{Barvinsky2007b}
Barvinsky, A.O.
\newblock Why there is something rather than nothing (out of everything)?
\newblock {\em Phys. Rev. Lett.} {\bf 2007}, {\em 99},~071301,
  \href{http://xxx.lanl.gov/abs/0704.0083}{{\normalfont [0704.0083]}}.

\bibitem[Robles-P{\'e}rez and Gonz{\'a}lez-D{\'\i}az(2014)]{RP2011b}
Robles-P{\'e}rez, S.; Gonz{\'a}lez-D{\'\i}az, P.F.
\newblock Quantum entanglement in the multiverse.
\newblock {\em JETP} {\bf 2014}, {\em 118},~34,
  \href{http://xxx.lanl.gov/abs/arXiv:1111.4128}{{\normalfont
  [arXiv:1111.4128]}}.

\bibitem[Robles-P{\'e}rez(2014)]{RP2014}
Robles-P{\'e}rez, S.J.
\newblock Creation of entangled universes avoids the Big Bang singularity.
\newblock {\em Journal of Gravity} {\bf 2014}, {\em 2014},~382675,
  \href{http://xxx.lanl.gov/abs/arXiv:1311.2379}{{\normalfont
  [arXiv:1311.2379]}}.

\bibitem[Chen \em{et~al.}(2017)Chen, Hu, and Yeom]{Chen2017}
Chen, P.; Hu, Y.C.; Yeom, D.H.
\newblock Fuzzy Euclidean wormholes in de Sitter space.
\newblock {\em JCAP} {\bf 2017}, {\em 07},~001.

\bibitem[Caderni and Martellini(1984)]{Caderni1984}
Caderni, N.; Martellini, M.
\newblock Third quantization formalism for Hamiltonian cosmologies.
\newblock {\em Int. J. Theor. Phys.} {\bf 1984}, {\em 23},~233.

\bibitem[Coleman(1988{\natexlab{a}})]{Coleman1988}
Coleman, S.
\newblock Black holes as red herrings: Topological fluctuations and the loss of
  quantum coherence.
\newblock {\em Nucl. Phys. B} {\bf 1988}, {\em 307},~867.

\bibitem[Coleman(1988{\natexlab{b}})]{Coleman1988b}
Coleman, S.
\newblock Why there is nothing rather than something? A theory of the
  cosmological constant.
\newblock {\em Nucl. Phys. B} {\bf 1988}, {\em 310},~643--668.

\bibitem[McGuigan(1988)]{McGuigan1988}
McGuigan, M.
\newblock Third quantization and the Wheeler-DeWitt equation.
\newblock {\em Phys. Rev. D} {\bf 1988}, {\em 38},~3031.

\bibitem[McGuigan(1989)]{McGuigan1989}
McGuigan, M.
\newblock Universe creation from the third quantized vacuum.
\newblock {\em Phys. Rev. D} {\bf 1989}, {\em 39},~2229.

\bibitem[McGuigan(1990)]{McGuigan1990}
McGuigan, M.
\newblock Universe decay and changing the cosmological constant.
\newblock {\em Phys. Rev. D} {\bf 1990}, {\em 41},~418.

\bibitem[Hawking(1990)]{Hawking1990}
Hawking, S.W.
\newblock Wormholes and Non-simply Connected Manifolds. In {\em Quantum
  Cosmology and Baby Universes}; Coleman, S.; Hartle, J.B.; Piran, T.;
  Weinberg, S., Eds.; World Scientific, London, UK,  1990; Vol.~7.

\bibitem[Robles-P{\'e}rez and Gonz{\'a}lez-D{\'\i}az(2010)]{RP2010}
Robles-P{\'e}rez, S.; Gonz{\'a}lez-D{\'\i}az, P.F.
\newblock Quantum state of the multiverse.
\newblock {\em Phys. Rev. D} {\bf 2010}, {\em 81},~083529,
  \href{http://xxx.lanl.gov/abs/arXiv:1005.2147v1}{{\normalfont
  [arXiv:1005.2147v1]}}.

\bibitem[Gonz{\'a}lez-D{\'\i}az(1992{\natexlab{a}})]{PFGD1992a}
Gonz{\'a}lez-D{\'\i}az, P.F.
\newblock Nonclassical states in quantum gravity.
\newblock {\em Phys. Lett. B} {\bf 1992}, {\em 293},~294.

\bibitem[Gonz{\'a}lez-D{\'\i}az(1992{\natexlab{b}})]{PFGD1992b}
Gonz{\'a}lez-D{\'\i}az, P.F.
\newblock Regaining quantum incoherence for matter fields.
\newblock {\em Phys. Rev. D} {\bf 1992}, {\em 45},~499.

\bibitem[Higuchi and Wald(1995)]{Higuchi1995}
Higuchi, A.; Wald, R.M.
\newblock Applications of a new proposal for solving the problem of time to
  some simple quantum cosmological models.
\newblock {\em Phys. Rev. D} {\bf 1995}, {\em 51},~544--561,
  \href{http://xxx.lanl.gov/abs/arXiv:gr-qc/9407038v1}{{\normalfont
  [arXiv:gr-qc/9407038v1]}}.

\bibitem[Barbour(2012)]{Barbour2011}
Barbour, J.
\newblock Shape Dynamics. An introduction.
\newblock  Quantum field theory and gravity; Finster, F.; M{\"u}ller, O.;
  Nardmann, M.; Tolksdorf, J.; Zeidler, E., Eds. Springer, Basel,  2012, pp.
  257--297,  \href{http://xxx.lanl.gov/abs/arXiv:1105.0183}{{\normalfont
  [arXiv:1105.0183]}}.

\bibitem[Griffiths and Podolsky(2009)]{Griffiths2009}
Griffiths, J.B.; Podolsky, J.
\newblock {\em Exact space-times in Einstein's general relativity}; Cambridge
  monographs on mathematical physics, Cambridge University Press, Cambridge,
  UK,  2009.

\bibitem[Biesiada and Rugh()]{Biesiada1994}
Biesiada, M.; Rugh, S.
\newblock Maupertuis principle, {W}heeler's superspace and an invariant
  criterion for local instability in general relativity,
  \href{http://xxx.lanl.gov/abs/arXiv:gr-qc/9408030}{{\normalfont
  [arXiv:gr-qc/9408030]}}.

\bibitem[Garay and Robles-P{\'e}rez(2018)]{Garay2018}
Garay, I.; Robles-P{\'e}rez, S.
\newblock Classical geodesics from the canonical quantisation of spacetime
  coordinates.
\newblock {\em submitted} {\bf 2018},
  \href{http://xxx.lanl.gov/abs/arXiv:1901.05171}{{\normalfont
  [arXiv:1901.05171]}}.

\bibitem[Pimentel and Mora(2001)]{Pimentel2001}
Pimentel, L.O.; Mora, C.
\newblock Third quantization of Brans-Dicke Cosmology.
\newblock {\em Phys. Lett. A} {\bf 2001}, {\em 280},~191--196,
  \href{http://xxx.lanl.gov/abs/gr-qc/0009026}{{\normalfont [gr-qc/0009026]}}.

\bibitem[Kim()]{Kim2012}
Kim, S.P.
\newblock Third quantization and quantum universes.
\newblock  \href{http://xxx.lanl.gov/abs/arXiv:1212.5355}{{\normalfont
  [arXiv:1212.5355]}}.

\bibitem[Ohkuwa and Ezawa(2013)]{Ohkuwa2013}
Ohkuwa, Y.; Ezawa, Y.
\newblock Third quantization of f(R)-type gravity II - General f(R) case.
\newblock {\em Class. Quant. Grav.} {\bf 2013}, {\em 20},~235015,
  \href{http://xxx.lanl.gov/abs/arXiv:1210.4719}{{\normalfont
  [arXiv:1210.4719]}}.

\bibitem[Calgani \em{et~al.}(2012)Calgani, Gielen, and Oriti]{Calgani2012}
Calgani, G.; Gielen, S.; Oriti, D.
\newblock Group field theory cosmology: a cosmological field theory of quantum
  geometry.
\newblock {\em Class. Quant. Grav.} {\bf 2012}, {\em 29},~105005.

\bibitem[Faizal(2014)]{Faizal2014}
Faizal, M.
\newblock Multiverse in the third quantized formalism.
\newblock {\em Comm. Theor. Phys.} {\bf 2014}, {\em 62},~697,
  \href{http://xxx.lanl.gov/abs/arXiv:1407.3118}{{\normalfont
  [arXiv:1407.3118]}}.

\bibitem[Balcerzak and Marosek(2019)]{Balcerzak2019}
Balcerzak, A.; Marosek, K.
\newblock Emergence of multiverse in third quantized varying constants
  cosmologies.
\newblock {\em Eur. Phys. J. C} {\bf 2019}, {\em 79},~563,
  \href{http://xxx.lanl.gov/abs/arXiv:1905.09671v3}{{\normalfont
  [arXiv:1905.09671v3]}}.

\bibitem[Balcerzak and Marosek(2020)]{Balcerzak2020}
Balcerzak, A.; Marosek, K.
\newblock Doubleverse entanglement in third quantized non-minimally coupled
  varying constants cosmologies.
\newblock {\em Eur. Phys. J. C} {\bf 2020}, {\em 80},~709,
  \href{http://xxx.lanl.gov/abs/arXiv:2003.06380v3}{{\normalfont
  [arXiv:2003.06380v3]}}.

\bibitem[Campanelli(2020)]{Campanelli2020}
Campanelli, L.
\newblock Creation of universes from the third-quantized vacuum.
\newblock {\em Phys. Rev. D} {\bf 2020}, {\em 102},~043514,
  \href{http://xxx.lanl.gov/abs/arXiv:2007.01732}{{\normalfont
  [arXiv:2007.01732]}}.

\bibitem[Robles-P{\'e}rez(2021)]{RP2021a}
Robles-P{\'e}rez, S.J.
\newblock Hartle-Hawking vacuum is full of Vilenkin's universe-antiuniverse
  pairs.
\newblock {\em submitted} {\bf 2021}.

\bibitem[Birrell and Davies(1982)]{Birrell1982}
Birrell, N.D.; Davies, P.C.W.
\newblock {\em Quantum fields in curved space}; Cambridge University Press,
  Cambridge, UK,  1982.

\bibitem[Mukhanov and Winitzki(2007)]{Mukhanov2007}
Mukhanov, V.F.; Winitzki, S.
\newblock {\em Quantum Effects in Gravity}; Cambridge University Press,
  Cambridge, UK,  2007.

\bibitem[Bander and Itzykson(1966)]{Bander1966}
Bander, M.; Itzykson, C.
\newblock Group theory and the hydrogen atom (II).
\newblock {\em Rev. Mod. Phys.} {\bf 1966}, {\em 38},~346.

\bibitem[Lewis(1968)]{Lewis1968}
Lewis, H.R.
\newblock Class of exact invariants for classical and quantum time dependent
  harmonic oscillators.
\newblock {\em J. Math. Phys.} {\bf 1968}, {\em 9},~1976.

\bibitem[Pedrosa(1987)]{Pedrosa1987}
Pedrosa, I.A.
\newblock Comment on ``Coherent states for the time-dependent harmonic
  oscillator''.
\newblock {\em Phys. Rev. D} {\bf 1987}, {\em 36},~1279.

\bibitem[Dantas \em{et~al.}(1992)Dantas, Pedrosa, and Baseia]{Dantas1992}
Dantas, C.M.A.; Pedrosa, I.A.; Baseia, B.
\newblock Harmonic oscillator with time-dependent mass and frequency and a
  perturbative potential.
\newblock {\em Phys. Rev. A} {\bf 1992}, {\em 45},~1320.

\bibitem[Song(2000)]{Song2000}
Song, D.Y.
\newblock Unitary relation between a harmonic oscillator of time-dependent
  frequency and a simple harmonic oscillator with and withuot an inverse-square
  potential.
\newblock {\em Phys. Rev. A} {\bf 2000}, {\em 62},~014103.

\bibitem[Kim and Page(2001)]{Kim2001}
Kim, S.P.; Page, D.N.
\newblock Classical and quantum action-phase variables for time-dependent
  oscillators.
\newblock {\em Phys. Rev. A} {\bf 2001}, {\em 64},~012104.

\bibitem[Park(2004)]{Park2004}
Park, T.J.
\newblock Canonical Transformations for Time-Dependent Harmonic Oscillators.
\newblock {\em Bull. Korean Chem. Soc.} {\bf 2004}, {\em 25}.

\bibitem[Robles-P{\'e}rez(2017)]{RP2017d}
Robles-P{\'e}rez, S.
\newblock Invariant vacuum.
\newblock {\em Phys. Lett. B} {\bf 2017}, {\em 774},~608--615,
  \href{http://xxx.lanl.gov/abs/arXiv:1706.05474}{{\normalfont
  [arXiv:1706.05474]}}.

\bibitem[Rajeev \em{et~al.}(2018)Rajeev, Chakraborty, and
  Padmanabhan]{Rajeev2018}
Rajeev, K.; Chakraborty, S.; Padmanabhan, T.
\newblock Inverting a normal harmonic oscillator: physical interpretation and
  applications.
\newblock {\em Gen. Rel. Grav.} {\bf 2018}, {\em 50},~116,
  \href{http://xxx.lanl.gov/abs/arXiv:1712.06617}{{\normalfont
  [arXiv:1712.06617]}}.

\bibitem[Olson and Ralph(2011)]{JOlson2011}
Olson, S.J.; Ralph, T.C.
\newblock Entanglement between the future and past in the quantum vacuum.
\newblock {\em Phys. Rev. Lett.} {\bf 2011}, {\em 106},~110404,
  \href{http://xxx.lanl.gov/abs/arXiv:1003.0720}{{\normalfont
  [arXiv:1003.0720]}}.

\bibitem[Feynman(1949)]{Feynman1949}
Feynman, R.P.
\newblock The theory of positrons.
\newblock {\em Phys. Rev.} {\bf 1949}, {\em 76},~749.

\bibitem[Robles-P{\'e}rez(2018)]{RP2018a}
Robles-P{\'e}rez, S.J.
\newblock Cosmological perturbations in the entangled inflationary universe.
\newblock {\em Phys. Rev. D} {\bf 2018}, {\em 97},~066018,
  \href{http://xxx.lanl.gov/abs/arXiv:1708.03860}{{\normalfont
  [arXiv:1708.03860]}}.

\bibitem[Robles-P{\'e}rez(2019)]{RP2019c}
Robles-P{\'e}rez, S.J.
\newblock Time reversal symmetry in cosmology and the creation of a
  universe-antiuniverse pair.
\newblock {\em Universe} {\bf 2019}, {\em 5},~150,
  \href{http://xxx.lanl.gov/abs/arXiv:1901.03387}{{\normalfont
  [arXiv:1901.03387]}}.

\bibitem[Robles-P{\'e}rez(2017)]{RP2017e}
Robles-P{\'e}rez, S.J.
\newblock Restoration of matter-antimatter symmetry in the multiverse {\bf
  2017}.
\newblock  \href{http://xxx.lanl.gov/abs/arXiv:1706.06304}{{\normalfont
  [arXiv:1706.06304]}}.

\bibitem[Brizuela \em{et~al.}(2019)Brizuela, Kiefer, Kr{\"a}mer, and
  Robles-P{\'e}rez]{Brizuela2019}
Brizuela, D.; Kiefer, C.; Kr{\"a}mer, M.; Robles-P{\'e}rez, S.
\newblock Quantum-gravity effects for excited states of inflationary
  perturbations.
\newblock {\em Phys. Rev. D} {\bf 2019},
  \href{http://xxx.lanl.gov/abs/arXiv:1903.01234}{{\normalfont
  [arXiv:1903.01234]}}.

\bibitem[Robles-P{\'e}rez(2019)]{RP2019b}
Robles-P{\'e}rez, S.J.
\newblock Quantum cosmology in the light of quantum mechanics.
\newblock {\em Galaxies} {\bf 2019}, {\em 7},~50.

\bibitem[Mukhanov(2008)]{Mukhanov2008}
Mukhanov, V.F.
\newblock {\em Physical foundations of cosmology}; Cambridge University Press,
  Cambridge, UK,  2008.

\bibitem[Kofman \em{et~al.}(1997)Kofman, Linde, and Starobinsky]{Koffman1997}
Kofman, L.; Linde, A.; Starobinsky, A.A.
\newblock Towards the theory of reheating after inflation.
\newblock {\em Phys. Rev. D} {\bf 1997}, {\em 56},~3258,
  \href{http://xxx.lanl.gov/abs/arXiv:hep-ph/9704452}{{\normalfont
  [arXiv:hep-ph/9704452]}}.

\bibitem[Boyle \em{et~al.}(2018)Boyle, Finn, and Turok]{Boyle2018}
Boyle, L.; Finn, K.; Turok, N.
\newblock CPT-Symmetric universe.
\newblock {\em Phys. Rev. Lett.} {\bf 2018}, {\em 121},~251301.

\bibitem[Tegmark(2007)]{Tegmark2007}
Tegmark, M.
\newblock The multiverse hierarchy. In {\em Universe or Multiverse}; Carr, B.,
  Ed.; Cambridge University Press, Cambridge, UK,  2007; chapter~7.

\bibitem[Alonso-Serrano and Jannes(2019)]{Alonso2019}
Alonso-Serrano, A.; Jannes, G.
\newblock Conceptual challenges on the road to the multiverse.
\newblock {\em Universe} {\bf 2019}, {\em 5},~212,
  \href{http://xxx.lanl.gov/abs/arXiv:1910.07375}{{\normalfont
  [arXiv:1910.07375]}}.

\bibitem[Scardigli \em{et~al.}(2011)Scardigli, Gruber, and Chen]{Scardigli2011}
Scardigli, F.; Gruber, C.; Chen, P.
\newblock Black hole remnants in the early universe.
\newblock {\em Phys. Rev. D} {\bf 2011}, {\em 83}.

\bibitem[Bouhmadi-Lopez \em{et~al.}(2011)Bouhmadi-Lopez, Chen, and
  Liu]{Bouhmadi2011}
Bouhmadi-Lopez, M.; Chen, P.; Liu, Y.
\newblock Cosmological imprints of a generalized Chaplygin gas model for the
  early universe.
\newblock {\em Phys. Rev. D} {\bf 2011}, {\em 84},~023505.

\bibitem[Morais \em{et~al.}(2018)Morais, Bouhmadi-Lopez, Kr{\"a}mer, and
  Robles-P{\'e}rez]{Morais2017}
Morais, J.; Bouhmadi-Lopez, M.; Kr{\"a}mer, M.; Robles-P{\'e}rez, S.
\newblock Pre-inflation from th emultiverse: can it solve the quadrupole
  problem in the cosmic microwave background?
\newblock {\em Eur. Phys. J. C} {\bf 2018}, {\em 78},~240,
  \href{http://xxx.lanl.gov/abs/arXiv:1711.05138}{{\normalfont
  [arXiv:1711.05138]}}.

\bibitem[Holman \em{et~al.}(2008)Holman, Mersini-Houghton, and
  Takahashi]{Mersini2008d}
Holman, R.; Mersini-Houghton, L.; Takahashi, T.
\newblock Cosmological avatars of the Landscape II.
\newblock {\em Phys. Rev. D} {\bf 2008}, {\em 77},~063511,
  \href{http://xxx.lanl.gov/abs/[arXiv:hep-th/0612142v1]}{{\normalfont
  [[arXiv:hep-th/0612142v1]]}}.

\bibitem[Mersini-Houghton(2008)]{Mersini2008b}
Mersini-Houghton, L.
\newblock Thoughts on defining the multiverse {\bf 2008}.
\newblock  \href{http://xxx.lanl.gov/abs/arXiv:0804.4280v1}{{\normalfont
  [arXiv:0804.4280v1]}}.

\bibitem[Holman \em{et~al.}(2008)Holman, Mersini-Houghton, and
  Takahashi]{Mersini2008c}
Holman, R.; Mersini-Houghton, L.; Takahashi, T.
\newblock Cosmological avatars of the Landscape I.
\newblock {\em Phys. Rev. D} {\bf 2008}, {\em 77},~063510,
  \href{http://xxx.lanl.gov/abs/[arXiv:hep-th/0611223v1]}{{\normalfont
  [[arXiv:hep-th/0611223v1]]}}.

\bibitem[Mersini-Houghton(2017)]{Mersini2017a}
Mersini-Houghton, L.
\newblock Predictions of the quantum landscape multiverse.
\newblock {\em Class. Quant. Grav.} {\bf 2017}, {\em 34},~047001.

\bibitem[Di~Valentino and
  Mersini-Houghton(2017{\natexlab{a}})]{DiValentino2017a}
Di~Valentino, E.; Mersini-Houghton, L.
\newblock Testing predictions of the quantum landscape multiverse 1: the
  Starobinsky inflationary potential.
\newblock {\em JCAP} {\bf 2017}, {\em 03},~002.

\bibitem[Di~Valentino and
  Mersini-Houghton(2017{\natexlab{b}})]{DiValentino2017b}
Di~Valentino, E.; Mersini-Houghton, L.
\newblock Testing predictions of the quantum landscape multiverse 2: the
  exponential inflationary potential.
\newblock {\em JCAP} {\bf 2017}, {\em 03},~020.

\bibitem[Alonso and Carmona(2018)]{AlonsoJL2018}
Alonso, J.; Carmona, J.
\newblock Before spacetime: a proposal of a framework for multiverse quantum
  cosmology based on three cosmological conjectures.
\newblock {\em Class. Quant. Grav.} {\bf 2018}, {\em 36},~185001,
  \href{http://xxx.lanl.gov/abs/arXiv:1812.02117}{{\normalfont
  [arXiv:1812.02117]}}.

\end{thebibliography}

%%%%%%%%%%%%%%%%%%%%%%%%%%%%%%%%%%%%%%%%%%
\end{document}